\documentclass[twocolumn,times]{aastex63}

\usepackage[utf8]{inputenc}
\usepackage{multirow}
\usepackage{amsmath}
\usepackage{newtxmath}
\usepackage{csquotes}

\usepackage{siunitx}
\newcommand*{\hourang}[2][]{%
    \ang[
        math-degree=\textsuperscript{h},
        text-degree=\textsuperscript{h},
        math-arcminute=\textsuperscript{m},
        text-arcminute=\textsuperscript{m},
        math-arcsecond=\textsuperscript{s},
        text-arcsecond=\textsuperscript{s},
        #1]{#2}%
}

\newcommand{\figref}[1]{Figure~\ref{#1}}
\newcommand{\secref}[1]{Section~\ref{#1}}
\newcommand{\tabref}[1]{Table~\ref{#1}}

\newcommand{\circa}{\ensuremath{{\sim}\mspace{2mu}}}
\newcommand{\ut}{UT}
\newcommand{\anutwometer}{ANU\thinspace\SI{2.3}{\meter}}
\newcommand{\chiron}{CHIRON}
\newcommand{\coralie}{CORALIE}
\newcommand{\feros}{FEROS}
\newcommand{\gaia}{Gaia}
\newcommand{\harps}{HARPS}
\newcommand{\hatsouth}{HATSouth}
\newcommand{\kepler}{Kepler}
\newcommand{\ktwo}{K2}
\newcommand{\lcogt}{LCOGT}
\newcommand{\minervaaus}{\textsc{Minerva}-Australis}

\newcommand{\pest}{PEST}
\newcommand{\pfs}{PFS}
\newcommand{\soar}{SOAR}
\newcommand{\tess}{TESS}
\newcommand{\tres}{TRES}
\newcommand{\fies}{FIES}

\newcommand{\planet}[2]{#1\thinspace #2}
\newcommand{\starone}{TOI-954}
\newcommand{\staronetic}{44792534}
\newcommand{\planetone}{\planet{\starone}{b}}
\newcommand{\startwoepic}{246193072}
\newcommand{\startwo}{K2-329}
\newcommand{\planettwo}{\planet{\startwo}{b}}
\newcommand{\toionenineseven}{HD\thinspace 221416}
\newcommand{\toioneninesevenb}{\planet{\toionenineseven}{b}}

\newcommand{\keltsixb}{\planet{KELT-6}{b}}
\DeclareSIPostPower{\nominal}{N}
\DeclareSIQualifier{\earth}{\ensuremath{\oplus}}
\DeclareSIQualifier{\jupiter}{J}
\DeclareSIQualifier{\planet}{p}
\DeclareSIQualifier{\etoile}{\ensuremath{\star}}
\DeclareSIQualifier{\sun}{\ensuremath{\odot}}
\DeclareSIUnit{\density}{\ensuremath{\mathnormal{\rho}}}
\DeclareSIUnit{\erg}{erg}
\DeclareSIUnit{\luminosity}{\ensuremath{\mathnormal{L}}}
\DeclareSIUnit{\magnitude}{mag}
\DeclareSIUnit{\mass}{\ensuremath{\mathnormal{M}}}
\DeclareSIUnit{\mas}{\milliarcsecond}
\DeclareSIUnit{\milliarcsecond}{mas}
\DeclareSIUnit{\parsec}{pc}
\DeclareSIUnit{\radius}{\ensuremath{\mathnormal{R}}}
\DeclareSIUnit{\year}{yr}
\newcommand{\bjdtdb}{\ensuremath{\mathrm{BJD}_\mathrm{TDB}}}
\newcommand{\feh}{\ensuremath{[\mathrm{Fe/H}]}}
\newcommand{\logg}{\ensuremath{\log g}}
\newcommand{\qmodplanet}{\ensuremath{Q'_{\mathrm{p}}}}
\newcommand{\tauep}{\ensuremath{\tau_{\mathrm{ep}}}}
\newcommand{\tempeff}{\ensuremath{T_{\mathrm{eff}}}}
\newcommand{\tempeq}{\ensuremath{T_{\mathrm{eq}}}}
\newcommand{\vsini}{\ensuremath{v \sin i}}
\newcommand{\gaiaG}{\ensuremath{G}}
\newcommand{\gaiaBP}{\ensuremath{G_\mathrm{BP}}}
\newcommand{\gaiaRP}{\ensuremath{G_\mathrm{RP}}}
\newcommand{\filterRc}{\ensuremath{R_\mathrm{c}}}
\newcommand{\filterzs}{\ensuremath{z_\mathrm{s}}}
\newcommand{\filterrp}{\ensuremath{r'}}
\newcommand{\astroimagej}{\textsc{AstroImageJ}}
\newcommand{\batman}{\textsc{Batman}}
\newcommand{\emcee}{\textsc{Emcee}}
\newcommand{\isochrones}{\textsc{Isochrones}}
\newcommand{\mist}{MIST}

\newcommand{\radvel}{\textsc{Radvel}}
\newcommand{\sklearn}{\textsc{Scikit-learn}}

\newcommand{\ToipPeriodRaw}{3.68}
\newcommand{\ToipMpMjRaw}{0.174_{-0.017}^{+0.018}}
\newcommand{\ToipRpRjRaw}{0.852_{-0.062}^{+0.053}}
\newcommand{\ToipEccUpper}{0.37}

\newcommand{\ToipPeriod}{$3.6849729$}
\newcommand{\ToipPeriodErr}{$_{-0.0000028}^{+0.0000027}$}
\newcommand{\ToipK}{$20.7$}
\newcommand{\ToipKErr}{$_{-1.9}^{+2.0}$}
\newcommand{\ToipSecosw}{$0.07$}
\newcommand{\ToipSecoswErr}{$_{-0.13}^{+0.06}$}
\newcommand{\ToipSesinw}{$-0.35$}
\newcommand{\ToipSesinwErr}{$_{-0.20}^{+0.49}$}
\newcommand{\ToipB}{$0.56$}
\newcommand{\ToipBErr}{$_{-0.24}^{+0.04}$}
\newcommand{\ToipRp}{$0.0462$}
\newcommand{\ToipRpErr}{$_{-0.0023}^{+0.0019}$}
\newcommand{\ToipAge}{$6.14$}
\newcommand{\ToipAgeErr}{$_{-0.30}^{+0.30}$}
\newcommand{\ToipDiluteTess}{0.0363}
\newcommand{\ToipDiluteTessErr}{_{-0.0045}^{+0.0046}}
\newcommand{\ToipDiluteHats}{(0.9}
\newcommand{\ToipDiluteHatsErr}{_{-0.7}^{+1.5}) \times 10^{-5}}
\newcommand{\ToipErrLcogtinflate}{1.809}
\newcommand{\ToipErrLcogtinflateErr}{_{-0.049}^{+0.052}}
\newcommand{\ToipGammaChiron}{$-8826.0$}
\newcommand{\ToipGammaChironErr}{$\pm 7.3$}
\newcommand{\ToipGammaCoralie}{$-7347.2$}
\newcommand{\ToipGammaCoralieErr}{$_{-2.9}^{+3.0}$}
\newcommand{\ToipGammaHarps}{$-7321.4$}
\newcommand{\ToipGammaHarpsErr}{$\pm 2.1$}
\newcommand{\ToipGammaPfs}{$-14.3$}
\newcommand{\ToipGammaPfsErr}{$_{-2.6}^{+2.9}$}
\newcommand{\ToipGammaMinerva}{$8.2$}
\newcommand{\ToipGammaMinervaErr}{$_{-4.8}^{+5.5}$}
\newcommand{\ToipJittergaia}{$0.026$}
\newcommand{\ToipJittergaiaErr}{$_{-0.019}^{+0.052}$}
\newcommand{\ToipMstar}{$1.201$}
\newcommand{\ToipMstarErr}{$_{-0.064}^{+0.066}$}
\newcommand{\ToipTeff}{$5710$}
\newcommand{\ToipTeffErr}{$_{-49}^{+53}$}
\newcommand{\ToipLogg}{$3.962$}
\newcommand{\ToipLoggErr}{$_{-0.024}^{+0.026}$}
\newcommand{\ToipFeh}{$0.215$}
\newcommand{\ToipFehErr}{$_{-0.051}^{+0.050}$}

\newcommand{\ToipJitterchiron}{$6.6$}
\newcommand{\ToipJitterchironErr}{$_{-4.7}^{+9.7}$}
\newcommand{\ToipJittercoralie}{$4.9$}
\newcommand{\ToipJittercoralieErr}{$_{-3.3}^{+4.2}$}
\newcommand{\ToipJitterharps}{$2.5$}
\newcommand{\ToipJitterharpsErr}{$_{-1.8}^{+4.3}$}
\newcommand{\ToipJitterpfs}{$6.8$}
\newcommand{\ToipJitterpfsErr}{$_{-2.2}^{+3.5}$}
\newcommand{\ToipJitterminerva}{$16.0$}
\newcommand{\ToipJitterminervaErr}{$_{-4.0}^{+5.7}$}
\newcommand{\ToipUone}{$0.25$}
\newcommand{\ToipUoneErr}{$_{-0.14}^{+0.16}$}
\newcommand{\ToipVone}{$0.29$}
\newcommand{\ToipVoneErr}{$_{-0.19}^{+0.20}$}
\newcommand{\ToipUtwo}{$0.483$}
\newcommand{\ToipUtwoErr}{$_{-0.034}^{+0.033}$}
\newcommand{\ToipVtwo}{$0.163$}
\newcommand{\ToipVtwoErr}{$_{-0.051}^{+0.053}$}
\newcommand{\ToipUthree}{$0.32$}
\newcommand{\ToipUthreeErr}{$\pm 0.31$}
\newcommand{\ToipVthree}{$0.199$}
\newcommand{\ToipVthreeErr}{$_{-0.058}^{+0.057}$}
\newcommand{\ToipRstar}{$1.892$}
\newcommand{\ToipRstarErr}{$_{-0.076}^{+0.077}$}
\newcommand{\ToipRhostar}{$0.249$}
\newcommand{\ToipRhostarErr}{$_{-0.021}^{+0.025}$}
\newcommand{\ToipLstar}{$3.43$}
\newcommand{\ToipLstarErr}{$_{-0.23}^{+0.24}$}
\newcommand{\ToipSemimajor}{$0.04963$}
\newcommand{\ToipSemimajorErr}{$_{-0.00090}^{+0.00089}$}
\newcommand{\ToipAor}{$5.63$}
\newcommand{\ToipAorErr}{$_{-0.17}^{+0.18}$}
\newcommand{\ToipTcjd}{$2458411.90651$}
\newcommand{\ToipTcjdErr}{$_{-0.00078}^{+0.00084}$}
\newcommand{\ToipTdur}{$5.09$}
\newcommand{\ToipTdurErr}{$_{-0.37}^{+0.93}$}
\newcommand{\ToipEcc}{$0.14$}
\newcommand{\ToipEccErr}{$_{-0.11}^{+0.18}$}
\newcommand{\ToipOmega}{$276$}
\newcommand{\ToipOmegaErr}{$_{-174}^{+9}$}
\newcommand{\ToipRpRj}{$0.852$}
\newcommand{\ToipRpRjErr}{$_{-0.062}^{+0.053}$}
\newcommand{\ToipMpSiniMj}{$0.174$}
\newcommand{\ToipMpSiniMjErr}{$_{-0.017}^{+0.018}$}
\newcommand{\ToipInc}{$84.4$}
\newcommand{\ToipIncErr}{$_{-0.6}^{+2.5}$}
\newcommand{\ToipMpMj}{$0.174$}
\newcommand{\ToipMpMjErr}{$_{-0.017}^{+0.018}$}
\newcommand{\ToipRhop}{$0.35$}
\newcommand{\ToipRhopErr}{$_{-0.07}^{+0.10}$}
\newcommand{\ToipIrrad}{$(1.896$}
\newcommand{\ToipIrradErr}{$_{-0.085}^{+0.090}) \times 10^9$}
\newcommand{\ToipTempeq}{$1526$}
\newcommand{\ToipTempeqErr}{$_{-164}^{+123}$}

\newcommand{\KtwopEccUpper}{$0.140$}
\newcommand{\KtwopPeriod}{$12.4551225$}
\newcommand{\KtwopPeriodErr}{$\pm 0.0000031$}
\newcommand{\KtwopPeriodRaw}{12.46}
\newcommand{\KtwopK}{$24.6$}
\newcommand{\KtwopKErr}{$_{-1.8}^{+1.6}$}
\newcommand{\KtwopSecosw}{$-0.23$}
\newcommand{\KtwopSecoswErr}{$_{-0.08}^{+0.13}$}
\newcommand{\KtwopSesinw}{$0.081$}
\newcommand{\KtwopSesinwErr}{$_{-0.094}^{+0.089}$}
\newcommand{\KtwopB}{$0.364$}
\newcommand{\KtwopBErr}{$_{-0.048}^{+0.039}$}
\newcommand{\KtwopRp}{$0.09679$}
\newcommand{\KtwopRpErr}{$_{-0.00057}^{+0.00056}$}
\newcommand{\KtwopAge}{$1.8$}
\newcommand{\KtwopAgeErr}{$_{-1.3}^{+2.2}$}
\newcommand{\KtwopGammaFeros}{$-16999.0$}
\newcommand{\KtwopGammaFerosErr}{$_{-4.2}^{+4.1}$}
\newcommand{\KtwopGammaHarps}{$-16982.9$}
\newcommand{\KtwopGammaHarpsErr}{$\pm 2.8$}
\newcommand{\KtwopGammaPfs}{$0.8$}
\newcommand{\KtwopGammaPfsErr}{$\pm 1.4$}
\newcommand{\KtwopJittergaia}{$0.039$}
\newcommand{\KtwopJittergaiaErr}{$_{-0.029}^{+0.080}$}
\newcommand{\KtwopMstar}{$0.901$}
\newcommand{\KtwopMstarErr}{$_{-0.049}^{+0.048}$}
\newcommand{\KtwopTeff}{$5282$}
\newcommand{\KtwopTeffErr}{$_{-39}^{+40}$}
\newcommand{\KtwopLogg}{$4.566$}
\newcommand{\KtwopLoggErr}{$_{-0.021}^{+0.013}$}
\newcommand{\KtwopFeh}{$0.098$}
\newcommand{\KtwopFehErr}{$_{-0.070}^{+0.065}$}
\newcommand{\KtwopJitterferos}{$17.0$}
\newcommand{\KtwopJitterferosErr}{$_{-2.8}^{+3.4}$}
\newcommand{\KtwopJitterharps}{$5.0$}
\newcommand{\KtwopJitterharpsErr}{$_{-3.1}^{+3.8}$}
\newcommand{\KtwopJitterpfs}{$3.7$}
\newcommand{\KtwopJitterpfsErr}{$_{-1.2}^{+1.7}$}
\newcommand{\KtwopJitterPest}{0.00232 \pm 0.00037}
\newcommand{\KtwopUone}{$0.535$}
\newcommand{\KtwopUoneErr}{$\pm 0.022$}
\newcommand{\KtwopVone}{$-0.002$}
\newcommand{\KtwopVoneErr}{$_{-0.045}^{+0.044}$}
\newcommand{\KtwopUtwo}{$0.542$}
\newcommand{\KtwopUtwoErr}{$_{-0.033}^{+0.033}$}
\newcommand{\KtwopVtwo}{$0.185$}
\newcommand{\KtwopVtwoErr}{$_{-0.059}^{+0.057}$}
\newcommand{\KtwopRstar}{$0.822$}
\newcommand{\KtwopRstarErr}{$_{-0.024}^{+0.026}$}
\newcommand{\KtwopRhostar}{$2.30$}
\newcommand{\KtwopRhostarErr}{$_{-0.16}^{+0.12}$}
\newcommand{\KtwopLstar}{$0.473$}
\newcommand{\KtwopLstarErr}{$_{-0.028}^{+0.033}$}
\newcommand{\KtwopSemimajor}{$0.1016$}
\newcommand{\KtwopSemimajorErr}{$_{-0.0019}^{+0.0018}$}
\newcommand{\KtwopAor}{$26.62$}
\newcommand{\KtwopAorErr}{$_{-0.61}^{+0.46}$}
\newcommand{\KtwopTcjd}{$2457773.157267$}
\newcommand{\KtwopTcjdErr}{$_{-0.000075}^{+0.000073}$}
\newcommand{\KtwopTdur}{$3.621$}
\newcommand{\KtwopTdurErr}{$_{-0.017}^{+0.019}$}
\newcommand{\KtwopEcc}{$0.0697$}
\newcommand{\KtwopEccErr}{$_{-0.040}^{+0.041}$}
\newcommand{\KtwopOmega}{$161$}
\newcommand{\KtwopOmegaErr}{$_{-30}^{+22}$}
\newcommand{\KtwopRpRj}{$0.774$}
\newcommand{\KtwopRpRjErr}{$_{-0.024}^{+0.026}$}
\newcommand{\KtwopRpRjRaw}{0.774_{-0.024}^{+0.026}}
\newcommand{\KtwopMpSiniMj}{$0.260$}
\newcommand{\KtwopMpSiniMjErr}{$_{-0.022}^{+0.020}$}
\newcommand{\KtwopInc}{$89.22$}
\newcommand{\KtwopIncErr}{$_{-0.09}^{+0.11}$}
\newcommand{\KtwopMpMj}{$0.260$}
\newcommand{\KtwopMpMjErr}{$_{-0.022}^{+0.020}$}
\newcommand{\KtwopMpMjRaw}{0.260_{-0.022}^{+0.020}}
\newcommand{\KtwopRhop}{$0.694$}
\newcommand{\KtwopRhopErr}{$_{-0.072}^{+0.070}$}
\newcommand{\KtwopIrrad}{$(6.22$}
\newcommand{\KtwopIrradErr}{$_{-0.22}^{+0.30}) \times 10^7$}
\newcommand{\KtwopTempeq}{$650$}
\newcommand{\KtwopTempeqErr}{$_{-70}^{+53}$}

\shorttitle{\planetone{} and \planettwo{}}
\shortauthors{Sha et al.}

\begin{document}

\title{%
\planetone{} and \planettwo{}: Short-Period Saturn-Mass Planets that Test
whether Irradiation Leads to Inflation}

\submitjournal{\aj}
\received{2020 January 15}
\revised{2020 October 27}
\accepted{2020 November 11}
\correspondingauthor{Lizhou Sha}
\email{slz@mit.edu}

\author[0000-0001-5401-8079]{Lizhou Sha}

\author[0000-0003-0918-7484]{Chelsea X.~Huang}
\altaffiliation{Juan Carlos Torres Fellow}

\author[0000-0001-5401-8079]{Avi Shporer}
\affiliation{Department of Physics and Kavli Institute for Astrophysics and Space Research, Massachusetts Institute of Technology, 77 Massachusetts Avenue, Cambridge, MA 02139, USA}

\author[0000-0001-8812-0565]{Joseph E.~Rodriguez}
\affiliation{Center for Astrophysics \textbar \ Harvard \& Smithsonian, 60 Garden Street, Cambridge, MA 02138, USA}

\author[0000-0001-7246-5438]{Andrew Vanderburg}
\altaffiliation{NASA Sagan Fellow}
\affiliation{Department of Astronomy, The University of Texas at Austin, Austin, TX 78705, USA}

\author[0000-0002-9158-7315]{Rafael Brahm} 
\affiliation{Facultad de Ingenier\'ia y Ciencias, Universidad Adolfo Ib\'a\~{n}ez, Av.\ Diagonal las Torres 2640, Pe\~{n}alol\'en, Santiago, Chile}
\affiliation{Millennium Institute for Astrophysics, Chile}
\author[0000-0002-1096-1433]{Janis Hagelberg}
\altaffiliation{SNSF Ambizione Fellow}
\affiliation{Geneva Observatory, University of Geneva, Chemin des Mailettes 51, 1290 Versoix, Switzerland}
\author[0000-0003-0593-1560]{Elisabeth C. Matthews}
\affiliation{Department of Physics and Kavli Institute for Astrophysics and Space Research, Massachusetts Institute of Technology, 77 Massachusetts Avenue, Cambridge, MA 02139, USA}
\author{Carl Ziegler}
\affil{Dunlap Institute for Astronomy and Astrophysics, University of Toronto, 50 St.\ George Street, Toronto, ON M5S 3H4, Canada}
\author[0000-0002-4881-3620]{John H. Livingston}
\affiliation{Department of Astronomy, University of Tokyo, 7-3-1 Hongo, Bunkyo-ku, Tokyo 113-0033, Japan}
\author[0000-0002-3481-9052]{Keivan G. Stassun}
\affiliation{Department of Physics \& Astronomy, Vanderbilt University, Nashville, TN 37235, USA}
\author[0000-0001-7294-5386]{Duncan J. Wright}
\affiliation{University of Southern Queensland, Centre for Astrophysics, West Street, Toowoomba, QLD 4350, Australia}
\author[0000-0002-5226-787X]{Jeffrey D. Crane}
\affiliation{The Observatories of the Carnegie Institution for Science, 813 Santa Barbara Street, Pasadena, CA, 91101, USA}
\author{Néstor Espinoza} 
\affiliation{Space Telescope Science Institute, 3700 San Martin Drive, Baltimore, MD 21218, USA}
\author{François Bouchy}
\affiliation{Geneva Observatory, University of Geneva, Chemin des Mailettes 51, 1290 Versoix, Switzerland}
\author[0000-0001-7204-6727]{Gáspár Á. Bakos}
\altaffiliation{Packard Fellow}
\affiliation{Department of Astrophysical Sciences, Princeton University, NJ 08544, USA}
\affiliation{MTA Distinguished Guest Fellow, Konkoly Observatory, Hungary}
\author[0000-0001-6588-9574]{Karen A.\ Collins}
\affiliation{Center for Astrophysics \textbar \ Harvard \& Smithsonian, 60 Garden Street, Cambridge, MA 02138, USA}
\author[0000-0002-4891-3517]{George Zhou}
\affiliation{Center for Astrophysics \textbar \ Harvard \& Smithsonian, 60 Garden Street, Cambridge, MA 02138, USA}
\author[0000-0001-6637-5401]{Allyson Bieryla}
\affiliation{Center for Astrophysics \textbar \ Harvard \& Smithsonian, 60 Garden Street, Cambridge, MA 02138, USA}
\author[0000-0001-8732-6166]{Joel D. Hartman}
\affil{Department of Astrophysical Sciences, Princeton University, NJ 08544, USA}
\author[0000-0001-9957-9304]{Robert A. Wittenmyer}
\affiliation{University of Southern Queensland, Centre for Astrophysics, West Street, Toowoomba, QLD 4350, Australia}
\author[0000-0002-5254-2499]{Louise D. Nielsen}
\affiliation{Geneva Observatory, University of Geneva, Chemin des Mailettes 51, 1290 Versoix, Switzerland}
\author[0000-0002-8864-1667]{Peter Plavchan}
\affiliation{George Mason University, 4400 University Drive MS 3F3, Fairfax, VA 22030, USA}
\author[0000-0001-6023-1335]{Daniel Bayliss}
\affiliation{Department of Physics, University of Warwick, Gibbet Hill Road, Coventry CV4 7AL, UK}
\author{Paula Sarkis} 
\affiliation{Max-Planck-Institut für Astronomie, Königstuhl 17, Heidelberg 69117, Germany }
\author[0000-0001-5603-6895]{Thiam-Guan Tan}
\affiliation{Perth Exoplanet Survey Telescope, Perth, Australia}
\author[0000-0001-5383-9393]{Ryan Cloutier}
\affiliation{Center for Astrophysics \textbar \ Harvard \& Smithsonian, 60 Garden Street, Cambridge, MA  02138, USA}
\author[0000-0002-9428-8732]{Luigi Mancini}
\affiliation{Department of Physics, University of Rome ``Tor Vergata,'' Via della Ricerca Scientifica 1, Roma I-00133, Italy}
\affiliation{Max-Planck-Institut für Astronomie, Königstuhl 17, Heidelberg 69117, Germany }
\affiliation{INAF -- Astrophysical Observatory of Turin, Via Osservatorio 20, Pino Torinese I-10025, Italy}
\author[0000-0002-5389-3944]{Andrés Jordán}
\affiliation{Facultad de Ingenier\'ia y Ciencias, Universidad Adolfo Ib\'a\~{n}ez, Av.\ Diagonal las Torres 2640, Pe\~{n}alol\'en, Santiago, Chile}
\affiliation{Millennium Institute for Astrophysics, Chile}
\author[0000-0002-6937-9034]{Sharon Wang}
\affiliation{The Observatories of the Carnegie Institution for Science, 813 Santa Barbara Street, Pasadena, CA, 91101, USA}
\author{Thomas Henning} 
\affiliation{Max-Planck-Institut für Astronomie, Königstuhl 17, Heidelberg 69117, Germany }
\author[0000-0001-8511-2981]{Norio Narita}
\affiliation{Komaba Institute for Science, The University of Tokyo, 3-8-1 Komaba, Meguro, Tokyo 153-8902, Japan}
\affiliation{JST, PRESTO, 3-8-1 Komaba, Meguro, Tokyo 153-8902, Japan}
\affiliation{Astrobiology Center, 2-21-1 Osawa, Mitaka, Tokyo 181-8588, Japan}
\affiliation{National Astronomical Observatory of Japan, 2-21-1 Osawa, Mitaka, Tokyo 181-8588, Japan}
\affiliation{Instituto de Astrof\'\i sica de Canarias, C. V\'\i a L\'actea S/N, E-38205 La Laguna, Tenerife, Spain}
\author[0000-0003-4464-1371]{Kaloyan Penev}
\affiliation{Department of Physics, University of Texas at Dallas, Richardson, TX 75080, USA}
\author{Johanna K. Teske}
\altaffiliation{Hubble Fellow}
\affiliation{The Observatories of the Carnegie Institution for Science, 813 Santa Barbara Street, Pasadena, CA, 91101, USA}
\author[0000-0002-7084-0529]{Stephen R. Kane}
\affiliation{Department of Earth and Planetary Sciences, University of California, Riverside, CA 92521, USA}
\author[0000-0003-3654-1602]{Andrew W. Mann}
\affiliation{Department of Physics and Astronomy, The University of North Carolina at Chapel Hill, Chapel Hill, NC 27599-3255, USA}
\author[0000-0003-3216-0626]{Brett C. Addison}
\affiliation{University of Southern Queensland, Centre for Astrophysics, West Street, Toowoomba, QLD 4350, Australia}
\author[0000-0002-6510-0681]{Motohide Tamura}
\affiliation{Department of Astronomy, University of Tokyo, 7-3-1 Hongo, Bunkyo-ku, Tokyo 113-0033, Japan}
\affiliation{Astrobiology Center, 2-21-1 Osawa, Mitaka, Tokyo 181-8588, Japan}
\affiliation{National Astronomical Observatory of Japan, 2-21-1 Osawa, Mitaka, Tokyo 181-8588, Japan}
\author[0000-0002-1160-7970]{Jonathan Horner}
\affiliation{University of Southern Queensland, Centre for Astrophysics, West Street, Toowoomba, QLD 4350, Australia}

\author[0000-0001-8362-3462]{Mauro Barbieri}
\affiliation{INCT, Universidad de Atacama, Calle Copayapu 485, Copiapó, Atacama, Chile}

\author[0000-0002-0040-6815]{Jennifer A. Burt}
\affiliation{Jet Propulsion Laboratory, California Institute of Technology, 4800 Oak Grove Drive, Pasadena, CA 91109, USA}
\author[0000-0002-2100-3257]{Mat\'ias R. D\'iaz}
\affiliation{Departamento de Astronom\'ia, Universidad de Chile, Camino El Observatorio 1515, Las Condes, Santiago, Chile}
\author{Ian J. M. Crossfield}
\affiliation{Department of Physics and Astronomy, University of Kansas, 1251 Wescoe Hall Dr., Lawrence, KS 66045, USA}
\affiliation{Department of Physics and Kavli Institute for Astrophysics and Space Research, Massachusetts Institute of Technology, 77 Massachusetts Avenue, Cambridge, MA 02139, USA}
\author[0000-0003-2313-467X]{Diana Dragomir}
\affiliation{Department of Physics and Astronomy, University of New Mexico, 1919 Lomas Blvd NE, Albuquerque, NM 87131, USA}
\author{Holger Drass} 
\affiliation{Center of Astro-Engineering UC, Pontificia Universidad Católica de Chile, Av.\ Vicuña Mackenna 4860, Macul, Santiago, Chile}
\author[0000-0002-9464-8101]{Adina~D.~Feinstein}
\altaffiliation{NSF Fellow}
\affil{Department of Astronomy and Astrophysics, University of
Chicago, 5640 S. Ellis Ave, Chicago, IL 60637, USA}
\author{Hui Zhang}
\affiliation{School of Astronomy and Space Science, Key Laboratory of Modern Astronomy and Astrophysics in Ministry of Education, Nanjing University, Nanjing 210046, Jiangsu, China}
\author{Rhodes Hart}
\affiliation{University of Southern Queensland, Centre for Astrophysics, West Street, Toowoomba, QLD 4350, Australia}
\author[0000-0003-0497-2651]{John F.\ Kielkopf}
\affiliation{Department of Physics and Astronomy, University of Louisville, Louisville, KY 40292, USA}
\author[0000-0002-4625-7333]{Eric L. N. Jensen}
\affiliation{Department of Physics \& Astronomy, Swarthmore College, Swarthmore, PA 19081, USA}
\author[0000-0001-7516-8308]{Benjamin T. Montet}
\affiliation{School of Physics, University of New South Wales, Sydney, NSW 2052, Australia}
\author{Gaël Ottoni} 
\affiliation{Geneva Observatory, University of Geneva, Chemin des Mailettes 51, 1290 Versoix, Switzerland}
\author[0000-0001-8227-1020]{Richard P. Schwarz}
\affiliation{Patashnick Voorheesville Observatory, Voorheesville, NY 12186, USA}
\author{Felipe Rojas}
\affiliation{Instituto de Astrof\'isica, Pontificia Universidad Cat\'olica de Chile, Av.\ Vicu\~na Mackenna 4860, Macul, Santiago, Chile}
\affiliation{Millennium Institute for Astrophysics, Chile}
\author{David Nespral}  %
\affiliation{Instituto de Astrof\'\i sica de Canarias, C. V\'\i a L\'actea S/N, E-38205 La Laguna, Tenerife, Spain}
\affiliation{Universidad de La Laguna, Dept.\ de Astrof\'\i sica, E-38206 La Laguna, Tenerife, Spain}
\author{Pascal Torres} 
\affiliation{Instituto de Astrof\'isica, Pontificia Universidad Cat\'olica de Chile, Av.\ Vicu\~na Mackenna 4860, Macul, Santiago, Chile}
\author[0000-0002-7830-6822]{Matthew W. Mengel}
\affiliation{University of Southern Queensland, Centre for Astrophysics, West Street, Toowoomba, QLD 4350, Australia}
\author{Stéphane Udry} 
\affiliation{Geneva Observatory, University of Geneva, Chemin des Mailettes 51, 1290 Versoix, Switzerland}
\author[0000-0003-2326-6488]{Abner Zapata}
\affiliation{Center of Astro-Engineering UC, Pontificia Universidad Católica de Chile, Av.\ Vicuña Mackenna 4860, Macul, Santiago, Chile}
\author[0000-0002-5254-7660]{Erin Snoddy}
\affiliation{Department of Physics \& Astronomy, Swarthmore College, Swarthmore, PA 19081, USA}
\author[0000-0002-4876-8540]{Jack Okumura}
\affiliation{University of Southern Queensland, Centre for Astrophysics, West Street, Toowoomba, QLD 4350, Australia}

\author{George R. Ricker}
\affiliation{Department of Physics and Kavli Institute for Astrophysics and Space Research, Massachusetts Institute of Technology, 77 Massachusetts Avenue, Cambridge, MA 02139, USA}

\author[0000-0001-6763-6562]{Roland K. Vanderspek}
\affiliation{Department of Physics and Kavli Institute for Astrophysics and Space Research, Massachusetts Institute of Technology, 77 Massachusetts Avenue, Cambridge, MA 02139, USA}

\author[0000-0001-9911-7388]{David W. Latham}
\affiliation{Center for Astrophysics \textbar \ Harvard \& Smithsonian, 60 Garden Street, Cambridge, MA 02138, USA}

\author[0000-0002-4265-047X]{Joshua N.\ Winn}
\affiliation{Department of Astrophysical Sciences, Princeton University, 4 Ivy Lane, Princeton, NJ 08544, USA}

\author[0000-0002-6892-6948]{Sara Seager}
\affiliation{Department of Physics and Kavli Institute for Astrophysics and Space Research, Massachusetts Institute of Technology, 77 Massachusetts Avenue, Cambridge, MA 02139, USA}
\affiliation{Department of Earth, Atmospheric and Planetary Sciences, Massachusetts Institute of Technology, Cambridge, MA 02139, USA}
\affiliation{Department of Aeronautics and Astronautics, Massachusetts Institute of Technology, Cambridge, MA 02139, USA}

\author[0000-0002-4715-9460]{Jon M. Jenkins}
\affiliation{NASA Ames Research Center, Moffett Field, CA 94035, USA}

\author[0000-0001-8020-7121]{Knicole D. Col\'{o}n}
\affiliation{NASA Goddard Space Flight Center, Exoplanets and Stellar Astrophysics Laboratory (Code 667), Greenbelt, MD 20771, USA}				

\author{Christopher E. Henze}
\affiliation{NASA Ames Research Center, Moffett Field, CA 94035, USA}

\author[0000-0002-8781-2743]{Akshata Krishnamurthy}
\affiliation{Department of Physics and Kavli Institute for Astrophysics and Space Research, Massachusetts Institute of Technology, 77 Massachusetts Avenue, Cambridge, MA 02139, USA}

\author[0000-0002-8219-9505]{Eric B. Ting}
\affiliation{NASA Ames Research Center, Moffett Field, CA 94035, USA}

\author{Michael Vezie}
\affiliation{Department of Physics and Kavli Institute for Astrophysics and Space Research, Massachusetts Institute of Technology, 77 Massachusetts Avenue, Cambridge, MA 02139, USA}

\author[0000-0001-6213-8804]{Steven Villanueva}
\altaffiliation{Pappalardo Fellow}
\affiliation{Department of Physics and Kavli Institute for Astrophysics and Space Research, Massachusetts Institute of Technology, 77 Massachusetts Avenue, Cambridge, MA 02139, USA}

\begin{abstract}
We report the discovery of two short-period Saturn-mass planets, one transiting the G subgiant
\starone{}
    (TIC\thinspace \staronetic{},
    $ V = 10.343 $,
    $ T = 9.78 $)
    observed in \tess{} sectors~4 and 5,
and one transiting the G dwarf \startwo{}
    (\added{EPIC\thinspace \startwoepic{}},
    $ V = 12.70 $,
    $ K = 10.67 $)
    observed in \ktwo{} campaigns~12 and 19.
We confirm and characterize these two planets with a variety of ground-based archival and follow-up observations, including photometry, reconnaissance spectroscopy, precise radial velocity, and high-resolution imaging.
Combining all available data,
we find that
\planetone{} has a radius of $\SI[parse-numbers=false]{\ToipRpRjRaw}{\radius\jupiter}$
and a mass of \SI[parse-numbers=false]{\ToipMpMjRaw}{\mass\jupiter}
and is in a \ToipPeriodRaw{}~day orbit,
while \planettwo{} has a radius of $\SI[parse-numbers=false]{\KtwopRpRjRaw}{\radius\jupiter}$
and a mass of \SI[parse-numbers=false]{\KtwopMpMjRaw}{\mass\jupiter}
and is in a \KtwopPeriodRaw~day orbit.
As \planetone{} is 30 times more irradiated than \planettwo{}
but more or less the same size,
these two planets provide an opportunity to test whether irradiation
leads to inflation of Saturn-mass planets and contribute to future comparative studies
that explore Saturn-mass planets at contrasting points in their lifetimes.
\end{abstract}


\section{Introduction}

Hot Saturns, with masses between
\SIlist{0.1;0.4}{\mass\jupiter}
and periods shorter than 20~days,
are the lower-mass cousins of hot Jupiters
(\SIrange[range-phrase=--]{0.4}{13}{\mass\jupiter}).
Like hot Jupiters,
hot Saturns' relatively large sizes make it possible to detect their transits with small, ground-based telescopes
\citep[e.g.][]{2018AJ....155..112B},
and their short periods and relatively high masses make it possible to measure
their masses with only a handful of radial velocity (RV) observations
\citep[e.g.][]{2017AJ....153..142P}.
Space-based observatories like NASA's Transiting Exoplanet Survey Satellite
\citep[\tess,][]{tess_mission_paper}
and the now-retired \kepler{} \citep{2010Sci...327..977B},
with their high photometric precision and nearly continuous observations,
are especially well suited to discover transiting hot Saturns en masse.

The occurrence rate of hot Saturns appears to be lower than other types of short-period exoplanets. 
\cite{petigura2018} found that hot Saturns are intrinsically rarer 
than both hot Jupiters and hot Neptunes,
even after accounting for selection effects.
This occurrence rate \enquote{valley} may be an indication that
hot Saturns are the 
smallest planets formed via runaway gas accretion.
An in-depth study of their population 
can help us further understand the divergent formation pathways of small planets and gas giants.

\begin{deluxetable*}{llrrcc}
    \tablecaption{Summary of Photometric Observations\label{tab:phot}}
    \tablecolumns{5}
    \tablehead{%
        \colhead{Instrument and Field}
        & \colhead{UT Date(s)}
        & \colhead{No.~Images\tablenotemark{a}}
        & \colhead{Cadence}
        & \colhead{Filter}
        & \colhead{Additional Parameter}\\
        & &
        & \colhead{(s)}
        &
    }
    \startdata
    \cutinhead{\starone{}}
    \tess{} sector~4 camera~2 CCD~2 & 2018 Oct 18--Nov 15     & 732   & 1800 & \tess{} & \multirow{2}{*}{$D = \ToipDiluteTess\ToipDiluteTessErr$\tablenotemark{b}} \\
    \tess{} sector~5 camera~2 CCD~1 & 2018 Nov 15--Dec 11     & 1074  & 1800 & \tess{} & \\
    \hatsouth{}                 & 2014 Sep 9--2015 Mar 6      & 22351 & 370\tablenotemark{c}  & $r$ & $D = \ToipDiluteHats\ToipDiluteHatsErr $\tablenotemark{b} \\
    \lcogt{} SAAO \SI{1}{\m}    & 2019 Nov 3                    & 396   & 30    & \filterzs{} & \multirow{3}{*}{$C = \ToipErrLcogtinflate\ToipErrLcogtinflateErr $\tablenotemark{d}} \\
    \lcogt{} SAAO \SI{1}{\m}   & 2019 Dec 21\tablenotemark{e}  & 299   & 30    & \filterzs{} & \\
    \lcogt{} SAAO \SI{1}{\m}   & 2020 Jan 12                   & 250   & 30    & \filterzs{} & \\
    \cutinhead{\startwo{}}
    \ktwo{} campaign~12 & 2016 Dec 15--2017 Mar 4    & 3314  & 1764 & \kepler{} & \ldots \\
    \ktwo{} campaign~19 & 2018 Aug 30--Sep 26 & 11595 & 59   & \kepler{} & \ldots \\
    \pest{}             & 2017 Nov 14                   & 117   & 120  & \filterRc{} & $ \mathrm{Jitter} = \KtwopJitterPest $\tablenotemark{f} \\
    PvdK $24''$\tablenotemark{e} & 2018 Jul 9                    & 78    & 90   & \filterrp{} & \ldots \\
    \enddata
    \tablenotetext{a}{Excluding frames flagged for instrumental or quality issues and outliers over $4\sigma$ (when out of transit).}
    \tablenotetext{b}{Dilution correction factor applied to the detrended light curve, defined as the ratio of contaminating flux to total flux.}
    \tablenotetext{c}{Multiplicative factor applied to the quoted noise of the LCOGT light curves.}
    \tablenotetext{d}{Estimated by taking the median time elapsed between consecutive exposures and rounding to the nearest \SI{10}{\second}.}
    \tablenotetext{e}{Not used for the global model fitting (\secref{sec:global_model}).}
    \tablenotetext{f}{Jitter is added in quadrature to the reported noise during the global model fitting.}
\end{deluxetable*}

The hot Saturns discovered thus far are in line with a broader trend that associates
the presence of short-period planets and of large planets
with higher host star metallicity
\citep{2016AJ....152..187M,2018PNAS..115..266D,petigura2018}.
In fact, \cite{petigura2018} found that hot Saturns
(roughly corresponding to what they called \enquote{hot sub-Saturns})
have the highest mean stellar metallicity
among all period and size bins.
Virtually no hot Saturn has been found to orbit any metal-poor ($\feh < -0.05$) star,
with the notable exceptions of
\toioneninesevenb{} \citep{2019AJ....157..245H}
and \keltsixb{} \citep{2014AJ....147...39C}.
This evidence suggests some kind of mechanism connected to high stellar metallicity
that leads to these short-period large planets.

Hot Saturns show a wide diversity in mean density,
ranging from \SIrange{0.09}{5}{\gram\per\cm\cubed}.
For planets of a similar size,
hot Saturns tend to have large scatter in mass
\citep{2017AJ....153..142P}.
Because they have lower surface gravities than typical hot Jupiters,
hot Saturns are some of the best targets for transmission spectroscopy observations \citep{wakeford2018}.
It is with this diversity and potential for future characterization in mind
that we search for new transiting hot Saturns.

In this paper, we report the discovery of two hot Saturns, \planetone{} and \planettwo{}.
As they orbit bright stars, we are able to confirm the two planets with precise RV measurements.
While the planets have similar sizes and masses and
both orbit high-metallicity G stars,
the two host stars are at different evolutionary stages;
\starone{} is evolved, while \startwo{} is on the main sequence.
Combining a more luminous host star and a smaller orbital distance,
\planetone{} is 30 times more irradiated than \planettwo{}.
This dramatic contrast allows us to probe a region of the parameter space
where theories on planetary inflation are poorly tested.

Our paper is organized as follows.
In \secref{sec:observation}, we describe the observations leading to the detection and confirmation of the two planets.
In \secref{sec:analysis}, we describe our data analysis procedures
and report our best estimates for the physical and orbital parameters for those systems.
Finally, in \secref{sec:discussion}, we discuss the two new discoveries in the context of other known hot Saturns,
focusing on reinflation and orbital eccentricity.

\section{Observations}
\label{sec:observation}

\subsection{Photometry}

\begin{figure*}
  \centering
  \begin{tabular}{cc}
    \multicolumn{2}{c}{(a) \tess{} Detrended Light Curve} \\
    \multicolumn{2}{c}{%
      \includegraphics[width=\textwidth]{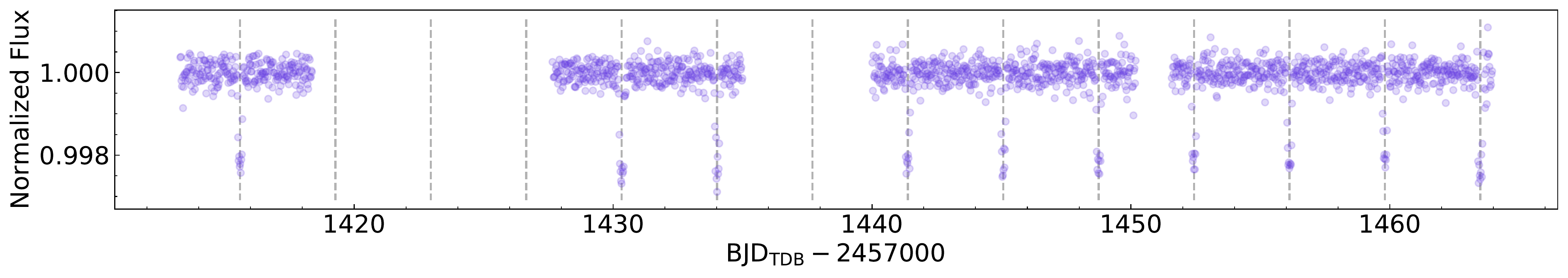}
    } \\
    (b) \tess{} Phase-Folded Light Curve
    &
    (d) \lcogt{} Detrended Light Curves \\
    \includegraphics[height=0.95\columnwidth]{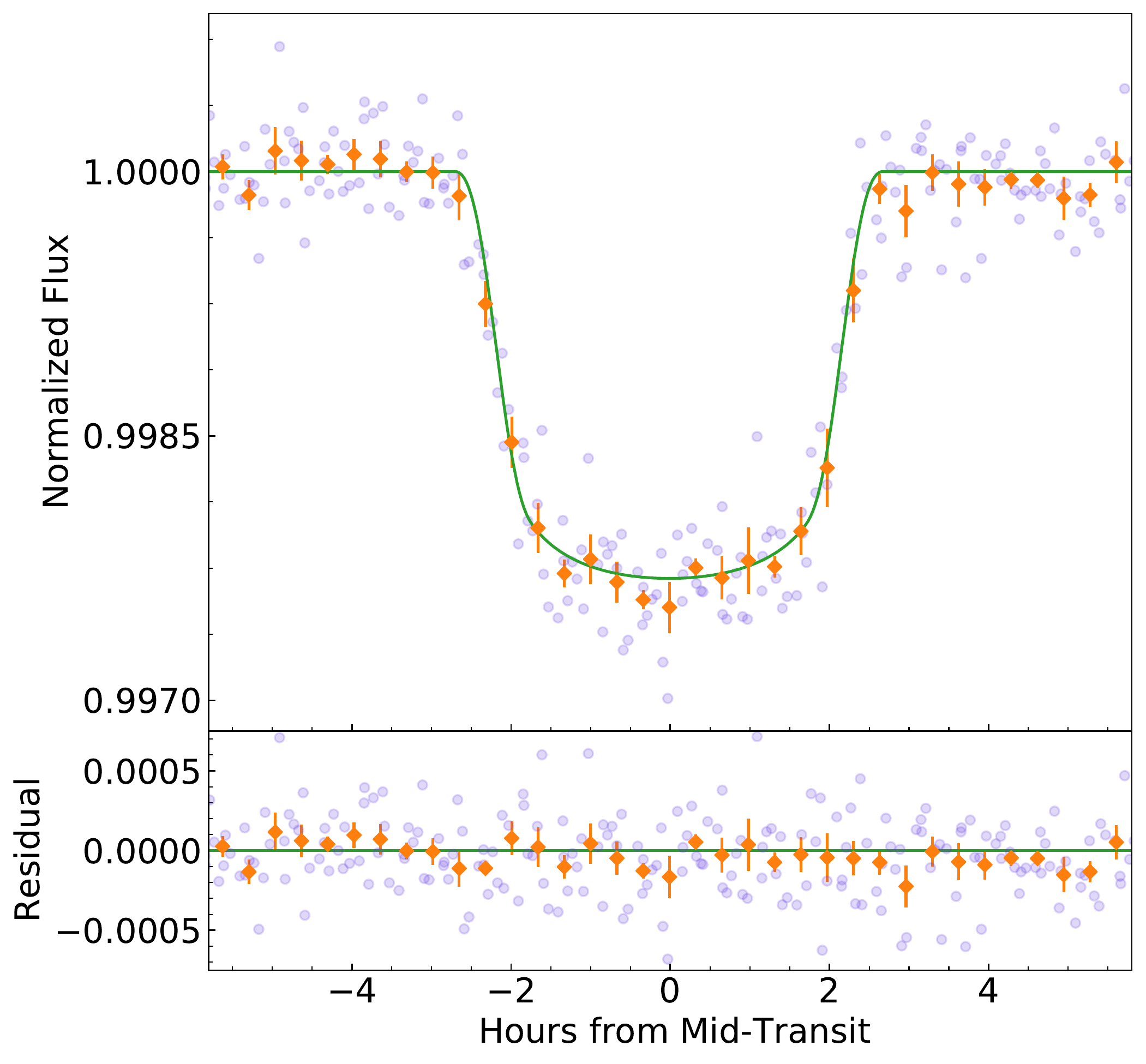}
    &
    \multirow{3}{*}[0.92\columnwidth]{%
      \hfill
      \includegraphics[width=\columnwidth]{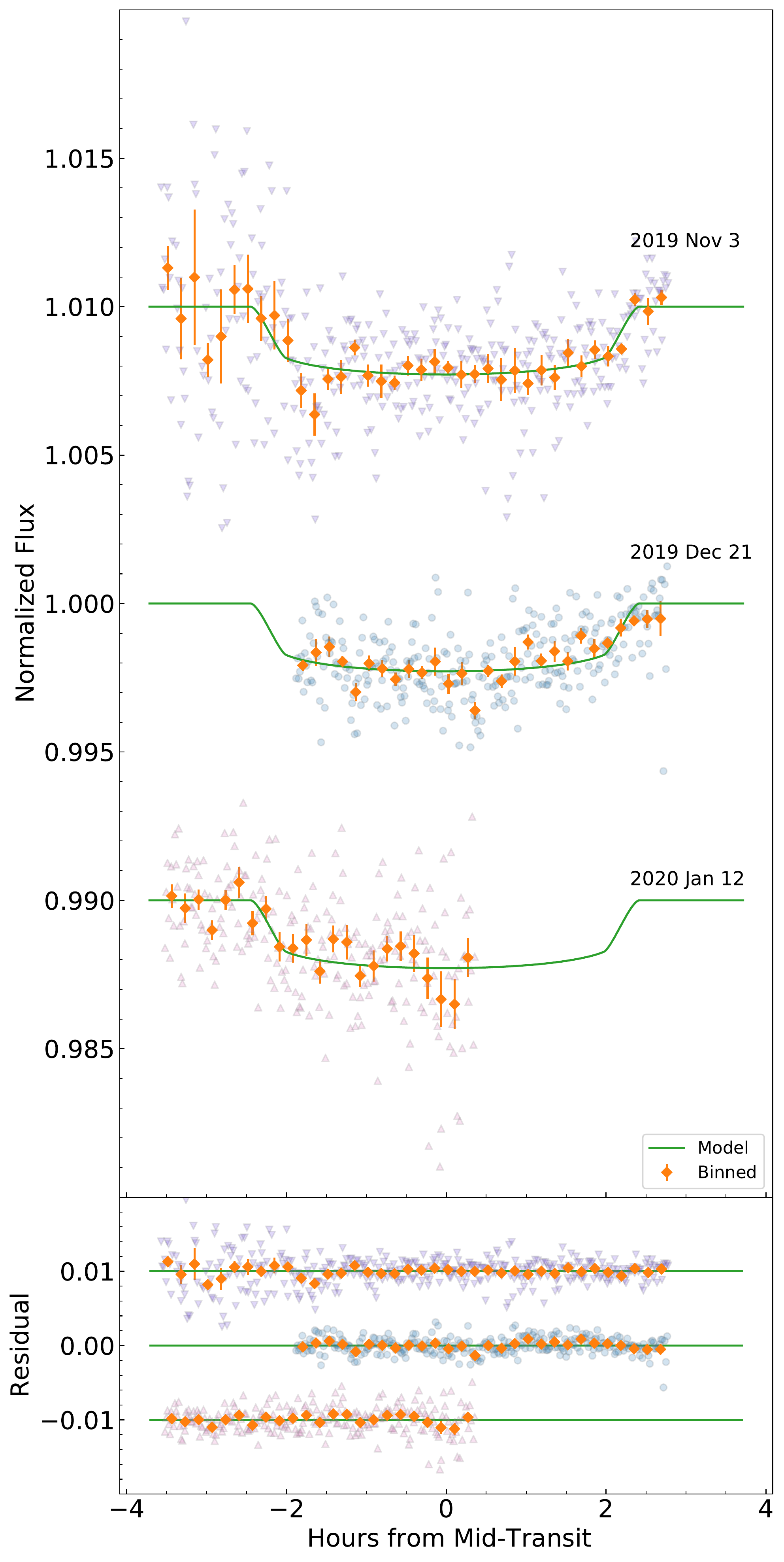}
    } \\
    (c) \hatsouth{} Phase-Folded Light Curve \\
    \hfill
    \includegraphics[width=\columnwidth]{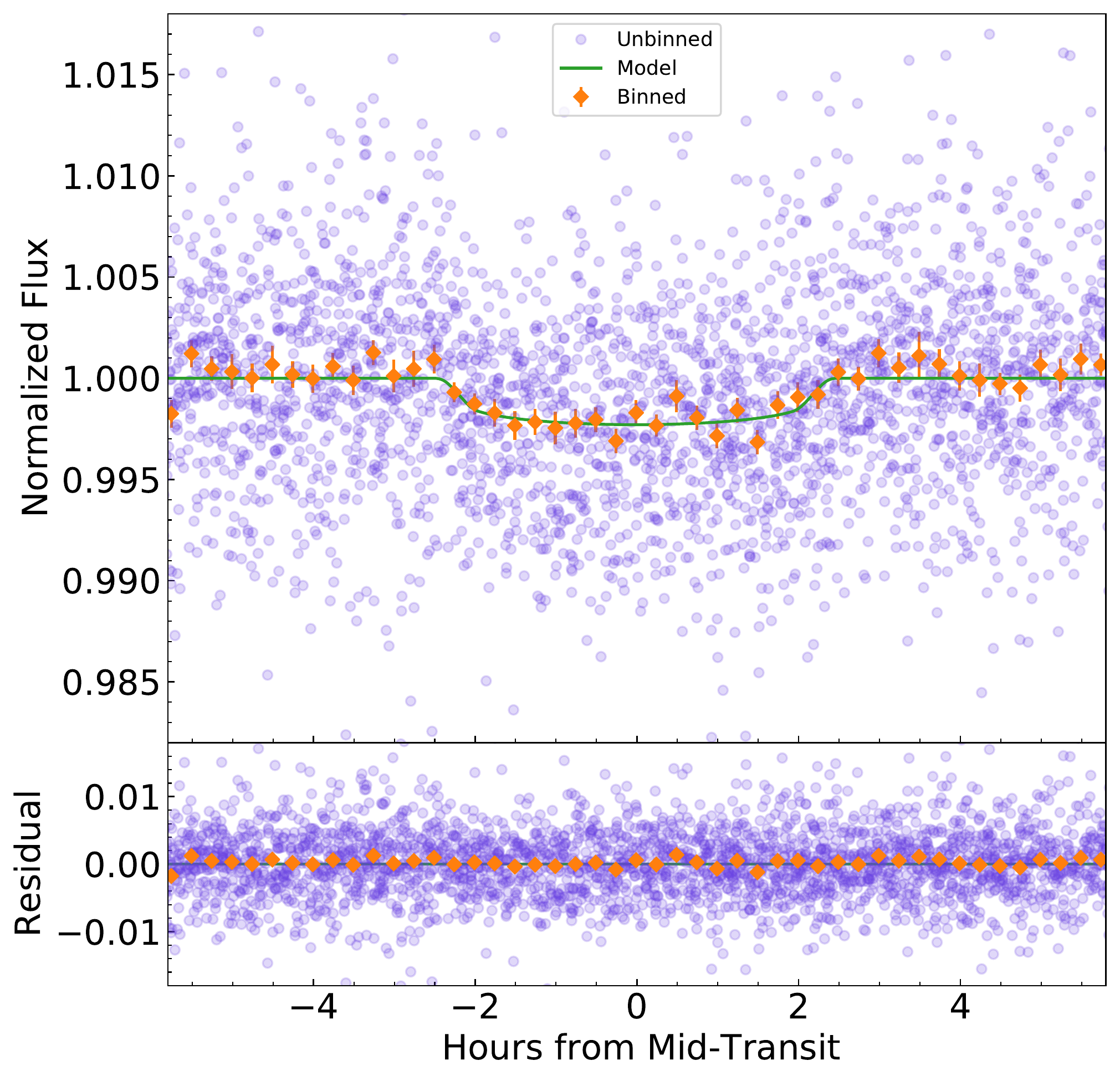} \\
  \end{tabular}
  \caption{Light curves of \starone{}.
    The model plotted is the MCMC best-fit solution to the global model.
    The binned points indicate the mean of each bin,
    with error bars representing the standard error of the mean.
    (a) \tess{} detrended 30~minute cadence light curve.
    (b) \tess{} detrended and phase-folded 30~minute cadence light curve
    after correcting for dilution (\secref{sec:phot_ground}),
    focusing on the transit.
    The bins are 20~minutes.
    (c) \hatsouth{} TFA detrended and phase-folded light curve
    after correcting for dilution (\secref{sec:phot_ground}),
    focusing on the transit.
    The bins are 15~minutes.
    The $y$-axes are restricted for the convenience of presentation,
    although the outliers are included in the per-bin calculations.
    (d) \lcogt{} detrended light curves.
    The bins are 10~minutes.
    }
  \label{fig:toi954_lc}
\end{figure*}

In this subsection, we describe the photometric observations we use to perform our analysis.
A summary of the observations can be found in \tabref{tab:phot}.
The light curves of \starone{} are shown in \figref{fig:toi954_lc},
and those of \startwo{} are shown in \figref{fig:k2p_lc}.

\subsubsection{\tess{} Photometry}

\tess{} observed \starone{} in sector~4 (\ut{} 2018 October 18--November 15) on camera~2 CCD~2
and in sector~5 (\ut{} 2018 November 15--December 11) on camera~2 CCD~1
with a 30 minute cadence in the full-frame images (FFIs).
The MIT Quick Look Pipeline
\citep[QLP;][]{qlp_rnaas_1,qlp_rnaas_2}
detected the candidate with a signal--to--pink noise ratio of $52.4$.
The candidate showed consistent transit depths in all five apertures used by QLP and
appeared to be on target in the difference image analysis.
The \tess{} Science Office released the candidate as a \tess{} object of interest (TOI)
after deeming it to have passed all of the vetting criteria
\citep{tess_toi_paper}.

We used the SPOC-calibrated FFIs \citep{Jenkins:2016},
obtained from the TESSCut service
\citep{tesscut},
to produce the detrended light curve used in this paper.
We used a 45 pixel aperture (a $7 \times 7$ square without the four pixels at its corners) centered on the target.
This aperture included an unresolved nearby star
TIC\thinspace 44792537
($\Delta T = 3.29$,
 \ang{;;48.5} separation,
 $\mathrm{PA} = \ang{153;;}$)
from the \tess{} Input Catalog
\citep[TIC\thinspace 8;][]{tic8},
which we would correct for by simultaneously fitting a dilution factor in the global model (\secref{sec:global_model}).
We rejected outliers due to spacecraft momentum dumps using the pointing quaternion time series.
We also rejected cadences affected by stray light from Earth between JD~\num{2458422.2} and \num{2458423.5} in orbit~15
\citep[sector 4;][]{tess_dr05_s04}.
No such stray light contamination affected camera~2 in Sector~5
\citep{tess_dr07_s05}.
Orbit~15 was also affected by an instrument anomaly that prevented data collection
from JD~\num{2458418.54} to \num{2458421.21}
\citep{tess_dr05_s04}.
We detrended the light curve by simultaneously modeling the spacecraft systematics, transits, and low-frequency variability following \cite{vanderburg_planetary_2016}.
As \tess{} observed the last transit of \planetone{} close to the spacecraft's perigee, the light curve was affected by excessive systematic noise, so we removed that transit from our subsequent analysis to obtain a more accurate transit depth.

\subsubsection{\ktwo{} Photometry}
\label{sec:ktwo_phot}

\begin{figure*}
  \centering
  (a) \ktwo{} Campaign 12 Long-Cadence Detrended Light Curve \\
  \includegraphics[width=\textwidth]{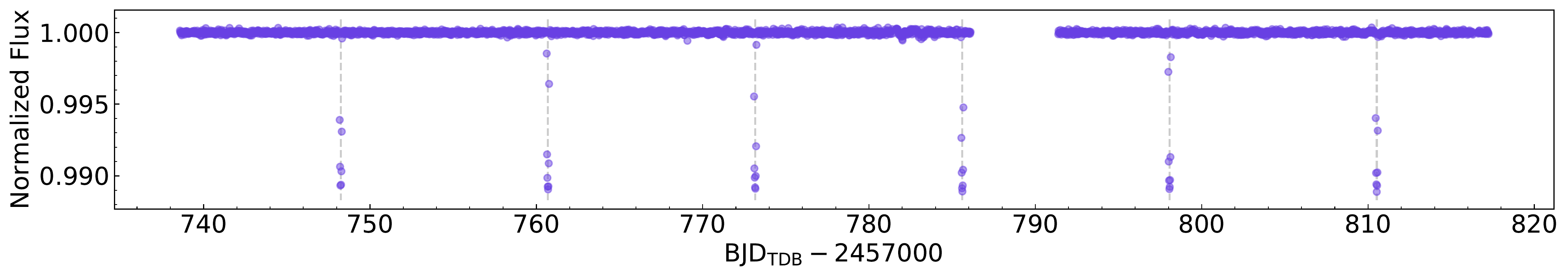} \\
  (b) \ktwo{} Campaign 19 Short-Cadence Detrended Light Curve
  \\
  \includegraphics[width=\textwidth]{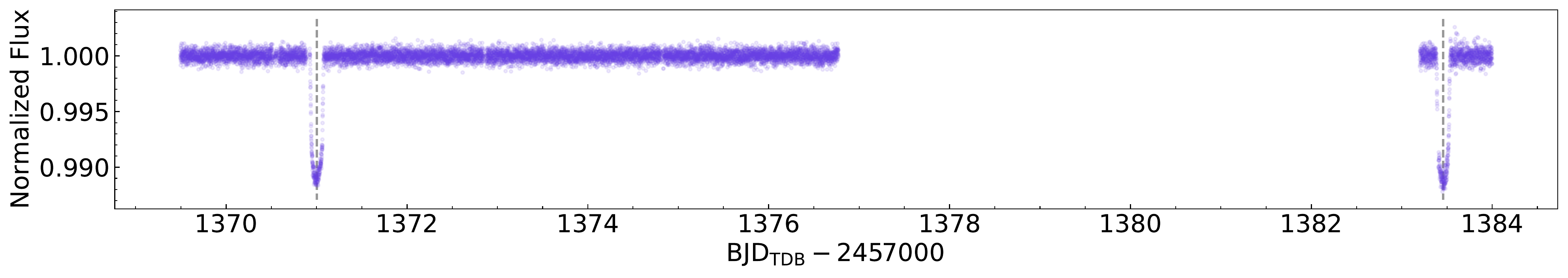} \\
  \hspace*{2em}
  (c) \ktwo{} Campaign 12 Long-Cadence Phase-Folded Light Curve
  \hfill
  (d) \pest{} Light Curve
  \hspace*{6em} \\
  \includegraphics[width=\columnwidth]{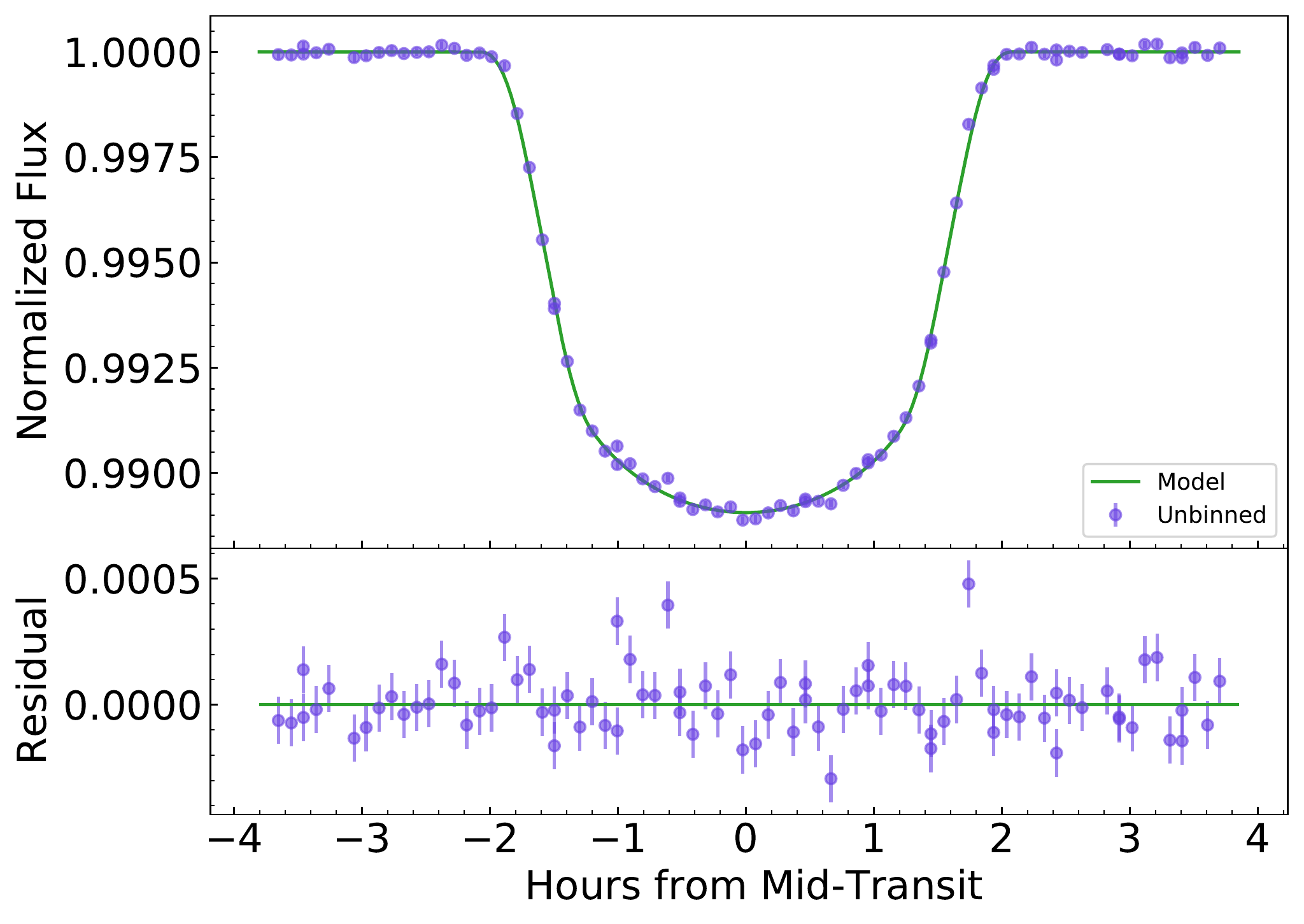}
  \hfill
  \includegraphics[width=\columnwidth]{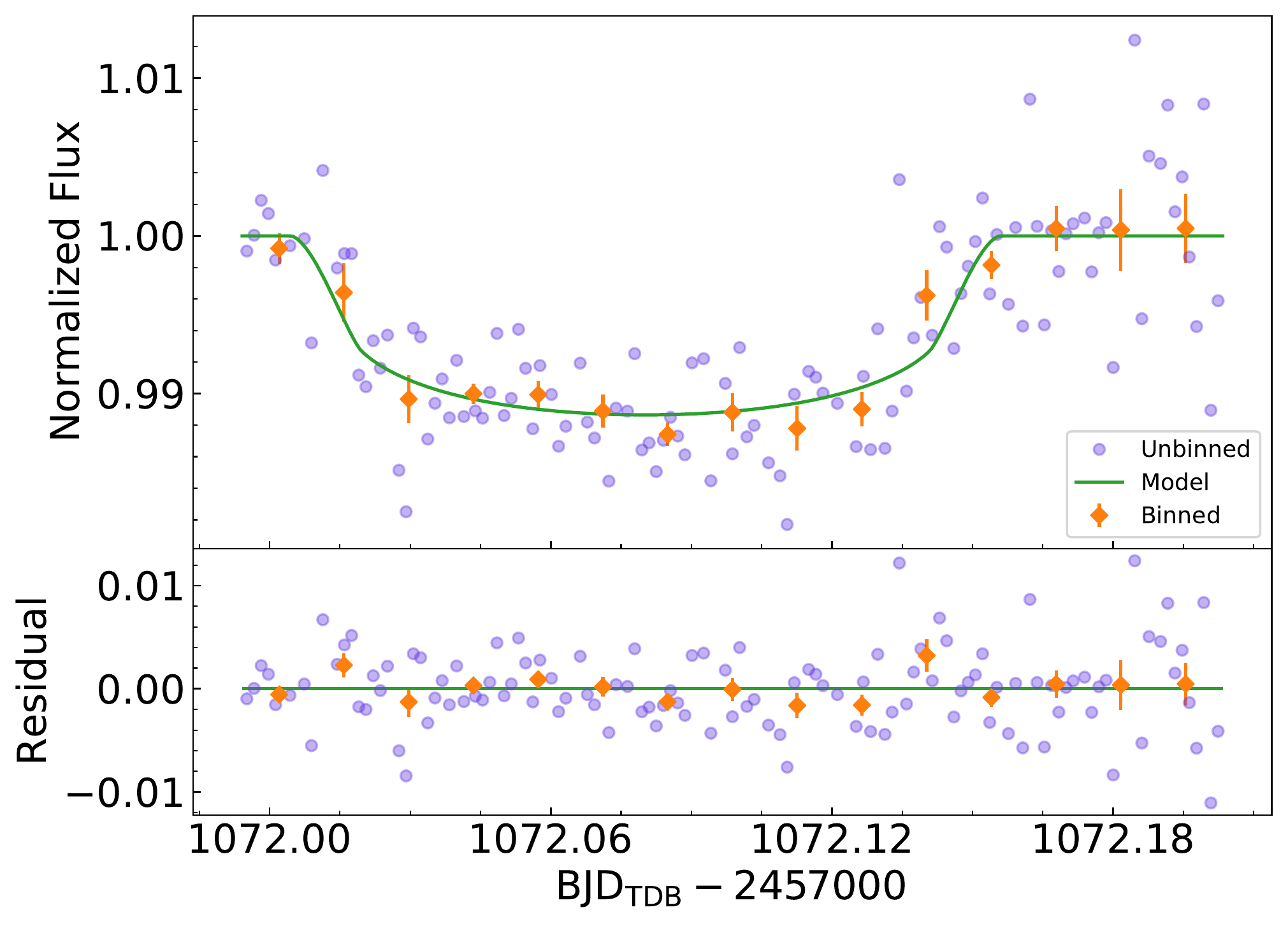} \\
  (e) \ktwo{} Campaign 19 Short-Cadence Detrended Light Curves \\
  \includegraphics[width=\columnwidth]{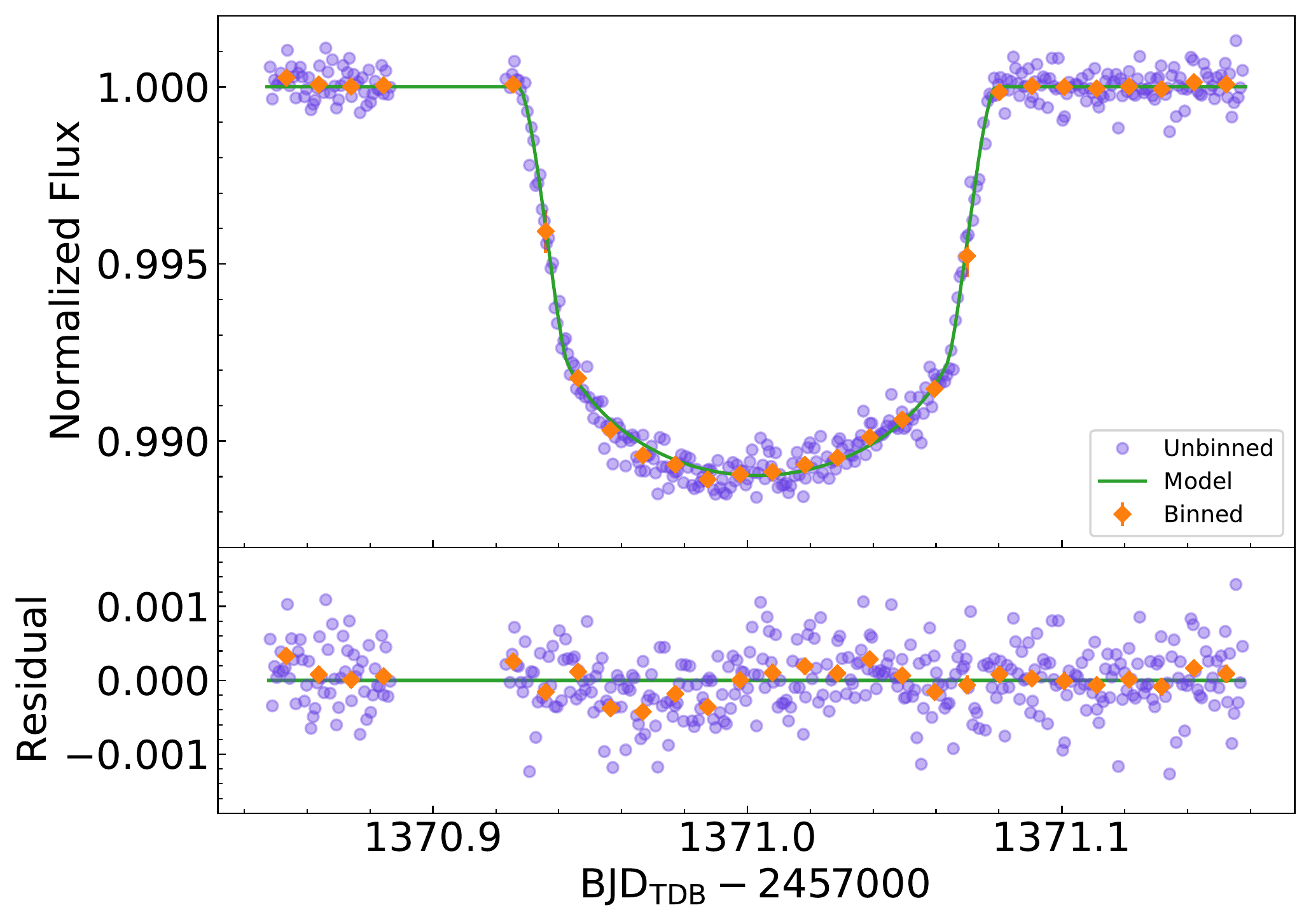}
  \hfill
  \includegraphics[width=\columnwidth]{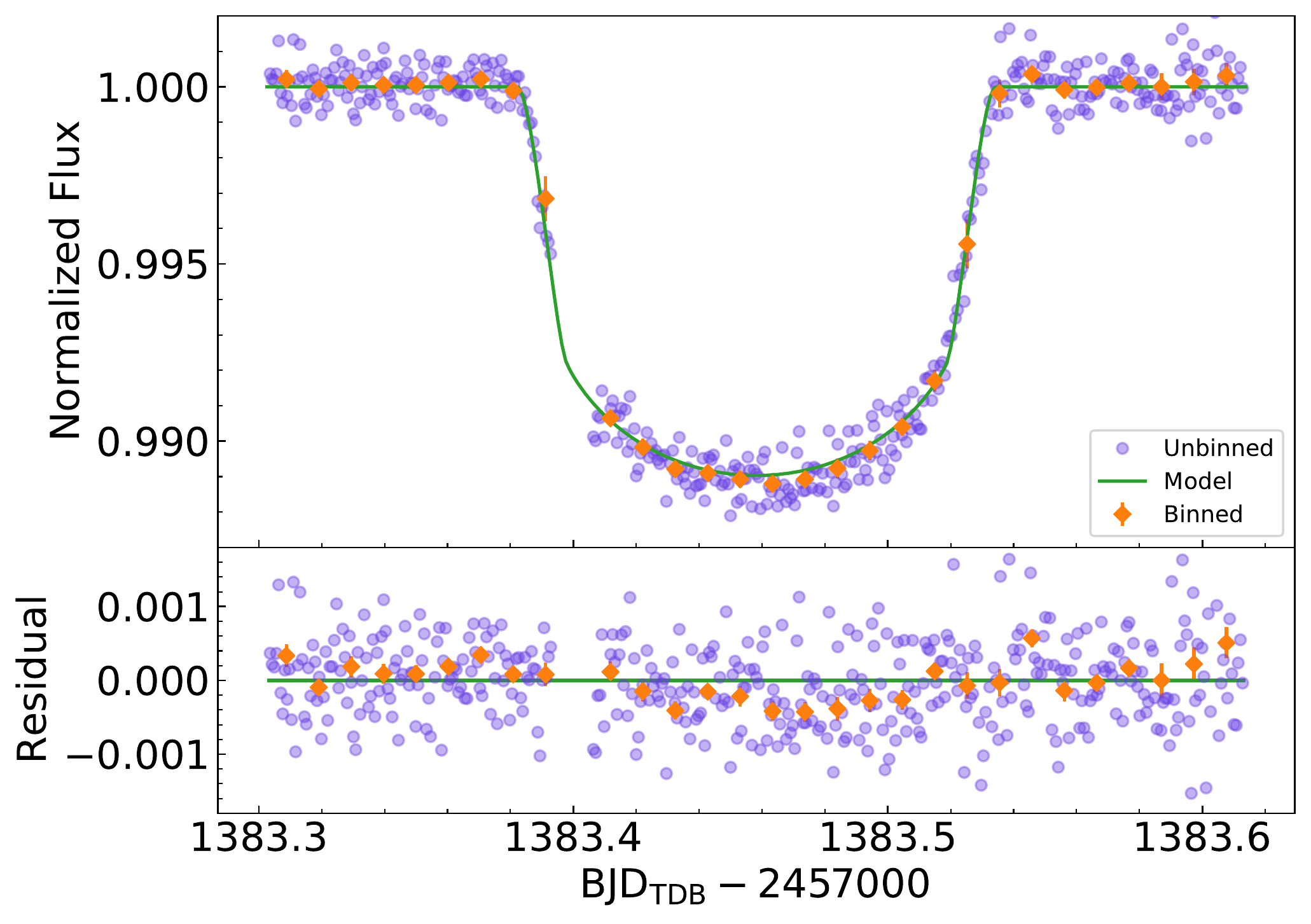} \\
  \caption{Light curves of \startwo{}.
    The model plotted is the MCMC best-fit solution to the global model.
    (a) \ktwo{} campaign~12 long-cadence detrended light curve.
    (b) \ktwo{} campaign~19 short-cadence detrended light curve.
    (c) \ktwo{} campaign~12 long cadence detrended and phase-folded light curve.
    The noise is estimated from the out-of-transit portions of the light curve.
    (d) \pest{} observations.
    The binned points indicate the mean of each bin (20~minutes),
    with error bars representing the standard error of the mean.
    (e) \ktwo{} Campaign~19 short cadence detrended light curves.
    The two transits covered by this campaign are shown side by side.
    The binned points indicate the mean of each bin (15~minutes).
    The error bars are the standard error of the mean multiplied by a factor $\beta$
    that accounts for the excess red noise (\secref{sec:ktwo_phot}).
    For the two transits, $\beta \approx 1.3$ and $1.6$, respectively.
  }
  \label{fig:k2p_lc}
\end{figure*}

The object \startwo{} was observed in \ktwo{} campaign~12 (\ut{} 2016 December 15--2017 March 4)
with the 29.4 minute long cadence as part of four GO programs%
\footnote{PIs: David Charbonneau, Andrew Howard, Dennis Stello, and Elisa Quintana.}.
It was also observed during campaign~19 (\ut{} 2018 August 30--September 26) as part of four long-cadence and one short-cadence (\SI{58.3}{\second}) guest observer programs%
\footnote{Short-cadence program PI: Andrew Vanderburg; long-cadence program PIs: Andrew Howard, Courtney Dressing, and Dennis Stello.}.

We initially identified the planet candidate in a box least-squares (BLS) search \citep{kovacs_bls_2002, vanderburg_planetary_2016}
of light curves produced by \cite{vanderburg_technique_2014} in \ktwo{} data from campaign~12.
After identifying the planet candidate, we rederived the campaign~12 light curve by simultaneously modeling the spacecraft systematics, transits, and low-frequency variability following \cite{vanderburg_planetary_2016}.

We took a more custom approach to reducing the campaign~19 data.
By the time \ktwo{} began executing this campaign, it was critically low on fuel,
so the spacecraft exhibited erratic pointing behavior and was only able to observe for about a month before completely exhausting its fuel reserves \citep[accessed on 2020 March 4]{k2_dr_c19}.
Nevertheless, \ktwo{} managed to achieve fairly typical pointing performance for about a week between \ut{} 2018 September 8 and 15,
during which time \planettwo{} transited once.
We reduced the data collected during this time interval as usual,
producing a first-pass light curve following \citet{vanderburg_technique_2014} and
refining the systematics correction with a simultaneous fit to the transit of \planettwo{}, systematics, and low-frequency variability \citep{vanderburg_planetary_2016}.

After \ut{} 2018 September 15, \ktwo{}'s thruster corrections became less effective,
and its pointing began to drift farther and less predictably than usual.
We managed to recover a second usable transit of \planettwo{} before the end of campaign~19
by performing a simplified systematics correction to a 0.8~day window of data surrounding \planettwo{}'s transit.
Here again, we simultaneously fit the transit of \planettwo{} with a model for stellar variability and \ktwo{} pointing systematics.
This time, however, we did not attempt to separate the stellar variability from the spacecraft's pointing drift systematics
and modeled both with an aggressive basis spline
(with knots spaced every 0.2~days instead of the typical 0.75~days used in a standard \ktwo{} reduction),
introducing a discontinuity when \ktwo{} fired its thrusters (which only happened once during the 0.8~day window).
We subtracted the best-fit spline from the original light curve to yield a systematics-free transit. 

We estimated the excess red noise affecting the two transits in the Campaign~19 short cadence data
following the method described by \cite{2008ApJ...683.1076W}.
We calculated the $\beta$ factor (the effective increase to flux uncertainty due to time-correlated noise)
to be $1.3$ for the first transit and $1.6$ for the second,
at a timescale comparable to the transit duration.
Thus, we multiplied our white-noise estimates for the two transits by their respective $\beta$ factors
to obtain the photometric uncertainties we used to calculate the global model posterior (\secref{sec:global_model}).

\subsubsection{Ground-based Follow-up Photometry}
\label{sec:phot_ground}

\begin{deluxetable}{DDDDDc}
  \tablecaption{\hatsouth{} Photometric Measurements of \starone{} \label{tab:hats_toi954}}
  \tabletypesize{\scriptsize}
  \tablehead{%
    \multicolumn2c{\bjdtdb}
    & \multicolumn2c{Raw Mag.}
    & \multicolumn2c{EPD Mag.}
    & \multicolumn2c{TFA Mag.}
    & \multicolumn2c{Uncertainty}
    & \colhead{Flag} \\
    \multicolumn2c{${} - \num{2400000}$}
  }
  \decimals
  \startdata
56983.74903 & 10.41281 & 10.35094 & 10.34490 & 0.00121 & 0 \\
56983.75455 & 10.40697 & 10.34580 & 10.34342 & 0.00121 & 1 \\
56983.75856 & 10.40240 & 10.33414 & 10.34393 & 0.00121 & 0 \\
56983.76259 & 10.39569 & 10.32704 & 10.33661 & 0.00120 & 1 \\
56983.76884 & 10.39573 & 10.34297 & 10.34667 & 0.00121 & 0 \\
56983.77290 & 10.40901 & 10.33885 & 10.34718 & 0.00121 & 0 \\
56983.77695 & 10.40787 & 10.34479 & 10.34520 & 0.00121 & 0 \\
56983.78253 & 10.40622 & 10.33925 & 10.33945 & 0.00121 & 1 \\
56983.78654 & 10.40170 & 10.33649 & 10.33636 & 0.00121 & 1 \\
56983.79106 & 10.40960 & 10.34238 & 10.34526 & 0.00121 & 0 \\
56983.79659 & 10.41026 & 10.34137 & 10.34823 & 0.00121 & 1 \\
56983.80074 & 10.41835 & 10.35022 & 10.34657 & 0.00122 & 0 \\
56983.80476 & 10.41969 & 10.34276 & 10.34678 & 0.00122 & 0 \\
56983.81011 & 10.41058 & 10.32626 & 10.34037 & 0.00121 & 0 \\
56983.81468 & 10.42444 & 10.36145 & 10.35338 & 0.00122 & 1 \\
56983.81878 & 10.42908 & 10.36117 & 10.35667 & 0.00122 & 0 \\
56983.82426 & 10.42465 & 10.35300 & 10.34349 & 0.00122 & 1 \\
56983.82834 & 10.42461 & 10.34070 & 10.34959 & 0.00122 & 1 \\
56983.83246 & 10.43064 & 10.34189 & 10.35611 & 0.00123 & 0 \\
56983.83846 & 10.42909 & 10.34239 & 10.34314 & 0.00122 & 0 \\
56983.84263 & 10.43798 & 10.34189 & 10.34984 & 0.00123 & 0 \\
56983.84664 & 10.43480 & 10.33644 & 10.34282 & 0.00123 & 0 \\
56983.85218 & 10.44484 & 10.34619 & 10.34190 & 0.00124 & 0 \\
56983.85624 & 10.45349 & 10.34244 & 10.34576 & 0.00124 & 0 \\
56983.86080 & 10.52435 & 10.33869 & 10.34762 & 0.00129 & 0 \\
56983.86621 & 10.48563 & 10.34870 & 10.34910 & 0.00130 & 0 \\
56983.87025 & 10.46183 & 10.33230 & 10.35159 & 0.00136 & 0 \\
56984.51946 & 10.47864 & 10.36408 & 10.34731 & 0.00129 & 0 \\
56984.52352 & 10.46848 & 10.35617 & 10.34055 & 0.00126 & 0 \\
56984.52758 & 10.46211 & 10.34552 & 10.33378 & 0.00125 & 0 \\
56984.53352 & 10.45185 & 10.32801 & 10.33251 & 0.00124 & 1 \\
56984.53812 & 10.44663 & 10.32807 & 10.33958 & 0.00123 & 0 \\
56984.54214 & 10.45214 & 10.34555 & 10.34681 & 0.00124 & 0 \\
56984.54819 & 10.44286 & 10.34677 & 10.33815 & 0.00123 & 0 \\
56984.55233 & 10.43931 & 10.33513 & 10.34743 & 0.00123 & 0 \\
56984.55643 & 10.44049 & 10.34325 & 10.33722 & 0.00123 & 1 \\
56984.56297 & 10.44801 & 10.35086 & 10.34400 & 0.00123 & 0 \\
56984.56709 & 10.44393 & 10.34679 & 10.34967 & 0.00123 & 1 \\
56984.57119 & 10.43688 & 10.34890 & 10.34800 & 0.00123 & 0 \\
56984.57646 & 10.44848 & 10.36341 & 10.35508 & 0.00123 & 1 \\
56984.58053 & 10.43158 & 10.35575 & 10.34443 & 0.00123 & 0 \\
\enddata
  \tablecomments{%
  Two detrended magnitudes are given:
  one using the external parameter decorrelation \citep[EPD;][]{2010ApJ...710.1724B} method
  and one using the TFA \citep{Kovacs:2005}.
  A nonzero flag value indicates that an observation is affected by anomalously high systematics and excluded from the global model. \\
  (This table is available in its entirety in machine-readable form.)}
\end{deluxetable}

The \hatsouth{} \citep{Bakos:2013} survey observed \starone{} before \tess{} launched,
for a total of \num{36396} observations in the $r$ band at an average cadence of 6~minutes
from \ut{} 2014 September 9 to 2015 March 6.
The star was slightly saturated for the \hatsouth{} observation,
but with the help of the trend-filtering algorithm \citep[TFA;][]{Kovacs:2005},
we were able to pick up the signal in the light curve.
While we were not able to use the \hatsouth{} light curve to confirm whether the transit is on target,
its long baseline helped us improve the precision of the planet's orbital period by an order of magnitude.
We listed the measurement data in \tabref{tab:hats_toi954}.

The object \starone{} was also observed as part of the \tess{} Follow-up Program (TFOP).
We attempted to observe an ingress on \ut{} 2019 August 3 from the \SI{0.7}{\m} PlaneWave CDK700 telescope at Maunakea Observatories, Hawaii,
stopping at twilight.
The limited precision of the light curve prevented us from either confirming or ruling out the ingress,
but we were able to rule out that the transit could have been caused by background eclipsing binaries within \ang{;;40}.
The light-curve time series can be found on the ExoFOP-TESS%
\footnote{\url{https://exofop.ipac.caltech.edu/tess/target.php?id=279741379}}
website.

A full transit of \starone{} was observed in the Pan-STARSS \filterzs{} band on
\ut{} 2019 November 3 and December 21 and 2020 January 12 using a \SI{1.0}{\m} telescope at the 
LCOGT South African Astronomical Observatory (SAAO) node in Sutherland, South Africa.
The LCOGT observations
were calibrated with the standard BANZAI pipeline,
and the light curves were extracted using \astroimagej{} \citep[AIJ;][]{Collins:2017}. 
The observation used a \ang{;;6} aperture, and recovered the expected transit signal.
The detrended light curves can also be found on the ExoFOP-TESS website.
The \ut{} 2019 December 21 data were not included in the global model fitting (\secref{sec:global_model})
because they did not contain enough out-of-transit observations to allow us to accurately measure the transit depth.

\begin{deluxetable}{DDD}
  \tablecaption{\pest{} Photometric Measurements of \startwo{} \label{tab:pest_k2p}}
  \tablehead{%
    \multicolumn2c{HJD$_{\mathrm{UTC}} - \num{2400000}$}
    & \multicolumn2c{Magnitude}
    & \multicolumn2c{Uncertainty}
  }
  \decimals
  \startdata
58071.9943414 & 12.4795 & 0.0028 \\
58071.9958805 & 12.4784 & 0.0027 \\
58071.9974061 & 12.4760 & 0.0027 \\
58071.9989486 & 12.4769 & 0.0026 \\
58072.0004769 & 12.4801 & 0.0026 \\
58072.0035502 & 12.4791 & 0.0026 \\
58072.0065972 & 12.4786 & 0.0026 \\
58072.0081422 & 12.4858 & 0.0026 \\
58072.0105639 & 12.4739 & 0.0026 \\
58072.0136167 & 12.4806 & 0.0025 \\
58072.0151562 & 12.4796 & 0.0026 \\
58072.0167014 & 12.4796 & 0.0026 \\
58072.0182275 & 12.4880 & 0.0026 \\
58072.0197626 & 12.4888 & 0.0025 \\
58072.0213111 & 12.4856 & 0.0026 \\
58072.0228323 & 12.4875 & 0.0026 \\
58072.0243530 & 12.4852 & 0.0025 \\
58072.0268029 & 12.4946 & 0.0025 \\
58072.0283283 & 12.4975 & 0.0026 \\
58072.0298796 & 12.4847 & 0.0025 \\
58072.0314200 & 12.4853 & 0.0025 \\
58072.0329524 & 12.4924 & 0.0026 \\
58072.0344828 & 12.4899 & 0.0025 \\
58072.0360143 & 12.4882 & 0.0025 \\
58072.0375664 & 12.4909 & 0.0025 \\
58072.0391149 & 12.4869 & 0.0026 \\
58072.0406562 & 12.4908 & 0.0025 \\
58072.0431032 & 12.4904 & 0.0025 \\
58072.0446452 & 12.4909 & 0.0025 \\
58072.0461702 & 12.4891 & 0.0025 \\
58072.0477101 & 12.4850 & 0.0025 \\
58072.0492365 & 12.4907 & 0.0025 \\
58072.0507722 & 12.4895 & 0.0025 \\
58072.0523084 & 12.4847 & 0.0025 \\
58072.0538298 & 12.4874 & 0.0025 \\
58072.0553637 & 12.4916 & 0.0025 \\
58072.0568884 & 12.4872 & 0.0025 \\
58072.0593443 & 12.4892 & 0.0025 \\
58072.0608676 & 12.4928 & 0.0025 \\
58072.0624086 & 12.4914 & 0.0025 \\
\enddata

  \tablecomments{%
  We present the observation times in HJD$_{\mathrm{UTC}}$ as originally reported;
  however, they have been converted to \bjdtdb{} before inclusion in the global model (\secref{sec:global_model}),
  following the methods described by \cite{2010PASP..122..935E}.
  The magnitudes reported in this table are not detrended. \\
  (This table is available in its entirety in machine-readable form.)
  }
\end{deluxetable}

On \ut{} 2017 November 14,
\startwo{} was observed by the Perth Exoplanet Survey Telescope (\pest{})
with 117 observations at \SI{120}{\s} cadence in the \filterRc{} band.
\pest{} is a $12''$ Meade LX200 SCT Schmidt--Cassegrain telescope
equipped with an SBIG ST-8XME camera located in a suburb of Perth, Australia.
The \pest{} pipeline automatically reduced and calibrated the images,
producing a light curve, which was normalized for transit model fitting
\citep{pest_pipeline}.
The transit arrived on time, and we were able to recover a full transit of the planet (\figref{fig:k2p_lc}(c)),
allowing us to improve the precision of the planet's orbital period.
We report the measurement data in \tabref{tab:pest_k2p}.

On \ut{} 2018 July 9,
\startwo{} was also observed by the $24''$ telescope
at the Peter van de Kamp Observatory of Swarthmore College, Pennsylvania.
We made 78 measurements in the \filterrp{} filter with an exposure time of \SI{90}{\second}.
We observed an egress of \planettwo{} on the expected target under good sky conditions.
We decided to not include this partially observed transit in our global model fitting (\secref{sec:global_model}).

\subsection{Spectroscopy}

We now describe the spectroscopic observations we use to confirm the planets of \starone{} and of \startwo{}.
A summary of those observations can be found in \tabref{tab:spec}.

\begin{deluxetable*}{ll@{\hspace{-2em}}rrcc@{\hspace{0pt}}r@{}lr@{}l}
    \tablecaption{Summary of Spectroscopic Observations\label{tab:spec}}
    \tablecolumns{10}
    \tabletypesize{\small}
    \tablehead{%
        \colhead{Instrument}
        & \colhead{UT Date(s)}
        & \colhead{No.~Spectra}
        & \colhead{Resolution\tablenotemark{a}}
        & \colhead{S/N Range}
        & \colhead{Wavelengths}
        & \multicolumn2c{Jitter}
        & \multicolumn2c{$\gamma$}\\
        & &
        & \colhead{$ \Delta\lambda / \lambda / 1000 $}
        & \colhead{(\SI{5000}{\angstrom})}
        & \colhead{(\si{\angstrom})}
        & \multicolumn2c{(\si{\meter\per\second})}
        & \multicolumn2c{(\si{\meter\per\second})}
    }
    \startdata
    \cutinhead{\starone{}}
    \anutwometer{}\tablenotemark{b} & 2019 Feb 18--22     & 3     & 23    & 49.6--75.8    & 3900--6700    & \multicolumn2c{\ldots} & \multicolumn2c{\ldots} \\
    \chiron{}\tablenotemark{c} & 2019 Feb 22--27     & 5     & 80    & 67.7--76.1    & 4100--8700    & \ToipJitterchiron & \ToipJitterchironErr & \ToipGammaChiron & \ToipGammaChironErr \\
    \coralie{}  & 2019 Aug 19--Sep 30 & 19    & 60    & 10.0--31.7    & 3900--6800    & \ToipJittercoralie & \ToipJittercoralieErr & \ToipGammaCoralie & \ToipGammaCoralieErr \\
    \harps{}    & 2019 Sep 27--30     & 4     & 115   & 33.6--80.4    & 3780--6910    & \ToipJitterharps & \ToipJitterharpsErr & \ToipGammaHarps & \ToipGammaHarpsErr \\
    \minervaaus{} & 2019 Sep 8--Nov 10 & 12 & 80 & \ldots\tablenotemark{d} & 5000--6300    & \ToipJitterminerva & \ToipJitterminervaErr & \ToipGammaMinerva & \ToipGammaMinervaErr \\
    \pfs{}      & 2019 Jul 11--Sep 15 & 8     & 130   & 45--75        & 3910--7340    & \ToipJitterpfs & \ToipJitterpfsErr & \ToipGammaPfs & \ToipGammaPfsErr \\
    \tres{}\tablenotemark{b} & 2019 Mar 1 & 1  & 44    & 30.3          & 3850--9096    & \multicolumn2c{\ldots} & \multicolumn2c{\ldots} \\
    \cutinhead{\startwo{}}
    \feros{}    & 2017 Oct 14--2018 Jul 15    & 27    & 48    & 45--55          & 3500--9200    & \KtwopJitterferos     & \KtwopJitterferosErr  & \KtwopGammaFeros      & \KtwopGammaFerosErr \\
    \fies{}\tablenotemark{b} & 2017 Aug 16 and 18         & 2     & 67    & \ldots             & 3650--9125    & \multicolumn2c{\ldots} & \multicolumn2c{\ldots} \\
    \harps{}    & 2017 Nov 6--2018 Sep 6      & 13    & 115   & 17--35          & 3780--6910    & \KtwopJitterharps     & \KtwopJitterharpsErr  & \KtwopGammaHarps      & \KtwopGammaHarpsErr \\
    \pfs{}      & 2018 May 24--Oct 26         & 10    & 130   & 30--90          & 3910--7340    & \KtwopJitterpfs     & \KtwopJitterpfsErr  & \KtwopGammaPfs      & \KtwopGammaPfsErr \\
    \tres{}\tablenotemark{b,c} & 2017 Sep 10 and 28 & 2 & 44    & 26.8--31.3      & 3850--9096    & \multicolumn2c{\ldots} & \multicolumn2c{\ldots} \\
    \enddata
    \tablecomments{%
        The jitter parameter is added in quadrature to the reported RV uncertainties in \tabref{tab:rv_toi954} and \tabref{tab:rv_k2p}.
        The $\gamma$ parameter is a constant offset added to RV measurements of a given RV instrument.
        The jitter and $\gamma$ values are empirically determined by a global model fit for each star system (see \secref{sec:global_model}).
    }
    \tablenotetext{a}{Approximate values of typical instrument performance.}
    \tablenotetext{b}{Reconnaissance spectroscopy; no RVs derived for the global model fitting (\secref{sec:global_model}).}
    \tablenotetext{c}{Spectra used to constrain the stellar parameters \tempeff{} and metallicity.}
    \tablenotetext{d}{The S/N estimate is unavailable; the rms scatter from the median RV model is reported instead in \secref{sec:prv}.}
\end{deluxetable*}

\subsubsection{Reconnaissance Spectroscopy}
\label{sec:recon_spec}

Three spectra of \starone{} were taken with the echelle spectrograph
on the Australia National University (ANU) \SI{2.3}{\meter} telescope for \starone{}
from \ut{} 2019 February 18 to 2019 February 22,
covering the wavelength region of \SIrange[range-phrase=--]{3900}{6700}{\angstrom} with a spectral resolution
$R \approx \num{23000}$.
These spectra can be found on the ExoFOP-TESS
website.
The observations suggested that there was no RV variation on the order of \SI{500}{\meter\per\second}.
We therefore moved on to higher-precision instruments for mass measurements of \planetone{}.

The Tillinghast Reflector Echelle Spectrograph
\citep[\tres{};][]{tres_design}
on the \SI{1.5}{\meter} telescope at
the Fred L. Whipple Observatory in Arizona
was used to obtain spectra for both
\starone{} (\ut{} 2019 March 1)
and \startwo{} (\ut{} 2017 September 10 and 28).
\tres{} is a fiber-fed echelle spectrograph with a spectral
resolution of $R \approx \num{44000}$ over the wavelength region of
\SIrange[range-phrase=--]{3850}{9100}{\angstrom}.
The observing strategy and data reduction process were described by
\cite{2012Natur.486..375B}.
Each spectrum was obtained from a combination of
three consecutive observations for optimal cosmic-ray
rejection, and the wavelength solution was provided by
bracketing ThAr hollow cathode lamp exposures.
The \tres{} spectra used in this paper can be found on the ExoFOP-TESS and ExoFOP-K2%
\footnote{\url{https://exofop.ipac.caltech.edu/k2/edit_target.php?id=246193072}}
websites.

We did not find evidence of strong stellar activity for either \starone{} or \startwo{} in the \tres{} spectra.
The \ion{Ca}{2}~H and K emission lines were absent.

The object \startwo{} was also observed with the FIbre-fed Echelle Spectrograph \citep[\fies;][]{Frandsen1999,Telting2014}
mounted at the \SI{2.56}{\meter} Nordic Optical Telescope (NOT) of Roque de los Muchachos Observatory (La Palma, Spain).
The observations%
\footnote{Part of the observing program 55-019, PI: D. Gandolfi}
were carried out on 2 nights (\ut{} 2017 August 16 and 18) using the high-resolution fiber,
which provides a resolving power of \num{67000} in the wavelength range \SIrange[range-phrase=--]{3650}{9125}{\angstrom}.
The observing strategy follows the one adopted for \tres{} observations.

\subsubsection{Precise RVs}
\label{sec:prv}

\begin{figure}
  \centering
  \includegraphics[width=\columnwidth]{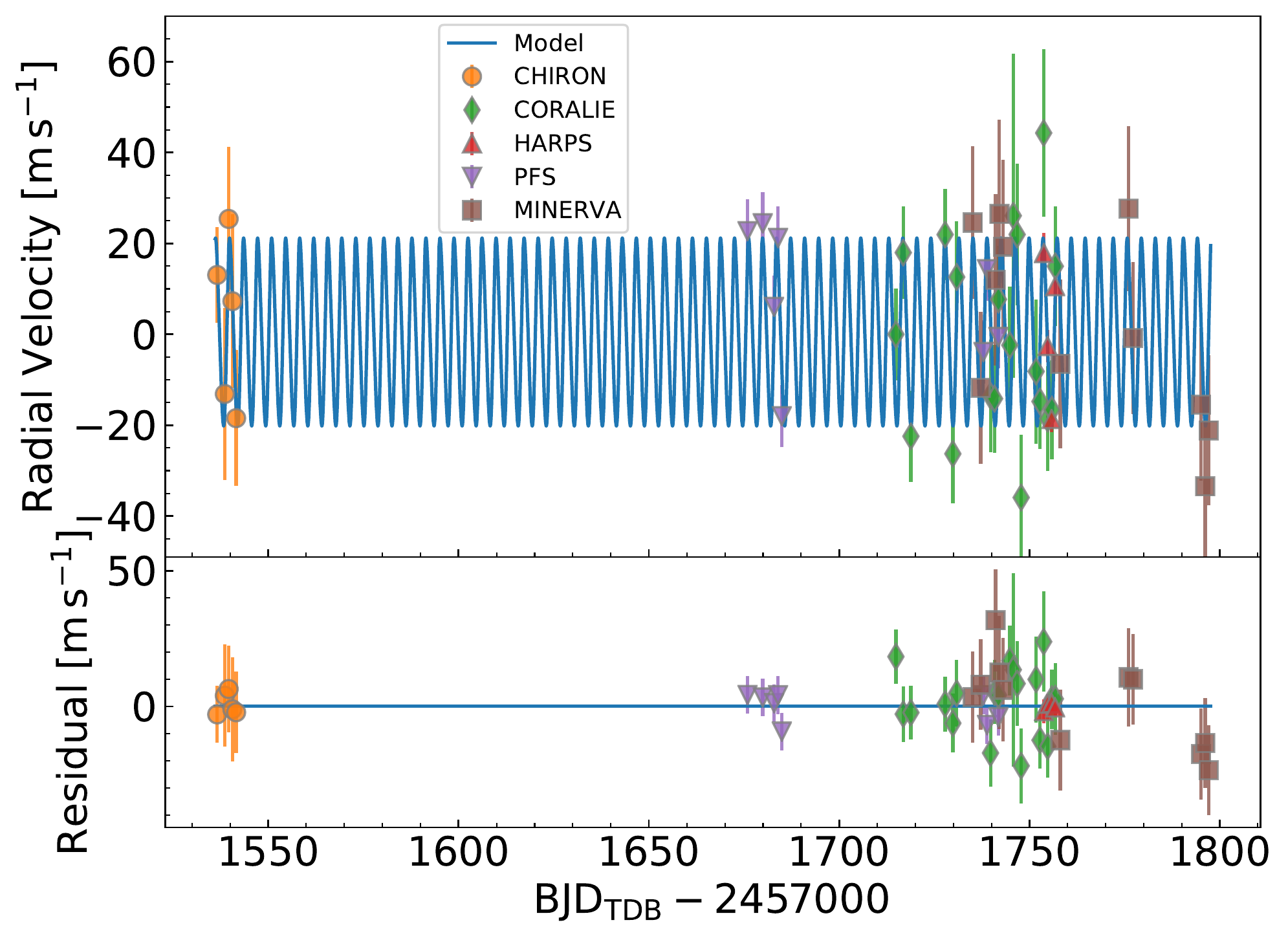}
  \includegraphics[width=\columnwidth]{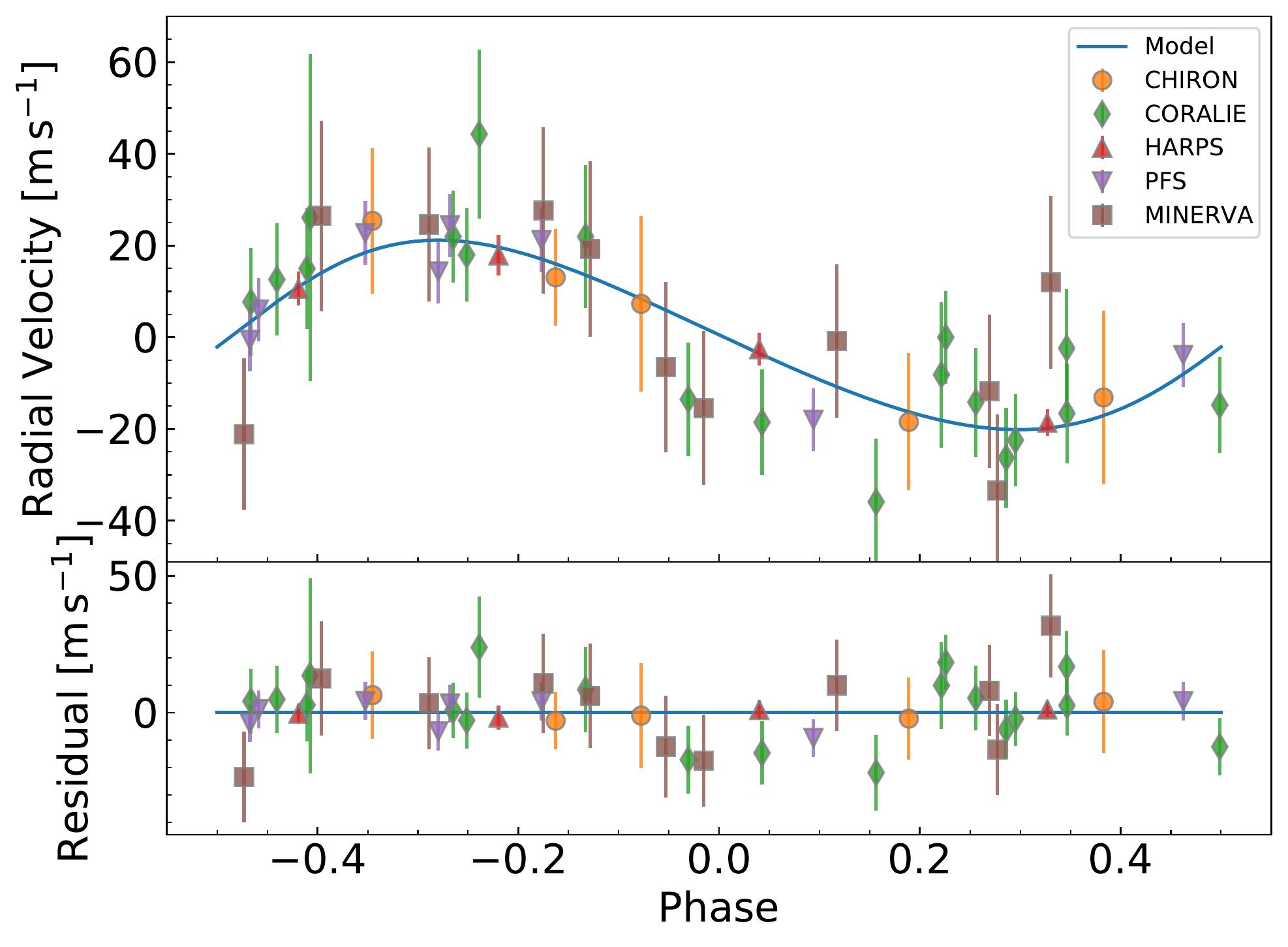}
  \caption{Precise RV measurements of \starone{}.
  The model plotted is the MCMC median of the global model.
  A empirically derived per-instrument offset $\gamma$ has been subtracted from the raw RV measurements.
  The error bars represent the reported uncertainty and the empirically derived per-instrument jitter, added in quadrature.
  These RV measurements are also listed in \tabref{tab:rv_toi954}.
  Top: RV measurements plotted against time.
  Bottom: phase-folded RV measurements.
  }
  \label{fig:toi954_rv}
\end{figure}

\begin{figure}
  \centering
  \includegraphics[width=\columnwidth]{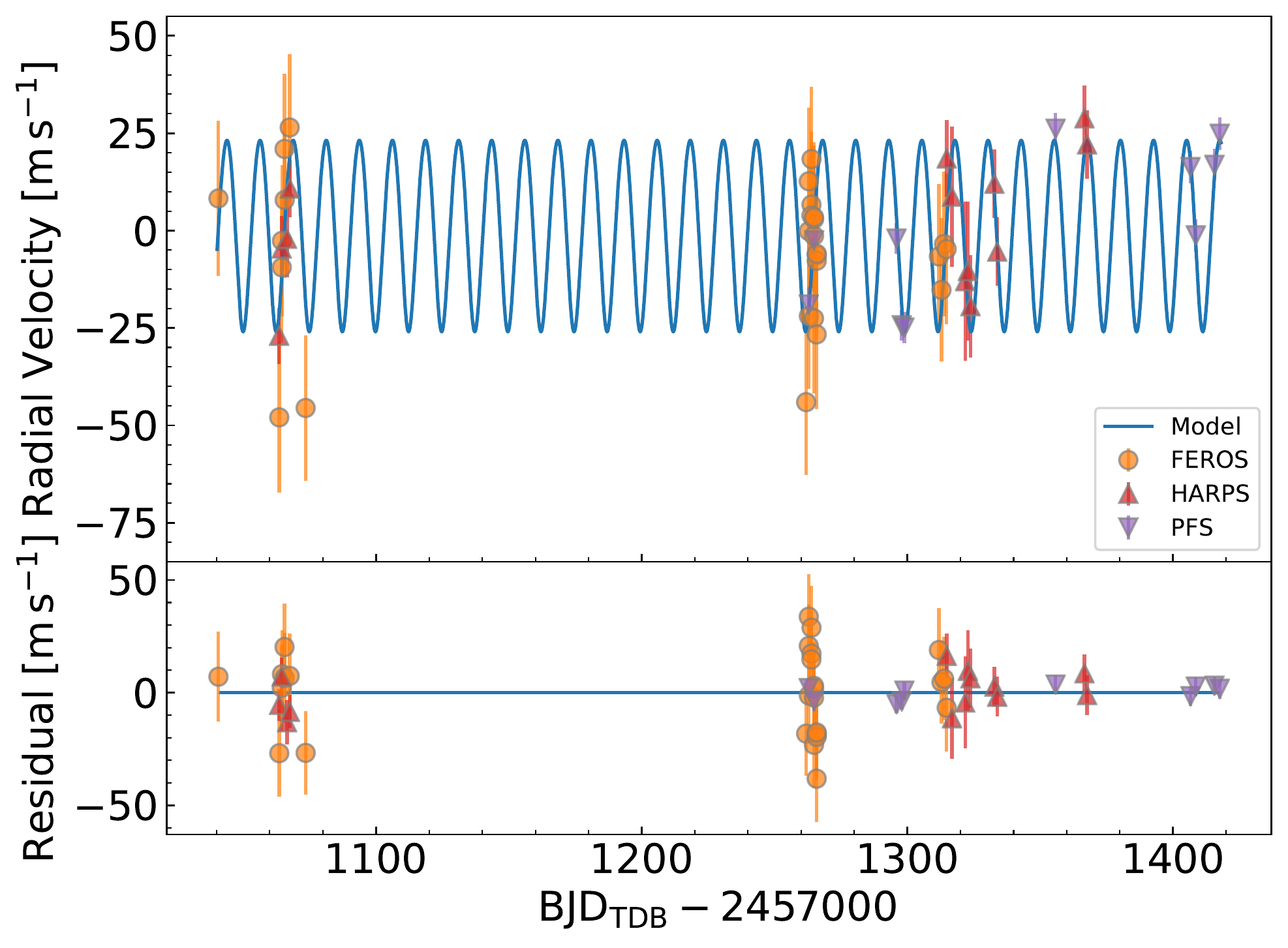}
  \includegraphics[width=\columnwidth]{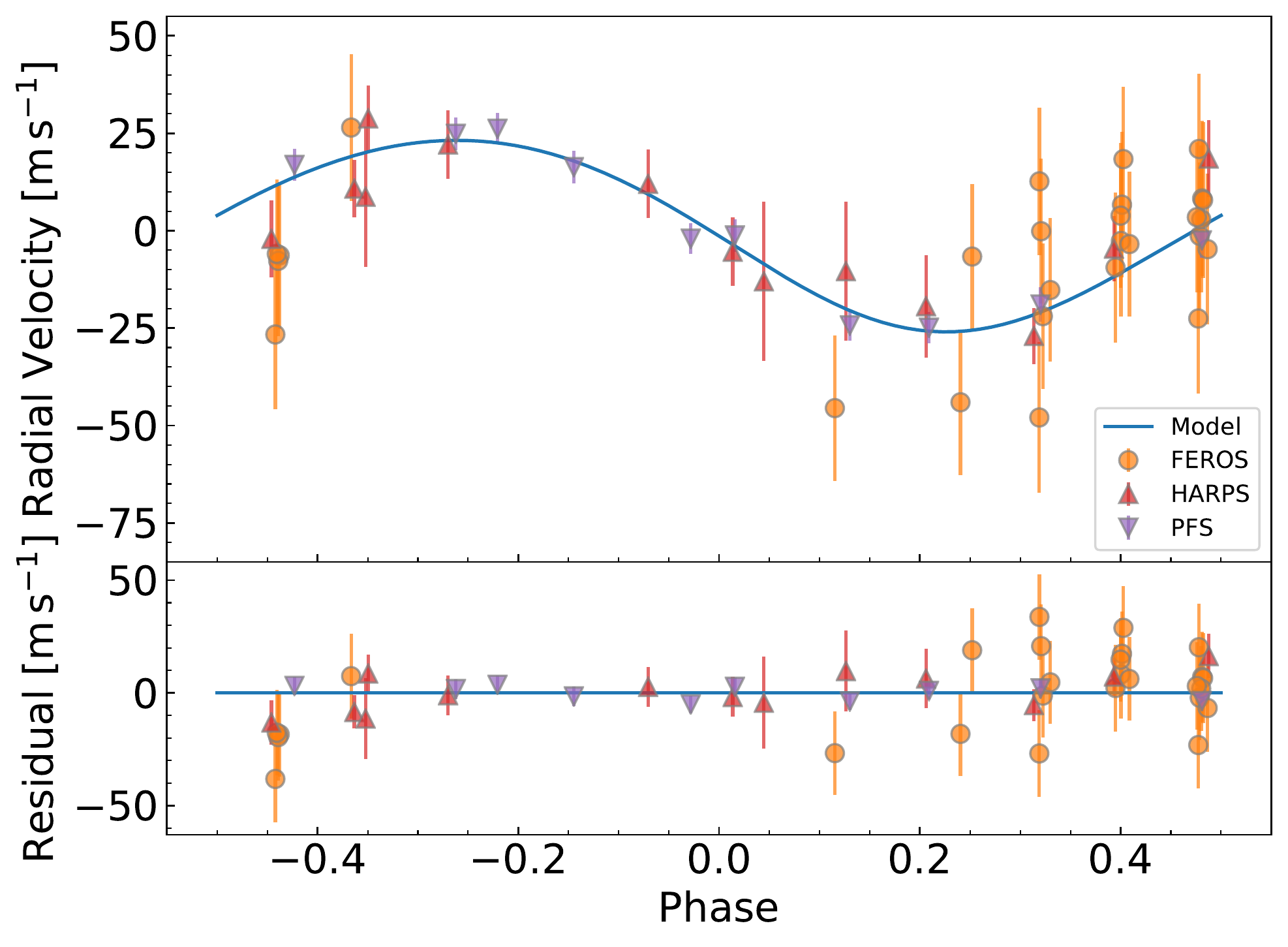}
  \caption{Precise RV measurements of \startwo{}.
  The model plotted is the MCMC median of the global model.
  A empirically derived per-instrument offset $\gamma$ has been subtracted from the raw RV measurements.
  The error bars represent the reported uncertainty and the empirically derived per-instrument jitter, added in quadrature.
  These RV measurements are also listed in \tabref{tab:rv_k2p}.
  Top: RV measurements plotted against time.
  Bottom: phase-folded RV measurements.
  }
  \label{fig:k2p_rv}
\end{figure}

The high-precision RV measurements used in this paper are presented in
\tabref{tab:rv_toi954} and \figref{fig:toi954_rv} for \starone{};
those for \startwo{} are presented in \tabref{tab:rv_k2p} and \figref{fig:k2p_rv}.
We now proceed to describe these observations in detail.

\begin{deluxetable}{DDDr}
  \tablecaption{Precise RV Measurements of \starone{}
    \label{tab:rv_toi954}}
  \tabletypesize{\scriptsize}
  \tablehead{%
    \multicolumn2c{BJD}
    & \multicolumn2c{RV}
    & \multicolumn2c{Uncertainty}
    & \colhead{Instrument} \\
    &
    & \multicolumn2c{(\si{\meter\per\second})}
    & \multicolumn2c{(\si{\meter\per\second})}
    &
    }
  \decimals
\startdata
2458536.59568 & -8812.91816 &  8.13894 & CHIRON \\
2458538.60658 & -8839.10797 & 17.70062 & CHIRON \\
2458539.60827 & -8800.58690 & 14.49605 & CHIRON \\
2458540.59466 & -8818.69516 & 18.01763 & CHIRON \\
2458541.57702 & -8844.42831 & 13.41980 & CHIRON \\
2458714.906319 & -7347.17 &  8.83 & CORALIE \\
2458716.832690 & -7329.20 &  9.01 & CORALIE \\
2458718.846938 & -7369.61 &  8.71 & CORALIE \\
2458727.837791 & -7325.22 &  8.84 & CORALIE \\
2458729.867972 & -7373.46 &  9.72 & CORALIE \\
2458730.875885 & -7334.57 & 11.22 & CORALIE \\
2458739.756150 & -7360.71 & 11.34 & CORALIE \\
2458740.811585 & -7361.33 & 10.84 & CORALIE \\
2458741.835606 & -7339.48 & 10.79 & CORALIE \\
2458744.829962 & -7349.56 & 11.96 & CORALIE \\
2458745.737422 & -7321.07 & 35.27 & CORALIE \\
2458746.749066 & -7325.21 & 14.84 & CORALIE \\
2458747.815505 & -7383.09 & 12.90 & CORALIE \\
2458751.740085 & -7355.34 & 15.11 & CORALIE \\
2458752.761760 & -7361.95 &  9.22 & CORALIE \\
2458753.728396 & -7302.87 & 17.82 & CORALIE \\
2458754.766691 & -7365.72 & 10.48 & CORALIE \\
2458755.885741 & -7363.77 &  9.77 & CORALIE \\
2458756.781573 & -7332.19 & 12.27 & CORALIE \\
2458753.799037 & -7303.56 & 3.63 & HARPS \\
2458754.756737 & -7324.01 & 2.46 & HARPS \\
2458755.813724 & -7340.08 & 1.49 & HARPS \\
2458756.749928 & -7310.85 & 2.72 & HARPS \\
2458735.11924305 &  32.749 &  5.2134 & MINERVA-Australis \\
2458737.17597222 & -3.6402 &  4.8895 & MINERVA-Australis \\
2458741.08650462 &  20.146 &  9.9089 & MINERVA-Australis \\
2458742.09328703 &  34.614 &  13.336 & MINERVA-Australis \\
2458743.08164351 &  27.424 &  10.512 & MINERVA-Australis \\
2458758.09859953 &  1.6529 &  9.5847 & MINERVA-Australis \\
2458776.07408564 &  35.803 &  8.5726 & MINERVA-Australis \\
2458777.15123842 &  7.3249 &  4.8943 & MINERVA-Australis \\
2458795.08677083 & -7.3261 &  5.2078 & MINERVA-Australis \\
2458796.16503472 & -25.319 &  4.4132 & MINERVA-Australis \\
2458797.08497685 & -12.976 &  4.3832 & MINERVA-Australis \\
2458798.21734953 & -38.352 &  4.4889 & MINERVA-Australis \\
2458675.92609 &   8.36 & 1.44 & PFS \\
2458679.92171 &  10.10 & 1.17 & PFS \\
2458682.90503 &  -8.29 & 1.08 & PFS \\
2458683.94257 &   6.87 & 1.37 & PFS \\
2458684.94079 & -32.33 & 0.90 & PFS \\
2458737.88787 & -18.21 & 1.34 & PFS \\
2458738.83838 &   0.00 & 1.27 & PFS \\
2458741.83181 & -14.90 & 1.13 & PFS \\
\enddata
  \tablecomments{
    The RVs given represent absolute relative motion to the solar system barycenter,
    except for those from \minervaaus{} and \pfs{}, where the mean relative motion has been subtracted.
    See \tabref{tab:spec} for the constant RV offsets derived by the global model for each instrument.
    \added{The \minervaaus{} measurement at \bjdtdb{} \num{2458798.2173} is excluded from the global model because it is a $> 4\sigma$ outlier.} \\
    (This table is available in machine-redable form.)
  }
\end{deluxetable}

\begin{deluxetable}{DDDr}
  \tablecaption{Precise RV Measurements of \startwo{}
    \label{tab:rv_k2p}}
  \tabletypesize{\scriptsize}
  \tablehead{%
    \multicolumn2c{BJD}
    & \multicolumn2c{RV}
    & \multicolumn2c{Uncertainty}
    & \colhead{Instrument} \\
    &
    & \multicolumn2c{(\si{\meter\per\second})}
    & \multicolumn2c{(\si{\meter\per\second})}
    &
    }
  \decimals
\startdata
2458040.70773863 & -16990.7 & 10.5 & FEROS \\
2458063.59792602 & -17046.9 &  9.3 & FEROS \\
2458064.54347944 & -17008.4 &  9.1 & FEROS \\
2458064.61015499 & -17001.6 &  9.5 & FEROS \\
2458065.63009163 & -16991.1 & 10.6 & FEROS \\
2458065.57747400 & -16978.0 &  9.0 & FEROS \\
2458067.51507559 & -16972.5 &  8.1 & FEROS \\
2458073.51680660 & -17044.5 &  7.7 & FEROS \\
2458261.90205817 & -17043.0 &  7.6 & FEROS \\
2458262.90253453 & -16999.1 &  7.5 & FEROS \\
2458262.88111421 & -16986.3 &  8.3 & FEROS \\
2458262.92396402 & -17020.9 &  7.6 & FEROS \\
2458263.90571009 & -16992.3 &  7.6 & FEROS \\
2458263.92368049 & -16980.6 &  7.5 & FEROS \\
2458263.88775467 & -16995.1 &  7.5 & FEROS \\
2458264.85218595 & -17021.5 &  8.9 & FEROS \\
2458264.87014076 & -17000.3 &  8.0 & FEROS \\
2458264.88809828 & -16996.0 &  8.2 & FEROS \\
2458264.83421933 & -16995.5 &  9.1 & FEROS \\
2458265.91063007 & -17005.3 &  8.2 & FEROS \\
2458265.85674146 & -17025.6 &  9.0 & FEROS \\
2458265.89267519 & -17006.7 &  9.3 & FEROS \\
2458265.87470911 & -17004.9 &  8.5 & FEROS \\
2458311.86733030 & -17005.6 &  7.7 & FEROS \\
2458312.83754233 & -17014.2 &  7.1 & FEROS \\
2458313.81862457 & -17002.4 &  7.5 & FEROS \\
2458314.78898939 & -17003.7 &  9.2 & FEROS \\
2458063.53094302 & -17009.9 &  5.2 & HARPS \\
2458064.52324815 & -16987.5 &  6.8 & HARPS \\
2458066.52560427 & -16985.0 &  8.6 & HARPS \\
2458067.55479447 & -16972.1 &  5.5 & HARPS \\
2458314.80212100 & -16964.4 &  8.5 & HARPS \\
2458316.79691151 & -16974.2 & 17.3 & HARPS \\
2458321.73828811 & -16995.9 & 19.8 & HARPS \\
2458322.75617163 & -16993.3 & 17.2 & HARPS \\
2458323.74986105 & -17002.3 & 12.2 & HARPS \\
2458332.75758515 & -16970.9 &  7.3 & HARPS \\
2458333.80634123 & -16988.3 &  7.3 & HARPS \\
2458366.64961358 & -16954.1 &  6.8 & HARPS \\
2458367.64086452 & -16960.8 &  7.3 & HARPS \\
2458262.89585 &  -18.20 & 2.40 & PFS \\
2458264.89722 &   -1.71 & 2.67 & PFS \\
2458295.92099 &   -1.29 & 1.44 & PFS \\
2458297.89449 &  -23.55 & 1.48 & PFS \\
2458298.87651 &  -24.14 & 1.51 & PFS \\
2458355.79845 &   26.91 & 1.51 & PFS \\
2458406.56625 &   17.06 & 2.02 & PFS \\
2458408.56428 &   -0.47 & 2.01 & PFS \\
2458415.55544 &   17.61 & 1.71 & PFS \\
2458417.55814 &   25.62 & 2.11 & PFS \\
\enddata

  \tablecomments{
    The RVs given represent absolute relative motion to the solar system barycenter,
    except for those from \pfs{}, where the mean relative motion has been subtracted.
    See \tabref{tab:spec} for the constant radial velocity offsets derived by the global model for each instrument. \\
    (This table is available in machine-redable form.)
  }
\end{deluxetable}

One of our main sources of precise RV data was 
the iodine-fed Planet Finder Spectrograph \citep[PFS;][]{Cra06,Cra08,Cra10}
on the \SI{6.5}{\m} Magellan II Telescope at Las Campanas Observatory in Chile.
In 2019 July and September,
\starone{} was observed for a total of eight RVs with a 20~minute exposure time for \tess{} follow-up.
The iodine data and iodine-free templates were taken through a \ang{;;0.3} slit, 
resulting in $R \approx \num{130000}$.
The mean internal uncertainty was \SI{1.2}{\m\per\s},
and the signal-to-noise ratio (S/N) ranged from 45 to 75 in the iodine region at the peak of the blaze.
Between 2018 May and October,
\startwo{} was observed for a total of 10 RVs for \ktwo{} follow-up.
The same \ang{;;0.3} slit was used with $3 \times 3$ binning, resulting in
$R \approx \num{43000}$.
The exposure times ranged from 33 to 50~minutes,
achieving an S/N of \circa 30--90 in the iodine region at the peak of the blaze
and a mean internal uncertainty of \SI{1.9}{\m\per\s}.
All PFS data were reduced with a custom IDL pipeline that flat-fielded, removed cosmic rays, and subtracted scattered light.
Further details about the iodine cell RV extraction method can be found in \cite{But96}.

The High Accuracy Radial velocity Planet Searcher
\citep[\harps{};][]{2003Msngr.114...20M}
also contributed a significant portion of the precise RV data used in this paper.
\harps{} is fiber-fed by the Cassegrain focus of the \SI{3.6}{\m} telescope at La Silla Observatory in Chile.
We obtained four spectra of \starone{} during consecutive nights,
\ut{} 2019 September 27--30, in good seeing conditions (\circa \ang{;;1.0}).
The exposure time was 20 minutes, leading to an S/N of 33.6--80.4 at \SI{5000}{\angstrom}.
\harps{} also observed \startwo{} for two pairs of consecutive nights in \ut{} 2017 November,
five nights in \ut{} 2018 July,
and a pair of consecutive nights each in \ut{} 2018 August and September,
for a total of 13 spectra.
We adopted exposure times of \SIlist{1500;1800}{\s} for \startwo{},
which resulted in spectra with an S/N per resolution element of 17--35.

The rest of the spectrographs observed either one of the two planets.
The \feros{} spectrograph \citep{1999Msngr..95....8K},
mounted on the MPG/ESO \SI{2.2}{\m} telescope at La Silla observatory in Chile,
observed 27 spectra of \startwo{} at $R \approx \num{48000}$
between \ut{} 2017 October 14 and 2018 July 15.
Each spectrum achieved an S/N of 50 per spectral resolution element with exposure times of \SI{1200}{\second}.
The instrumental drift was determined via comparison with a simultaneous fiber illuminated with a ThAr+Ne lamp.
The data were processed with the CERES suite of echelle pipelines \citep{2017PASP..129c4002B},
which produce RVs and bisector spans in addition to reduced spectra.

The final two spectrographs observed \starone{} only.
We obtained a total of seven spectra
using the \chiron{} echelle spectrograph \citep{2013PASP..125.1336T}
on the SMARTS \SI{1.5}{\m} telescope located at the Cerro Tololo Inter-American Observatory (CTIO), Chile,
between \ut{} 2019 February 22 and 27.
\chiron{} was fed via an image slicer and a fiber bundle,
yielding a resolving power of $R \approx \num{80000}$ over the wavelength range of \SIrange[range-phrase=--]{4100}{8700}{\angstrom}.
Our observations were obtained at an exposure time of \SI{900}{\s},
achieving an average S/N of 65 over the Mg\,b lines.
The RVs were derived via a least-squares deconvolution
between the observed spectra and synthetic nonrotating spectral templates generated via the ATLAS9 stellar models
\citep{Castelli:2004}.

Last but not least, we observed \starone{} with the fiber-fed spectrograph \coralie{}
\cite[$R \approx 60000$,][]{2001Msngr.105....1Q}
on the Swiss \SI{1.2}{\m} Euler telescope located at La Silla Observatory (ESO, Chile).
We acquired 19 RV measurements between \ut{} 2019 August 19 and September 29,
with the first \coralie{} fiber on the star and the second one
connected to a Fabry--P\'{e}rot etalon for simultaneous wavelength calibration,
yielding an S/N of 10.0--31.7 over the wavelength range of \SIrange{3900}{6800}{\angstrom}.
The RVs were computed for each epoch by cross-correlating
with a G2 mask using the standard \coralie{} Data Reduction Software \citep[DRS;][]{Pepe2002}, which also produced
various line-profile diagnostics such as cross-correlating bisector span, FWHM, and contrast.
The typical RV uncertainty achieved for the star was \SI{11}{\m\per\s}.

We observed \starone{} with the \minervaaus{} telescope array \citep{2019PASP..131k5003A,2020arXiv200107345A}
at Mt.\ Kent Observatory in Queensland, Australia,
for 12 RV measurements between \ut{} 2019 September 8 and November 10.
\minervaaus{} is a set of PlanetWave CDK700 telescopes connected by fibers
to a single KiwiSpec R4-100 spectrograph \citep{2012SPIE.8446E..88B},
yielding a resolution of $R \approx 80,000$ with wavelength coverage from \SIrange{5000}{6300}{\angstrom}.
We calculated the radial velocities using least-squares analysis,
correcting for instrumental drift with simultaneous observations of a ThAr lamp.
We measured an RMS scatter of \SI{18}{\m\per\s} from the median Markov Chain Monte Carlo (MCMC) solution to the global model (\secref{sec:global_model}).
We discarded one RV measurement made on \ut{} 2019 November 10 (\bjdtdb{} \num{2458798.2173})
for the global model fitting (\secref{sec:global_model})
because it was a $> 4\sigma$ outlier.

\subsection{High Spatial Resolution Imaging}

\begin{figure}
    \centering
    \includegraphics[width=\columnwidth]{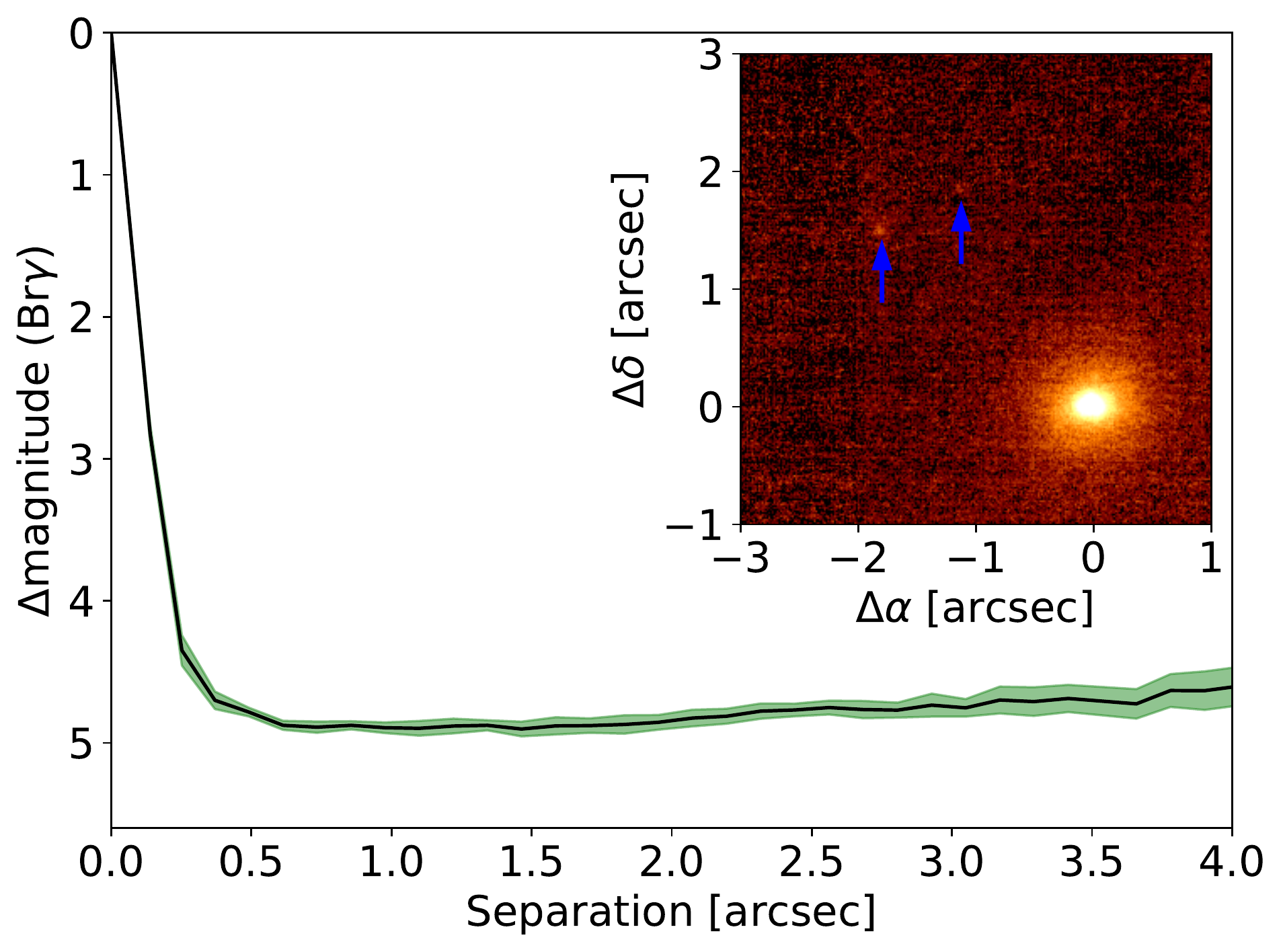}
    \caption{The VLT/NaCo image of \starone{} and sensitivity to background companions at close separations.
    The sensitivity is shown in the main plot, and the inset shows the central portion of the image itself,
    with the star offset so that the companions can be more clearly seen.
    Both visual companions are highlighted with blue arrows.
    The northern companion is a marginal detection of just $3\sigma$,
    but the southern companion is well above our detection threshold.}
    \label{fig:ao_naco}
\end{figure}

We collected AO images of \starone{} with VLT/NAOS--CONICA \cite[NaCo; ][]{lenzen_naos-conica_2003,rousset_naos_2003}
on \ut{} 2019 September 14 to search for nearby companions.
Nine exposures were collected in the Br$\gamma$ filter, each with an integration time of \SI{30}{\second}.
The telescope was dithered by \ang{;;2} between each individual image.
We used a custom IDL code to process the data following standard practice;
bad pixels were removed, data were flat-fielded, a sky background was constructed from the dithered frames and subtracted, and the images were finally aligned and coadded.
When visually inspecting the data, we found two visual companions:
a firm detection of a companion at \ang{;;2.3} and a marginal detection of a companion at \ang{;;2.1}, at approximately $3\sigma$.
Both companions are to the NE of the target and mutually separated by \SI{781}{\milliarcsecond}.

We calculate our sensitivity to background stars by injecting model companions into the data and scaling their brightness until they are detected at $5\sigma$.
This process is repeated at a range of angles and radii, and the final sensitivity is averaged azimuthally.
The brighter companion is masked out during this sensitivity calculation.
A plot of the sensitivity to companions is shown in \figref{fig:ao_naco}, which also includes a thumbnail image of the target with the two companions highlighted.

\begin{figure}
    \centering
    \includegraphics[width=\columnwidth]{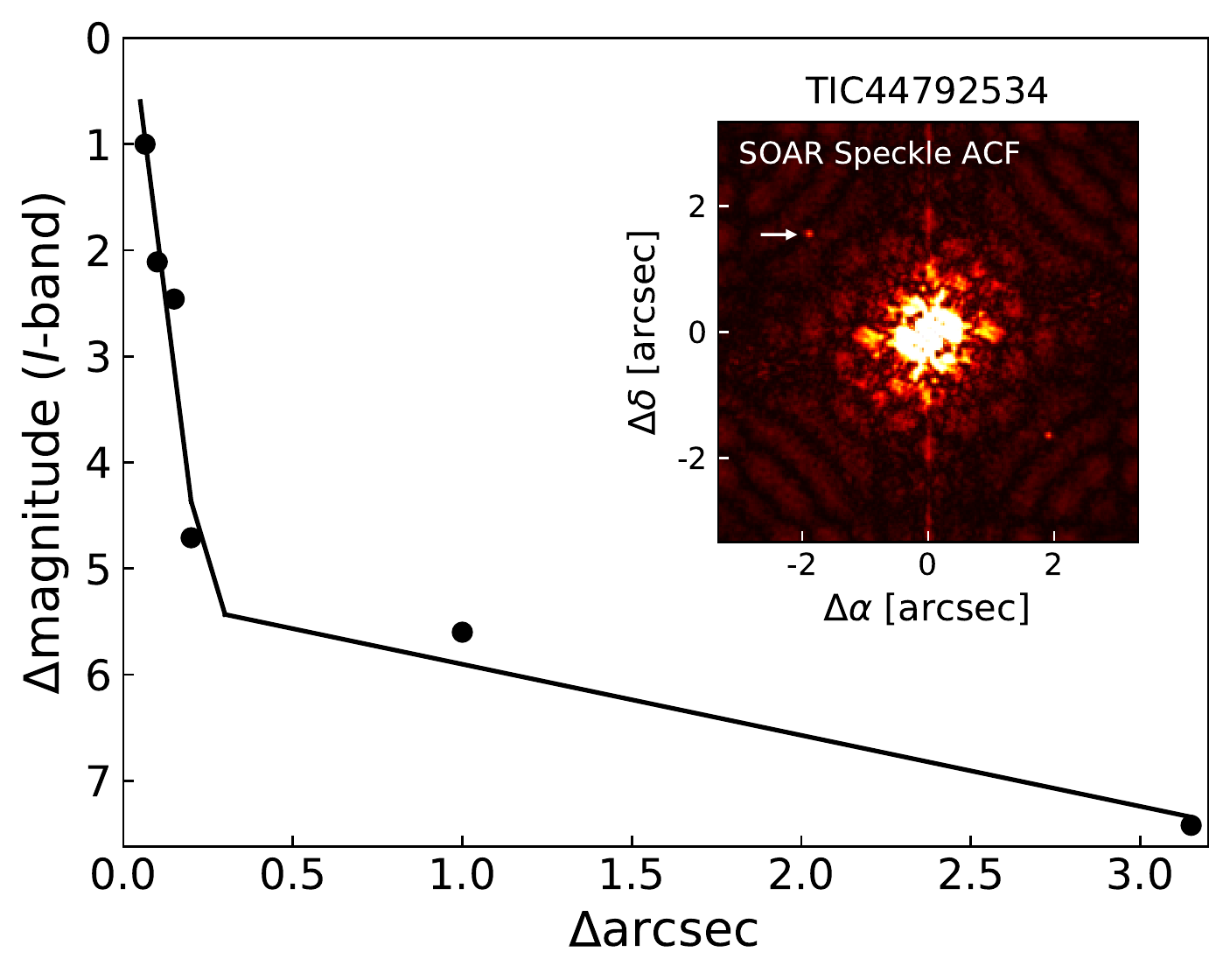}
    \caption{\soar{} observation of \starone{}.
    The $5\sigma$ detection sensitivity is plotted
    with the speckle imaging ACF inset.
    The companion is mirrored in the ACF,
    but its true position ($\mathrm{PA} = \ang{50;;}$) is marked with an arrow.}
    \label{fig:speckle_soar}
\end{figure}

We also searched for nearby sources to \starone{} with \soar{} speckle imaging \citep{2018PASP..130c5002T}
on \ut{} 2019 August 12,
observing in a similar visible bandpass as \tess{}.
Further details of the observations are available from
\cite{2020AJ....159...19Z}.
Confirming the finding from NaCo,
a faint companion ($\delta m_{I} = 6.2$) was detected at a separation of \ang{;;2.35}.
The contamination from the star is negligible, implying a planetary radius correction factor of $1.002$ due to dilution of the transit depth.
The $5\sigma$ detection sensitivity and the speckle autocorrelation function (ACF) from the \soar{} observation are plotted in \figref{fig:speckle_soar}.

\begin{figure}
  \centering
  (a) 2017 August 14 \\
  \includegraphics[width=\columnwidth]{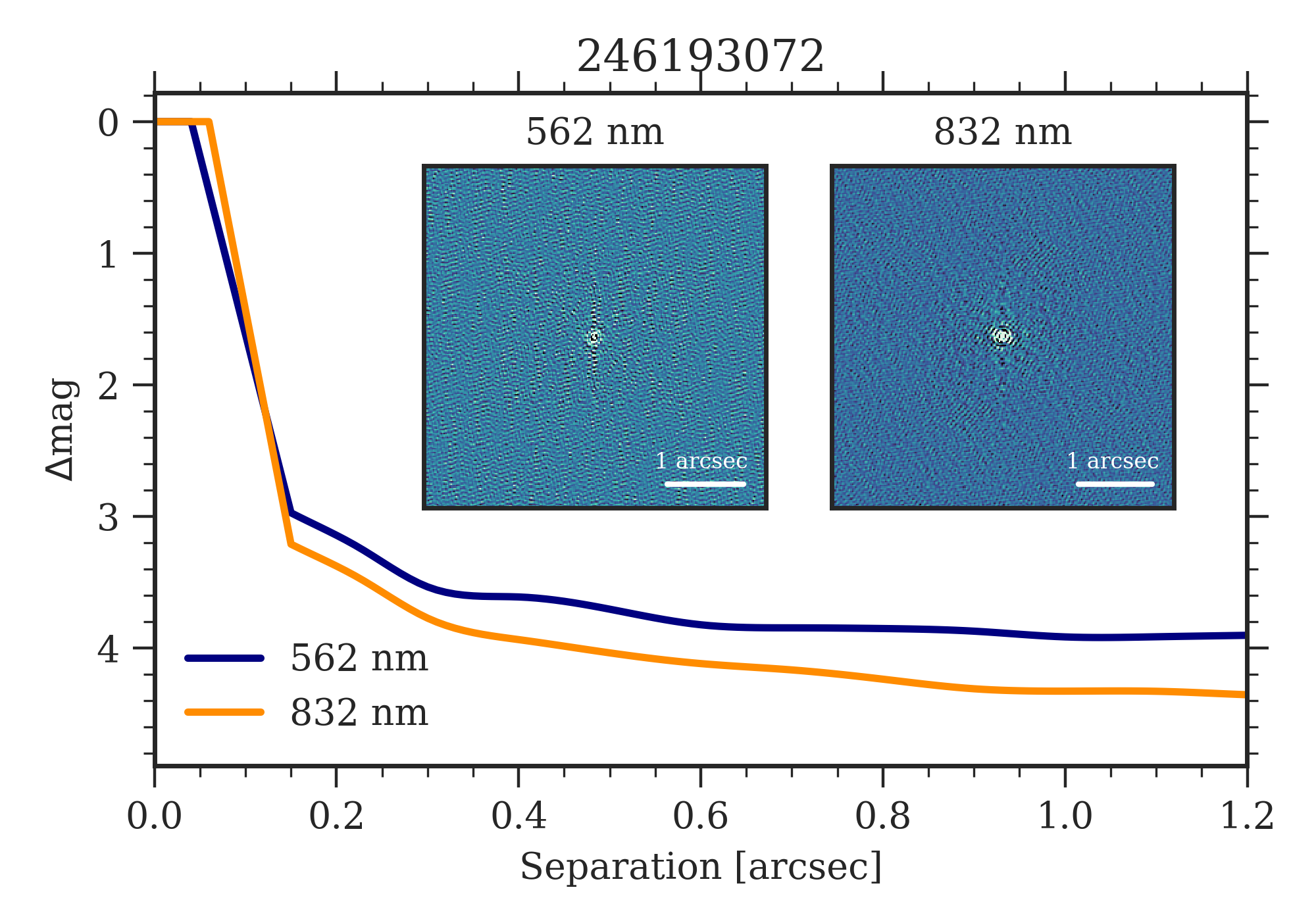}
  (b) 2017 September 5 \\
  \includegraphics[width=\columnwidth]{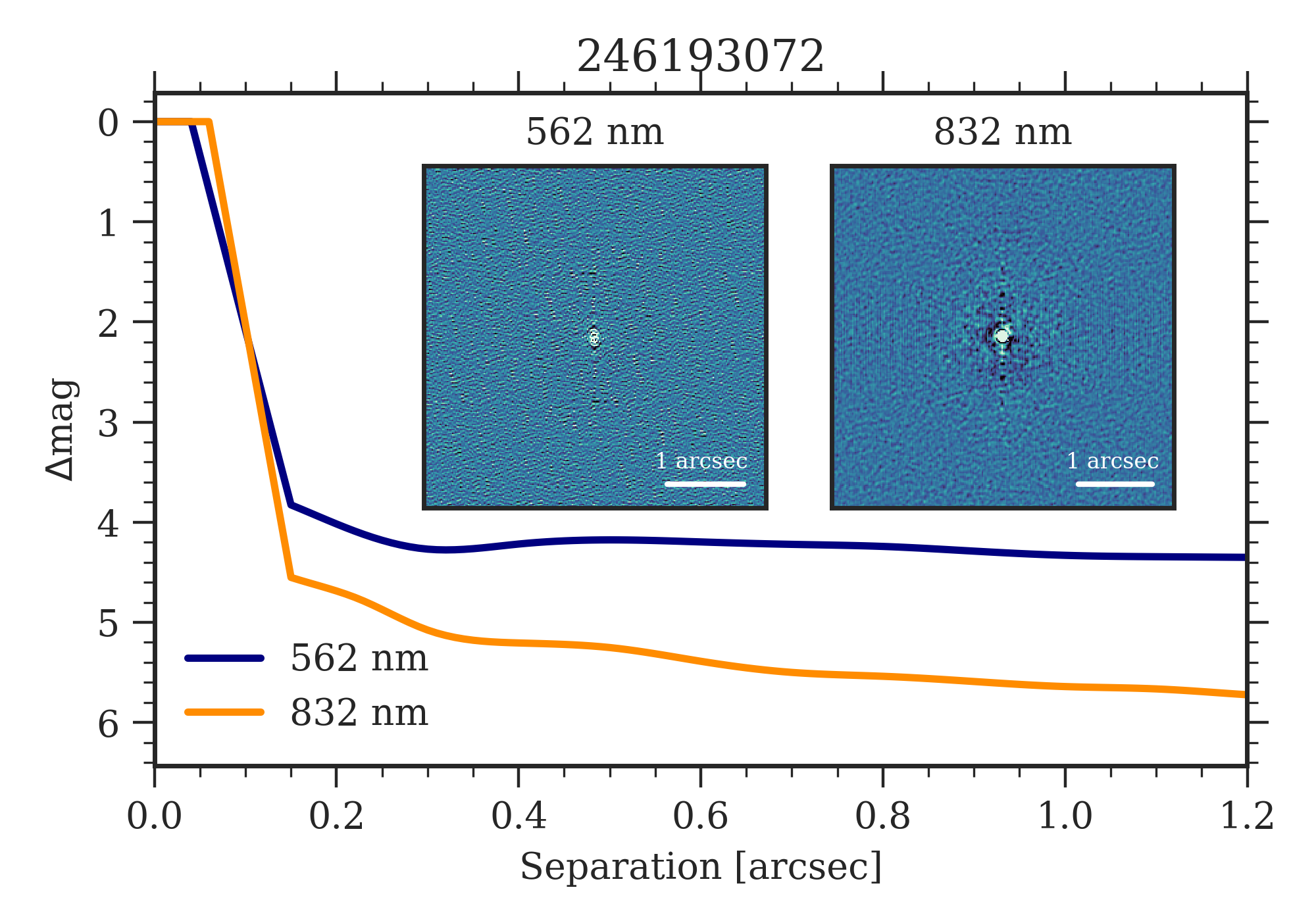}
  \caption{WIYN observations of \startwo{}
    on \ut{} 2017 August 14 (top)
    and \ut{} 2017 September 5 (bottom).
    The $5\sigma$ contrast curves are plotted with the reconstructed images.
    No companions are detected.
  }
  \label{fig:speckle_wiyn}
\end{figure}

We acquired speckle imaging data of \startwo{}
with the NN-Explore Exoplanet Stellar Speckle Imager
\citep[NESSI;][]{2018PASP..130e4502S}
installed on the \SI{3.5}{\m} telescope at WIYN Observatory
on 2 nights, \ut{} 2017 August 14 and September 5.
The observations were in two passbands, \SIlist{562;832}{nm}.
Plots of the contrast curves with the reconstructed images inset
can be found in \figref{fig:speckle_wiyn}.
The conditions were slightly more favorable on September 5,
which led to better contrast in both channels.
No companions were detected.

\section{Analysis}
\label{sec:analysis}

\subsection{Confirmation of \planetone{} and \planettwo{}}

Our follow-up observations rule out common astrophysical false positives that may be mistaken for a transiting planet.
The transit signals for both \planetone{} and \planettwo{} are confirmed to be on target by
ground-based observations.
We have also detected strong RV signals matching
the transit-derived orbital periods for both planets.
We can rule out that the \planetone{} signal is contaminated by the stars
at a separation of \circa \ang{;;2} because the \pfs{} slit (\ang{;;0.3}) excludes those sources.
Speckle imaging did not detect any stars that could contaminate the signal of \planettwo{}.
Therefore, we confirm the two planets with high confidence.

\subsection{Stellar Parameters}

\subsubsection{Reconnaissance Spectra}

We used the Stellar Parameter Classification tool
\citep[SPC;][]{2012Natur.486..375B}
to extract stellar parameters
such as \tempeff, \logg, metallicity, and $v \sin i$
from the \tres{} spectra.
We were able to determine that
\starone{} is slightly evolved off the main sequence
and that \startwo{} is a main-sequence star with
$\tempeff = \SI{5359 \pm 50}{\kelvin}$,
$\logg = 4.6 \pm 0.1$,
$\feh = 0.16 \pm 0.08$,
and $\vsini = \SI{1.9 \pm 0.5}{\km\per\s}$.

Because we only obtained a relatively low S/N spectrum for \starone{} from \tres{},
we opted to use the stellar parameters derived from the average of seven \chiron{} spectra.
The \chiron{} spectra were calibrated against a library of \circa \num{10000} observed spectra
classified by the SPC pipeline, interpolated via a gradient-boosting regressor.
We derived
$\tempeff = \SI{5710 \pm 51}{\kelvin}$,
$\logg = 4.00 \pm 0.05$,
$\feh = 0.21 \pm 0.05$,
and $\vsini = \SI{5.6 \pm 0.5}{\km\per\s}$
for \starone{}
from the \chiron{} spectra. 

As a sanity check, we also determined the stellar parameters of \startwo{} from the coadded \fies{} spectra.
We followed the method outlined in \citet{Gandolfi2017} and used a customized IDL software suite
that fitted spectral features sensitive to different photospheric parameters with ATLAS~9 model atmospheres \citep{Castelli:2004}.
The results corroborated the stellar parameters derived from the \tres{} spectra to within $1\sigma$.

The \tempeff{} and \feh{} of both stars are used as Gaussian priors in our global model fitting
(\secref{sec:global_model}).
The \logg{} values, however, are not used to constrain the global model
and thus serve as an independent check on the stellar density as constrained by the transit light curve.

\subsubsection{Spectral Energy Distribution}
\label{sec:sed}

\begin{figure}[ht]
    \centering
    (a) \starone{} \\
    \includegraphics[width=\columnwidth]{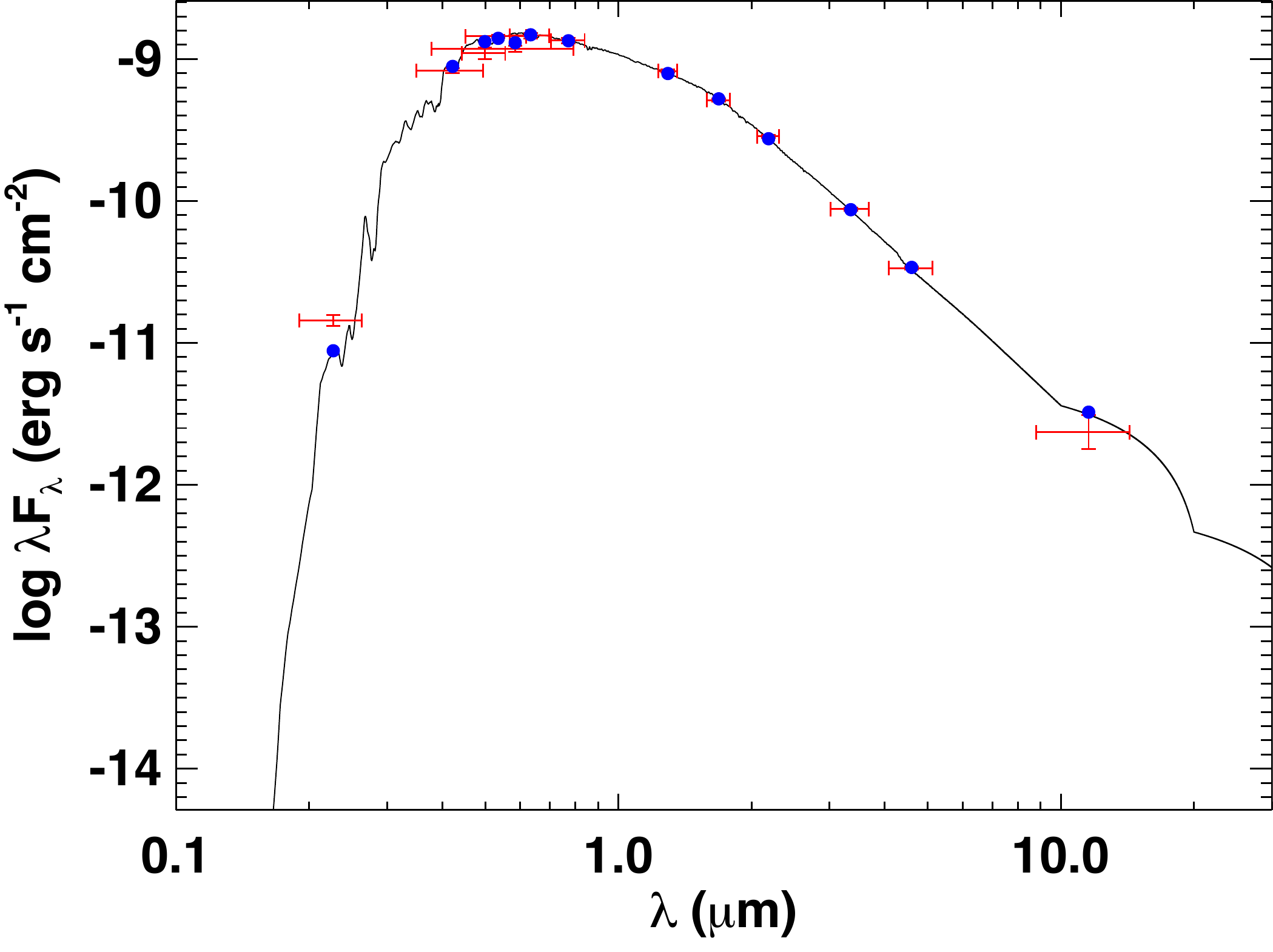} \\
    (b) \startwo{} \\
    \includegraphics[width=\columnwidth]{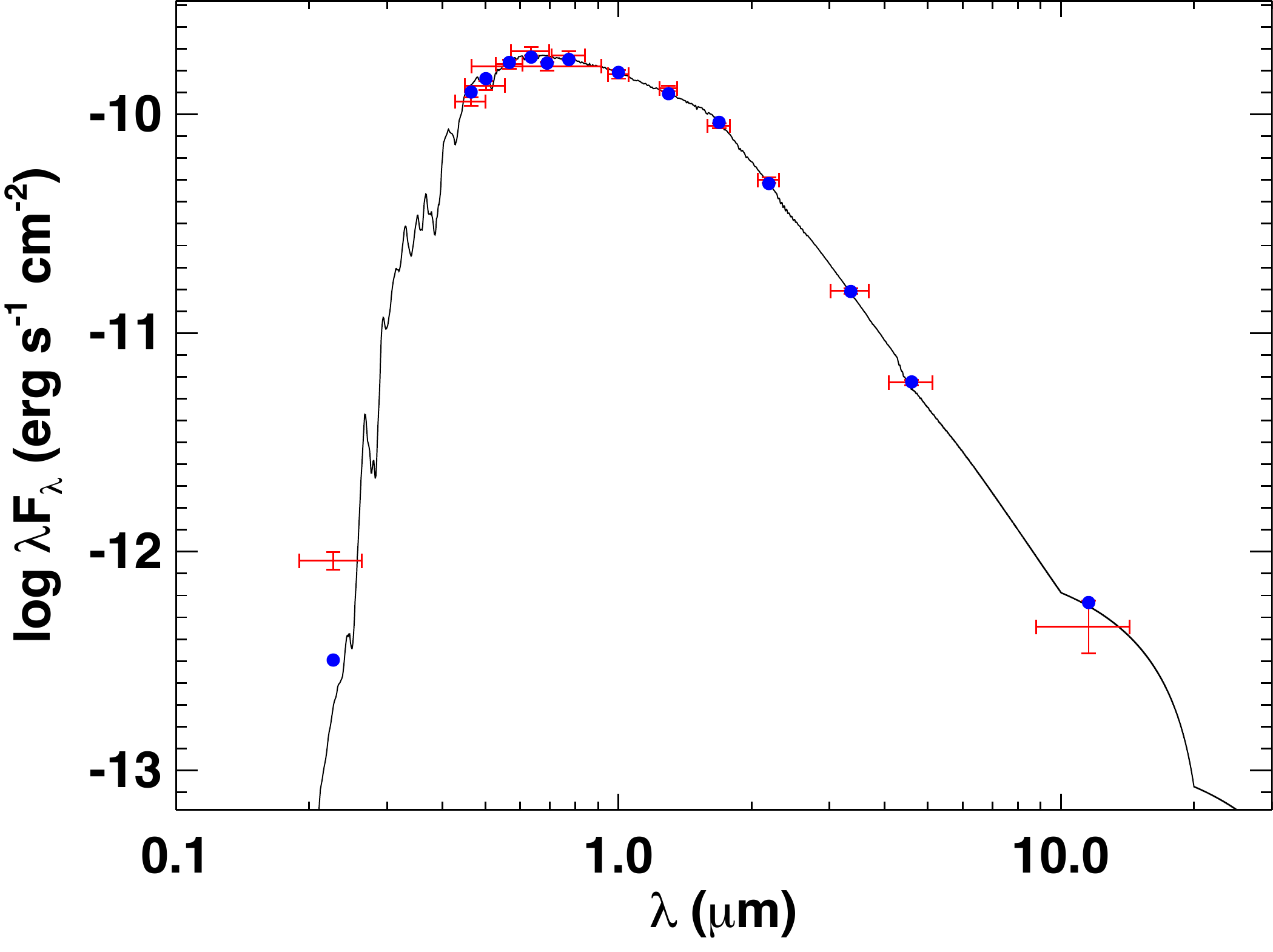} \\
    \caption{The SED for \starone{} (top) and \startwo{} (bottom).
    Red symbols represent the observed photometric measurements,
    where the horizontal bars represent the effective width of the passband.
    Blue symbols are the model fluxes from the best-fit Kurucz atmosphere model (black). }
    \label{fig:sed}
\end{figure}

As an independent check on the derived stellar parameters,
we performed an analysis of the broadband spectral energy distribution (SED)
together with the \gaia{} parallax in order to determine an empirical measurement of the stellar radius, following the procedures described by \citet{Stassun:2016} and \citet{Stassun:2017,Stassun:2018}.
We obtained the $B_\mathrm{T} V_\mathrm{T}$ magnitudes from Tycho-2,
the \emph{BVgri} magnitudes from APASS,
the $JHK_S$ magnitudes from the Two Micron All Sky Survey (2MASS),
the W1--W3 magnitudes from the Wide-field Infrared Survey Explorer,
the \gaiaG{} magnitude from \gaia{},
and the Galaxy Evolution Explorer near-UV flux.
Together, the available photometry spanned the full stellar SED over the wavelength range \SIrange[range-phrase=--]{0.2}{10}{\micro\meter}
(see \figref{fig:sed}).

We performed separate fits for \starone{} and \startwo{}
using Kurucz stellar atmosphere models
with the priors on effective temperature (\tempeff), surface gravity (\logg), and metallicity (\feh) from the reconnaissance spectroscopic values.
The remaining free parameter was the extinction ($A_V$),
which we limited to the maximum line-of-sight extinction from the \citet{Schlegel:1998} dust maps.
The resulting fits were very good (\figref{fig:sed}),
with a reduced $\chi^2 = 2.3$ and a best-fit extinction of $A_V = 0.06 \pm 0.06$ for \starone{}
and a reduced $\chi^2 = 1.58$ and a best-fit $A_V = 0.09 \pm 0.03$ for \startwo{}.
We adopted these $A_V$ values as Gaussian priors for bolometric corrections in our global model
(\secref{sec:global_model}).
The NUV flux of \starone{} implies a modest level of chromospheric activity.

Integrating the unextincted model SEDs gave a bolometric flux of
$F_{\mathrm{bol}}
= \SI[allow-number-unit-breaks=true,parse-numbers=false]{(1.975 \pm 0.093) \times 10^{-9}}{\erg\per\s\per\cm\squared}$
at Earth for \starone{} and
$F_{\mathrm{bol}}
= \SI[allow-number-unit-breaks=true,parse-numbers=false]{(2.880 \pm 0.033) \times 10^{-10}}{\erg\per\s\per\cm\squared}$
at Earth
for \startwo{}.
Taking the $F_{\mathrm{bol}}$ and \tempeff{} together with the \gaia{} parallax,
adjusted by \SI[retain-explicit-plus]{+0.08}{\milliarcsecond} to account for the systematic offset reported by \citet{StassunTorres:2018},
gave stellar radii of \SI{1.898 \pm 0.058}{\radius\sun} for \starone{}
and \SI{0.804 \pm 0.017}{\radius\sun} for \startwo{}.

\subsection{Global Modeling}
\label{sec:global_model}

For each of the \starone{} and the \startwo{} planetary systems,
we used the \emcee{} Python package
\citep{emcee_paper}
to perform a global MCMC model fit for
planet properties,
orbital parameters,
and stellar properties.
We ran the MCMC sampler over \num{500000} iterations for each system,
with 350 walkers for \starone{}
and 250 walkers for \startwo{}.
We calculated the integrated autocorrelation time,
or roughly the number of iterative steps it takes for a walker chain to generate an independent sample from the posterior distribution,
following \cite{goodman2010} and empirically found it to be in the range of \numrange{7500}{20000} steps for the \starone{} system model
and \numrange{6000}{11000} steps for the \startwo{} system model.
Based on this autocorrelation timescale,
we discarded the first \num{100000} iterations as burn-in.

\begin{deluxetable}{l@{\hspace{52pt}}r@{}l@{\hspace{32pt}}r@{}l}
  \tablecaption{Quadratic Limb-darkening Parameters \label{tab:limbdark}}
  \tabletypesize{\footnotesize}
  \tablecolumns{5}
  \tablehead{
    \colhead{Filter}
    & \multicolumn2c{$u_1$}
    & \multicolumn2c{$u_2$}
  }
  \startdata
  \cutinhead{\starone}
  \tess{} & \ToipUone & \ToipUoneErr & \ToipVone & \ToipVoneErr \\
  $r$ & \ToipUtwo & \ToipUtwoErr & \ToipVtwo & \ToipVtwoErr \\
  \filterzs & \ToipUthree & \ToipUthreeErr & \ToipVthree & \ToipVthreeErr \\
  \cutinhead{\startwo}
  \kepler{} & \KtwopUone & \KtwopUoneErr & \KtwopVone & \KtwopVoneErr \\
  \filterRc{} & \KtwopUtwo & \KtwopUtwoErr & \KtwopVtwo & \KtwopVtwoErr \\
  \enddata
  \tablecomments{In the MCMC global modeling fit, we used the triangle sampling parameterization $(q_1, q_2)$ of the quadratic limb-darkening law by \cite{2013MNRAS.435.2152K}.
  We used the least-squares values
  tabulated by \cite{2017AnA...600A..30C} for \tess{}
  and \cite{2013AnA...552A..16C} for all other filters
  as Gaussian priors,
  taking the average deviation between the least-squares
  and the flux-conservation methods as the standard error.
  Here we report the posterior distribution (16th, 50th, and 84th percentiles) of the
  conventional $(u_1, u_2)$ quadratic limb-darkening parameters,
  converted from the MCMC samples of $(q_1, q_2)$.
  }
\end{deluxetable}

For each system, we modeled the detrended and normalized light curves
(after rejecting out-of-transit outliers $> 4\sigma$) with the \batman{} package
\citep{batman_paper}.
In each model, we constrained the impact parameter so that the planet transits
(i.e. $ |b| <  1 + \si{\radius\planet} / \si{\radius\etoile}$),
and the radius of the planet is additionally constrained to be within 0.4 of the stellar radius.
We also restricted the eccentricity to be less than $0.9$ because the Kepler solver
in \batman{} may fail at extremely high eccentricities.
For each transit in the \lcogt{}, \pest{}, and \ktwo{} campaign~19 data,
we also simultaneously fitted and subtracted a weighted least-squares linear trend across the transit
from the model residuals before calculating the $\chi^2$ for the model posterior.

We added a variety of additional parameters to improve the light curves' fit to the data;
these additional parameters are reported in \tabref{tab:phot}.
We included a dilution factor for the \tess{} light curve
with a Gaussian prior centered at $0.0460$ and a width of $0.0046$,
which is the theoretical dilution from the unresolved star in our aperture
calculated using \tess{} magnitudes in TIC\thinspace 8.
As the TFA detrending method tended to make the transit depth shallower than the true transit in \hatsouth{} light curves,
we included an additional dilution factor in the model
with a flat prior in $(0, 1)$.
To account for possible underestimation of the noise of \lcogt{} and \pest{} measurements,
we multiplied the quoted \lcogt{} noise by a factor $\geq 1$
and added a nonnegative jitter term in quadrature to the quoted \pest{} noise
as free parameters in their respective global model.
We did not add any additional jitter to either the \tess{} or \ktwo{} light curves
because the global model consistently preferred a value of zero during trial runs.

We note that the MCMC posterior solution for the \starone{} system prefers a value much lower than the prior for the \tess{} light-curve dilution factor.
We believe that this apparent discrepancy arises because the dilution-corrected \tess{} light curve
and the \lcogt{} light curves imply slightly different transit depths for the planet,
which is captured by the global model in the larger uncertainties
for the radius of \planetone{} it reports.

We used the triangular sampling of the quadratic limb-darkening law coefficients recommended by \cite{2013MNRAS.435.2152K},
and constrained the coefficients using Gaussian priors. The priors were centered on values interpolated using the gradient-boosting regressor in the \sklearn{} package from the
least-squares fitted values tabulated by \cite{2017AnA...600A..30C} for \tess{}
and \cite{2013AnA...552A..16C} for all other filters.
The posterior distributions of the quadratic limb-darkening parameters are reported in \tabref{tab:limbdark}.

We modeled the corresponding RVs with \radvel{}
\citep{radvel_paper} in the joint fit.
For each spectroscopic instrument, we introduced a constant offset term $\gamma$ to the reported RVs
and added a jitter term in quadrature with the reported errors,
allowing the model to adjust each freely.
The $\gamma$ terms are unconstrained,
while the jitter terms are constrained to be nonnegative.
Those additional RV parameters are reported in \tabref{tab:spec}.
We did not find any statistically significant long-term trends in the RV measurements of either star.

To simultaneously constrain stellar properties,
we fitted for initial stellar mass, initial stellar metallicity, and age in our global model.
Those three parameters served as independent variables of the \mist{} isochrones \citep{2016ApJ...823..102C,2016ApJS..222....8D},
which we interpolated with the \isochrones{} Python package
\citep{isochrones_python}.
We constrained the initial stellar mass to not exceed the valid range for the \mist{} isochrones
$[0.1, 300]\,\si{\mass\sun}$
and the stellar age to $[10^{-4}, 14]\,\si{\giga\year}$.
The isochrone interpolation produced the current mass, metallicity (\feh{}), \tempeff{}, and \logg{} of the star,
as well as the predicted \gaia{} absolute magnitudes.
The transit light curves implicitly constrained the mass and \logg{}.
To account for the theoretical uncertainty of the underlying stellar evolutionary model, we added a random Gaussian noise of 5\% to the current stellar mass at each iterative step.
We used the values derived from the reconnaissance spectra
(\chiron{} for \starone{}, \tres{} for \startwo{}; see \secref{sec:recon_spec})
as Gaussian priors for the metallicity and \tempeff{}.

We used the three Gaia magnitudes \gaiaG, \gaiaRP, and \gaiaBP{}
to constrain the absolute magnitudes obtained from the isochrone interpolation.
To apply bolometric corrections to the absolute magnitudes,
we randomly drew values of $A_V$ extinctions from a normal distribution
based on the results of the independent SED analysis described in \secref{sec:sed}.
The isochrone-derived absolute magnitudes were then compared to those
computed from the photometric and parallax measurements in \gaia{} DR2.
We added an additional jitter, constrained to within $[0, 0.2)$, in quadrature
to the three \gaia{} magnitudes as a free parameter to the model to account for additional systematic uncertainties.
As with the independent SED analysis,
we added a systematic offset of \SI[retain-explicit-plus]{+0.082 \pm 0.033}{\mas}
reported by \cite{StassunTorres:2018} to the \gaia{} parallaxes.

\begin{deluxetable*}{lr@{}lr@{}lr}
  \tablecaption{Stellar Parameters\label{tab:stars}}
  \tabletypesize{\normalsize}
  \tablehead{%
    \colhead{Parameter \hfill Unit}
    & \multicolumn2c{\starone{}}
    & \multicolumn2c{\startwo{}}
    & \colhead{Source}
  }
  \startdata
  \sidehead{Identifying Information}
  $\alpha$ R.A. (J2015.5) \dotfill h:m:s  & \multicolumn2c{\hourang{04;07;45.854}} & \multicolumn2c{\hourang{23;24;32.491}} & \gaia{} DR2 \\
  $\delta$ Decl.~(J2015.5) \dotfill d:m:s & \multicolumn2c{\ang{-25;12;31.69}} & \multicolumn2c{\ang{-05;09;50.92}} & \gaia{} DR2 \\
  2MASS ID \dotfill             & \multicolumn2c{J04074585--2512312} & \multicolumn2c{J23243247--0509507} & TIC 8 \\
  TIC ID \dotfill               & \multicolumn2c{\staronetic{}} & \multicolumn2c{301258470} & TIC 8 \\
  \added{EPIC ID} \dotfill               & \multicolumn2c{\ldots} & \multicolumn2c{\startwoepic{}} & EPIC \\
  Space observations \dotfill   & \multicolumn2c{\tess{} sectors 4 and 5} & \multicolumn2c{\ktwo{} campaigns 12 and 19} & \\
  Parallax\tablenotemark{a} \dotfill              mas                   & $4.242$ & $\pm 0.046$ & $4.341$ & $\pm 0.053$ & \gaia{} DR2 \\
  $\mu_\alpha$, R.A. proper motion \ldots         \si{\mas\per\year}    & $-5.787$ & $\pm 0.042$ & $14.169$ & $\pm 0.063$ & \gaia{} DR2 \\
  $\mu_\delta$, decl.\ proper motion \ldots       \si{\mas\per\year}    & $-29.623$ & $\pm 0.045$ & $-12.389$ & $\pm 0.052$ & \gaia{} DR2 \\
  \sidehead{Photometric Properties}
  \tess{} \dotfill mag & $9.779$ & $\pm 0.006$ & $11.904$ & $\pm 0.006$ & TIC\thinspace 8 \\
  \kepler{} \dotfill mag & \multicolumn2c{\ldots} & 12.463 & & EPIC \\
  $B$ \dotfill mag & $11.153$ & $\pm 0.063$ & $13.542$ & $\pm 0.070$ & TIC\thinspace 8 \\
  $V$ \dotfill mag & $10.343$ & $\pm 0.005$ & $12.697$ & $\pm 0.080$ & TIC\thinspace 8 \\
  $J$ \dotfill mag & $ 9.180$ & $\pm 0.024$ & $11.174$ & $\pm 0.023$ & TIC\thinspace 8 \\
  $H$ \dotfill mag & $ 8.891$ & $\pm 0.026$ & $10.798$ & $\pm 0.028$ & TIC\thinspace 8 \\
  $K$ \dotfill mag & $ 8.779$ & $\pm 0.023$ & $10.668$ & $\pm 0.023$ & TIC\thinspace 8 \\
  \gaiaG{} \dotfill mag & $10.2321$ & $\pm 0.0002$ & $12.4462$ & $\pm 0.0003$ & \gaia{} DR2 \\
  \gaiaBP{} \dotfill mag & $10.5978$ & $\pm 0.0006$ & $12.9023$ & $\pm 0.0015$ & \gaia{} DR2 \\
  \gaiaRP{} \dotfill mag & $ 9.7328$ & $\pm 0.0005$ & $11.8537$ & $\pm 0.0013$ & \gaia{} DR2 \\
  Jitter$_G$\tablenotemark{b} \dotfill mag & \ToipJittergaia & \ToipJittergaiaErr & \KtwopJittergaia & \KtwopJittergaiaErr & Sampled \\
  \sidehead{Stellar Properties}
  \tempeff{}, effective temperature \dotfill      K                     & \ToipTeff & \ToipTeffErr  & \KtwopTeff & \KtwopTeffErr & Derived \\
  \feh{}, metallicity \dotfill                    dex                   & \ToipFeh & \ToipFehErr & \KtwopFeh & \KtwopFehErr & Sampled \\
  \logg{}, surface gravity \dotfill               dex                   & \ToipLogg & \ToipLoggErr & \KtwopLogg & \KtwopLoggErr & Derived \\
  $v \sin i$, rotational speed \dotfill \si{\km\per\s} & $5.6$ & $\pm 0.5$ & $1.9$ & $\pm 0.5$ & Spectroscopy \\
  $M_\star$, mass \dotfill                        \si{\mass\sun}        & \ToipMstar & \ToipMstarErr & \KtwopMstar & \KtwopMstarErr & Sampled \\
  $R_\star$, radius \dotfill                      \si{\radius\sun}      & \ToipRstar & \ToipRstarErr & \KtwopRstar & \KtwopRstarErr & Derived \\
  $\rho_\star$, density \dotfill                  \si{\gram\per\cm\cubed} & \ToipRhostar & \ToipRhostarErr & \KtwopRhostar & \KtwopRhostarErr & Derived \\
  $L_\star$, luminosity \dotfill                  $L_\odot$             & \ToipLstar & \ToipLstarErr & \KtwopLstar & \KtwopLstarErr & Derived \\
  Age \dotfill                                    Gyr                   & \ToipAge & \ToipAgeErr & \KtwopAge & \KtwopAgeErr & Sampled \\
  \enddata
  \tablecomments{The values from this work are the medians and 16th and 84th percentiles of the MCMC-derived posterior distribution.
    All solar-scaled units in this paper
    are defined and calculated as the recommended nominal values adopted in IAU 2015 Resolution B3
    \citep{iau2015b3}.
  }
  \tablenotetext{a}{Adjusted by a systematic offset of \SI[retain-explicit-plus]{+0.082 \pm 0.033}{\mas} reported by \cite{StassunTorres:2018}.}
  \tablenotetext{b}{Jitter$_G$ is added in quadrature to the reported noise of the \gaiaG, \gaiaBP, and \gaiaRP{} magnitudes during the global model fitting.}
  \tablerefs{%
    Sampled: this work.
    Derived: calculated from the sampled parameters with the MIST isochrone models \citep{2016ApJS..222....8D,2016ApJ...823..102C}.
    Spectroscopy: derived from reconnaissance spectra only; not part of the global model.
    EPIC: \cite{epic}.
    \gaia{} DR2: \cite{gaia_dr2}.
    TIC\thinspace 8: \cite{tic8}.
  }
\end{deluxetable*}

\begin{deluxetable*}{lr@{}lr@{}l}
  \tabletypesize{\normalsize}
  \tablecaption{Planetary and System Parameters\label{tab:planets}}
  \tablehead{%
    \colhead{Parameter \hfill Unit}
    & \multicolumn2c{\planetone{}}
    & \multicolumn2c{\planettwo{}}
  }
  \startdata
  \sidehead{Sampled Parameters}
  $T_c$, time of conjunction  \dotfill            BJD                   & \ToipTcjd & \ToipTcjdErr & \KtwopTcjd & \KtwopTcjdErr \\
  $P$, period \dotfill                            day                   & \ToipPeriod & \ToipPeriodErr & \KtwopPeriod & \KtwopPeriodErr \\
  $K$, RV semiamplitude \dotfill                  \si{\m\per\s}         & \ToipK & \ToipKErr & \KtwopK & \KtwopKErr \\
  $\sqrt{e}\sin\omega$ \dotfill                                         & \ToipSesinw & \ToipSesinwErr & \KtwopSesinw & \KtwopSesinwErr \\
  $\sqrt{e}\cos\omega$ \dotfill                                         & \ToipSecosw & \ToipSecoswErr & \KtwopSecosw & \KtwopSecoswErr \\
  $ b \equiv a \cos{i} / R_\star $ \dotfill                             & \ToipB & \ToipBErr & \KtwopB & \KtwopBErr \\
  $ R_p / R_\star $ \dotfill                                            & \ToipRp & \ToipRpErr & \KtwopRp & \KtwopRpErr \\
  \sidehead{Derived Parameters}
  $T_{14}$, total transit duration \dotfill       hour                  & \ToipTdur & \ToipTdurErr & \KtwopTdur & \KtwopTdurErr \\
  $e$, eccentricity \dotfill                                            & \ToipEcc & \ToipEccErr & \KtwopEcc & \KtwopEccErr \\
  $e$, eccentricity (95th percentile)       \dotfill                    & \ToipEccUpper & & \KtwopEccUpper & \\
  $\omega$, argument of periastron \ldots         degree                & \ToipOmega & \ToipOmegaErr & \KtwopOmega & \KtwopOmegaErr \\
  $a$, semimajor axis \dotfill                    au                    & \ToipSemimajor & \ToipSemimajorErr & \KtwopSemimajor & \KtwopSemimajorErr \\
  $ a / R_\star $ \dotfill                                              & \ToipAor & \ToipAorErr & \KtwopAor & \KtwopAorErr \\
  $i$, inclination \dotfill                       degree                & \ToipInc & \ToipIncErr & \KtwopInc & \KtwopIncErr \\
  $\si{\mass\planet} \sin i$, minimum mass \dotfill \si{\mass\jupiter}  & \ToipMpSiniMj & \ToipMpSiniMjErr & \KtwopMpSiniMj & \KtwopMpSiniMjErr \\
  \si{\mass\planet}, mass \dotfill                \si{\mass\jupiter}    & \ToipMpMj & \ToipMpMjErr & \KtwopMpMj & \KtwopMpMjErr \\
  \si{\radius\planet}, radius \dotfill            \si{\radius\jupiter}  & \ToipRpRj & \ToipRpRjErr & \KtwopRpRj & \KtwopRpRjErr \\
  \si{\density\planet}, density \dotfill \si{\gram\per\cm\cubed}        & \ToipRhop & \ToipRhopErr & \KtwopRhop & \KtwopRhopErr \\
  Stellar irradiation \dotfill \si{\erg\per\s\per\cm\squared}           & \ToipIrrad & \ToipIrradErr & \KtwopIrrad & \KtwopIrradErr \\
  \tempeq{}, equilibrium temperature\tablenotemark{a} \dotfill K        & \ToipTempeq & \ToipTempeqErr & \KtwopTempeq & \KtwopTempeqErr \\
  \enddata
  \tablecomments{%
    The values given are the medians and 16th and 84th percentiles of the MCMC-derived marginalized posterior distribution.
    All Jupiter-scaled units in this paper
    are defined and calculated as the recommended nominal values adopted in IAU 2015 Resolution B3
    \citep{iau2015b3}.
    The Jupiter radius \si{\radius\jupiter} is taken to be the equatorial radius.}
  \tablenotetext{a}{Here \tempeq{} is calculated assuming no atmospheric redistribution
    and a uniformly random distribution of Bond albedo in the interval $[0, 0.7)$.}
\end{deluxetable*}

The resulting posterior distributions show an excellent fit with the observations
(Figures~\ref{fig:toi954_lc}--\ref{fig:k2p_rv}).
The final median and 16th, 84th percentile values of the stellar and planetary system parameters
are reported in Tables~\ref{tab:stars} and \ref{tab:planets}.
Representative MCMC samples of the posterior distributions for the two models
are reported separately in Tables~\ref{tab:mcmc_toi954} and \ref{tab:mcmc_k2p}.

We also ran an independent global analysis of the available
photometric and RV measurements of \startwo{} using EXOFASTv2
\citep{Eastman:2013,Eastman:2019}.
The results of the EXOFASTv2 analysis were consistent ($< 1 \sigma$) with the analysis described above in this section.

\section{Discussion}
\label{sec:discussion}

\begin{figure*}
    \centering
    (a) \hspace{0.5\textwidth} (b) \\
    \includegraphics[width=0.49\textwidth]{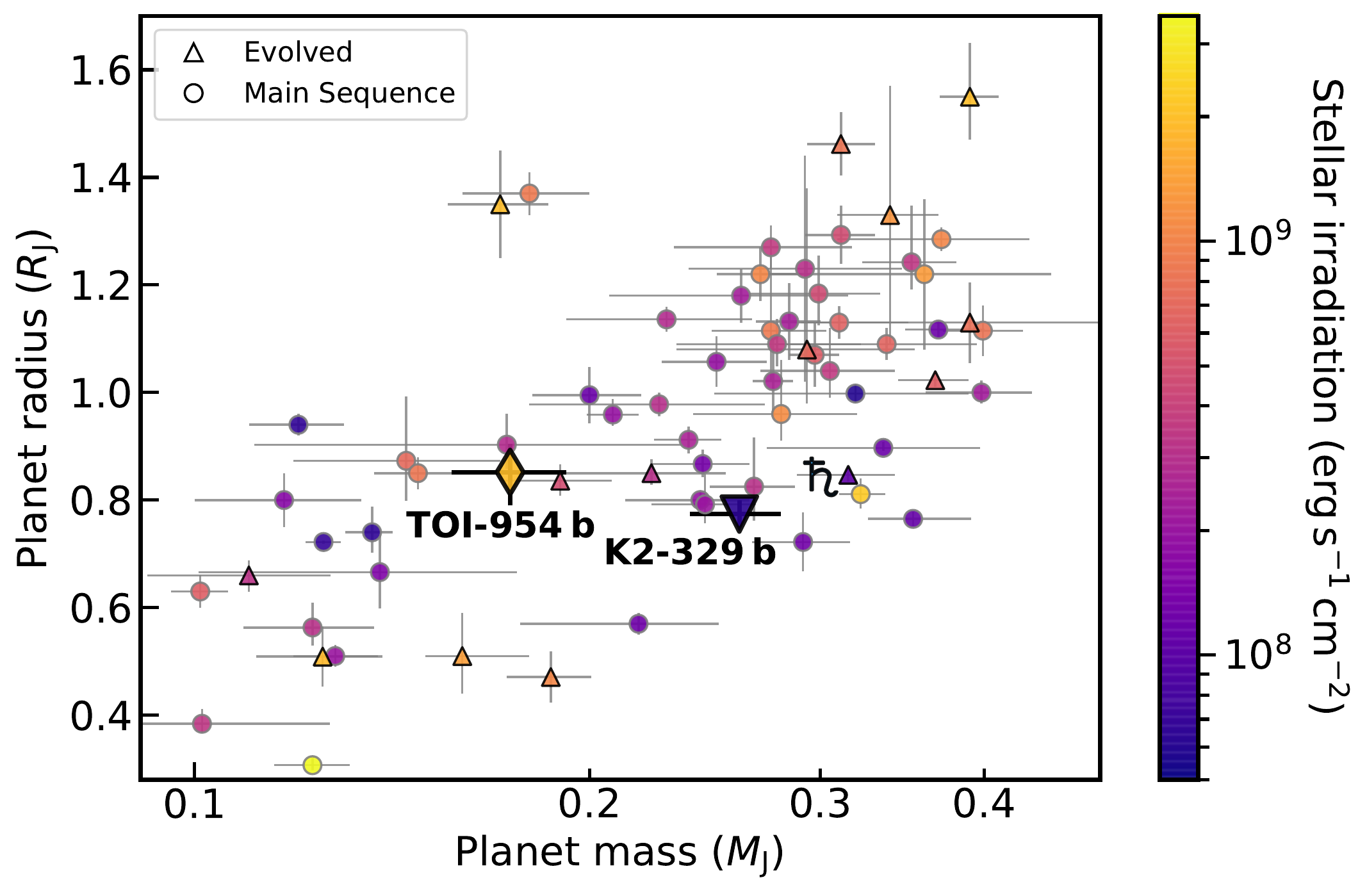}
    \hfill
    \includegraphics[width=0.49\textwidth]{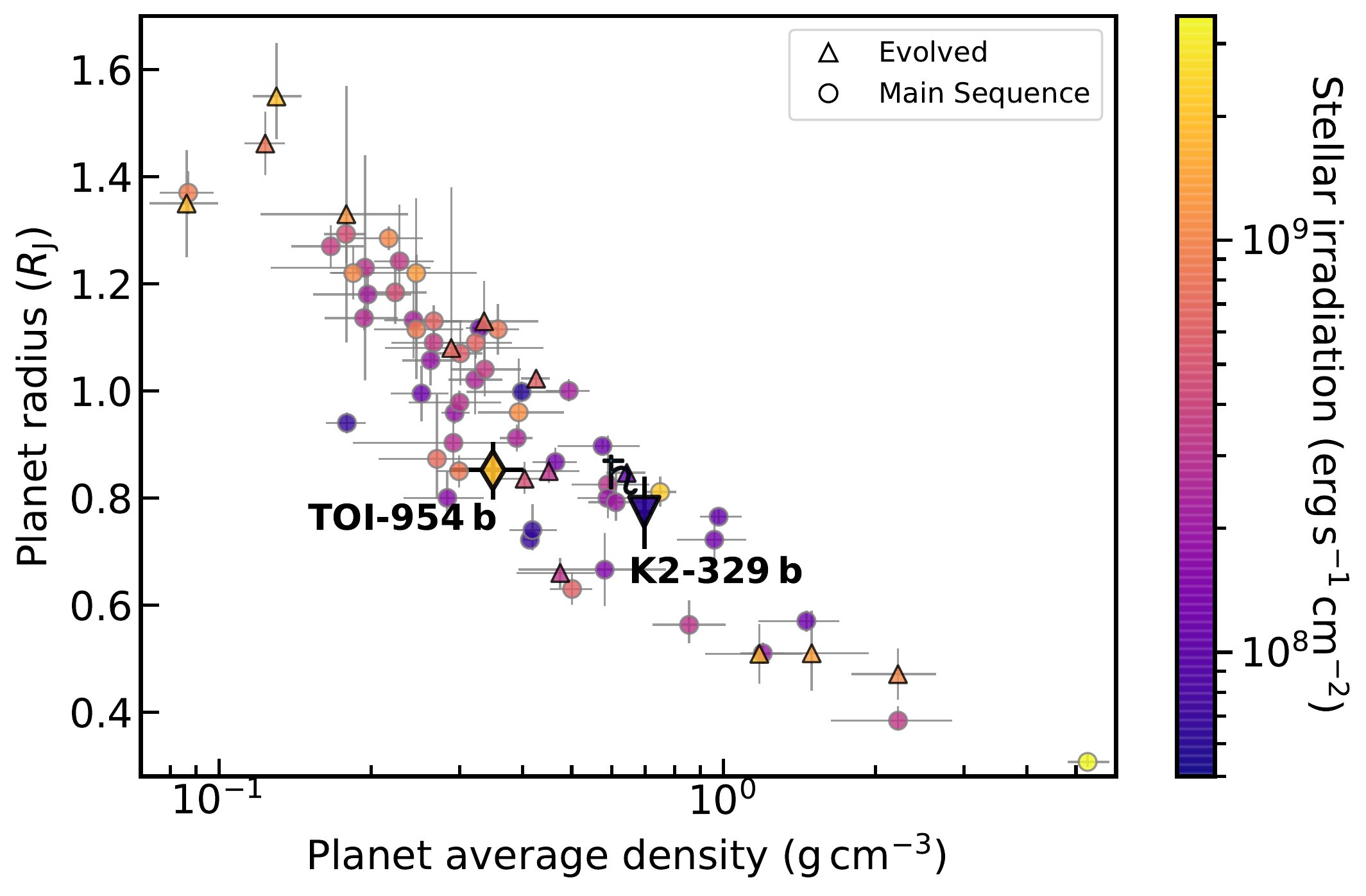} \\
    (c) \hspace{0.5\textwidth} (d) \\
    \includegraphics[width=0.49\textwidth]{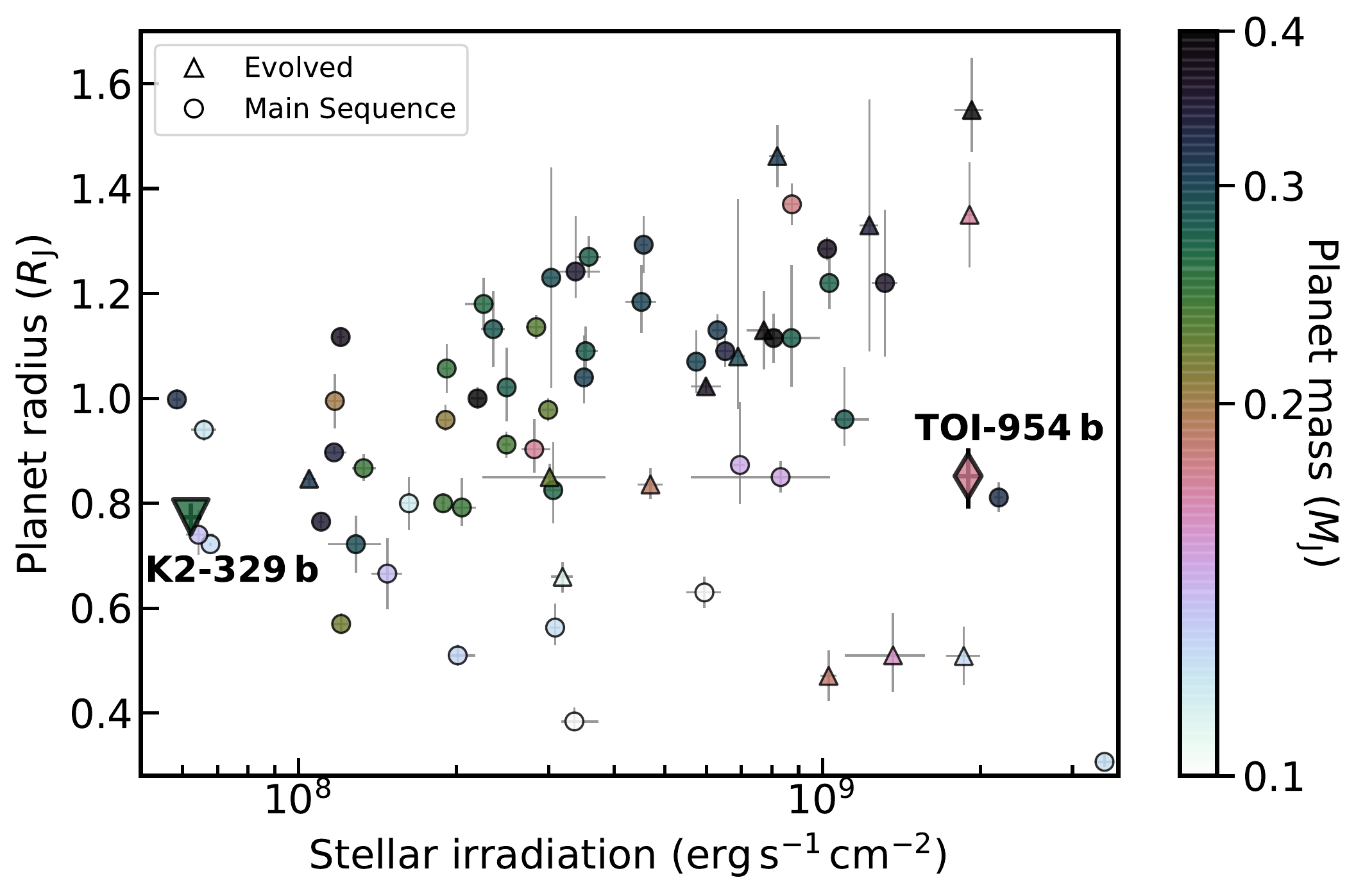}
    \hfill
    \includegraphics[width=0.49\textwidth]{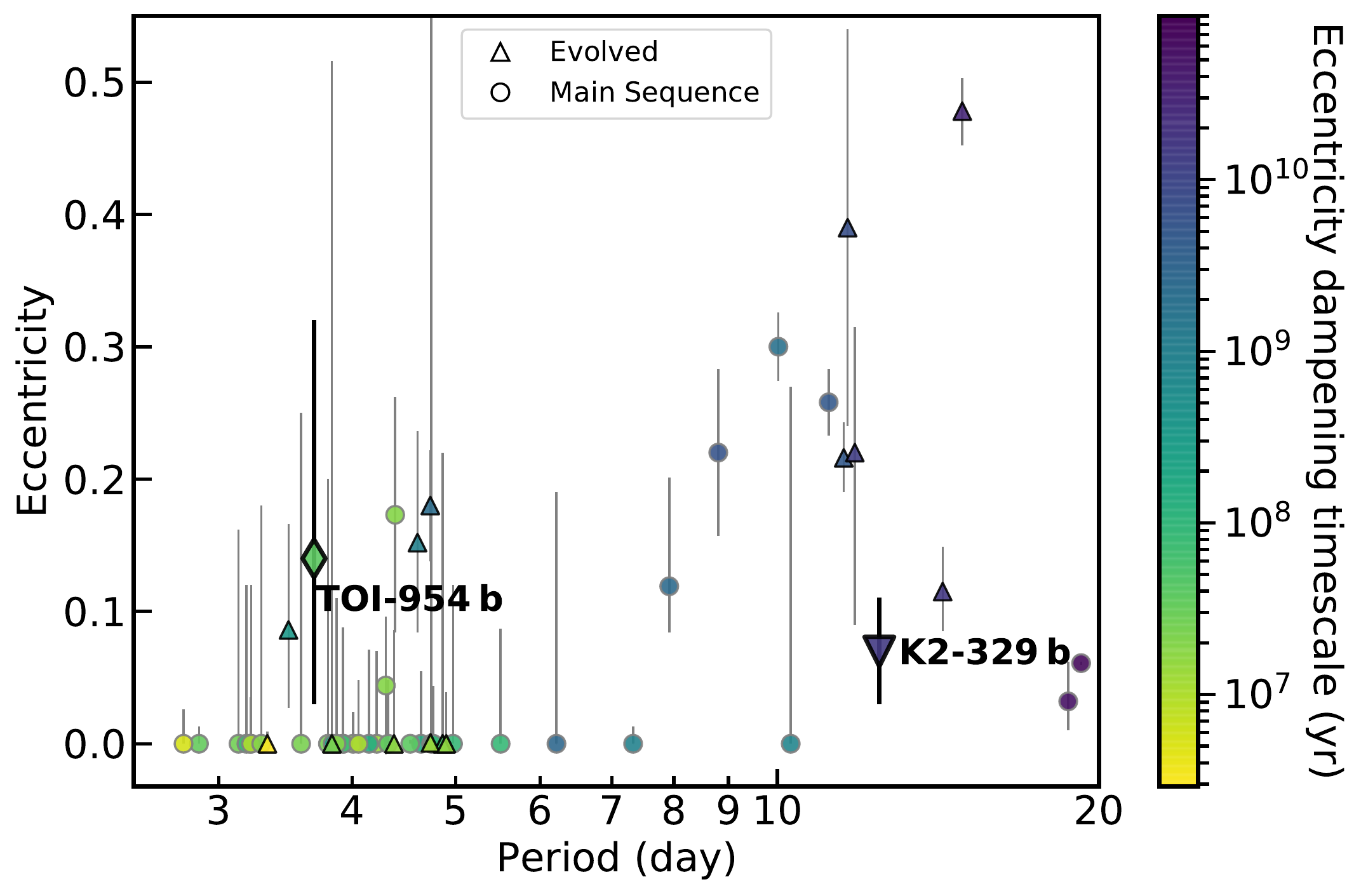} \\
    \caption{Planets \planetone{} and \planettwo{} compared to planets of similar mass (measured to better than 50\%) and period
    ($0.1 \leq \si{\mass\planet}/\si{\mass\jupiter} \leq 0.4$,
     $P < 20\ \mathrm{days}$).
    The planet parameters were retrieved from the NASA Exoplanet Archive on 2020 October 16 \citep{akeson2013}, preferring those by  \cite{2017A&A...602A.107B} where possible.
    Upward triangles represent planets around stars that are off the main sequence,
    while circles represent planets around stars on the main sequence.
    The evolutionary stage of the stars is determined by comparing the absolute \gaia{} \gaiaG{} magnitude and the color using the \gaia{} \gaiaBP{} and \gaiaRP{} bands with the theoretical MIST terminal-age main-sequence isochrones.
    Saturn is included in panels (a) and (b) for comparison.
    In panel (d), only planets with quoted uncertainties in eccentricity are plotted.
    Most of the plotted eccentricity upper limits are at the $2\sigma$ significance level, but a few are at $1\sigma$ or even $3\sigma$.
    We choose to present the upper limits in the literature as is,
    because it is impossible to convert the upper limits to a uniform significance level
    without access to their underlying posterior distributions.
    The eccentricity dampening timescale (\tauep) is the characteristic time for tidal effects to circularize a planet's orbit.
    It is calculated according to Equation~(1) in \cite{dobbsdixon_spin_2004}
    with the quality factor $Q'_{p}$ assumed to be $10^5$.
    }
    \label{fig:population}
\end{figure*}

In order to understand how \planetone{} and \planettwo{} fit into the landscape of known planets,
we compare them to transiting planets of similar size, mass, and period
($0.1 \leq \si{\mass\planet}/\si{\mass\jupiter} \leq 0.5$%
 {$0.1 \leq \si{\mass\planet}/\si{\mass\jupiter} \leq 0.4$},
 $P < \SI{20}{\day}$)
with mass measurements better than 50\% and radius measurements better than 20\%
in \figref{fig:population}.
This includes all \enquote{hot} giant planets around Saturn's mass.

We use the parameters from the new planetary systems table on the NASA Exoplanet Archive website
\citep[accessed on 2020 October 16]{akeson2013}.
Unlike the traditional confirmed planets table,
the planetary systems table presents all available planetary and stellar parameters from the literature,
so we are free to choose sets of parameters that produce more uniform results.

We use the following procedures to choose between different sets of parameters for a given planet.
\begin{enumerate}
  \item 
  We prefer the parameters from \cite{2017A&A...602A.107B}, if available,
  because they provide eccentricity and mass measurements from HARPS-N for the largest number of planets under consideration.
  \item
  We prefer parameters with quoted uncertainties on eccentricity;
  that is, the fit allows the eccentricity to vary rather than assuming a circular orbit.
  \item
  If there is more than one set of parameters with similar quoted uncertainties,
  we use the one with the \enquote{default parameter set} flag;
  that is, the one presented in the traditional confirmed planets table.
\end{enumerate}
These procedures yield 65 planets that satisfy our selection criteria,
50 of which have quoted uncertainties on their eccentricity measurements.

\subsection{Stellar Irradiation and Reinflation}
\label{sec:reinflation}

Both of our planets are around the lower ranges of mass and size for gas giants.
The planet \planetone{} has about two-thirds of the mass of \planettwo{}, but its size is 10\% larger.
\figref{fig:population}(a) plots the measured masses and sizes and the derived stellar irradiation of the selected planets,
highlighting the two Saturn-sized planets from this work.
Figures~\ref{fig:population}(b) and (c) cast the same information in different axes.

For our sample of hot Saturns,
there is a weak positive correlation between a planet's mass and size, but the scatter is large.
A weak positive correlation has also been noted by
\cite{2015ApJ...810L..25H}
and \cite{2017ApJ...834...17C}
at the smaller end of gas giants,
although they each used subtly different binning under which our planet population does not neatly fall.
We consider further investigation of this weak correlation to be outside the scope of this work,
and we make no further attempt to compare our results quantitatively.

Neither planet is appreciably inflated compared to planets within the same mass bin,
even though \planetone{} receives 30 times the stellar irradiation of \planettwo{}.
We note that the first hot Saturn discovered by \tess{}, \toioneninesevenb{},
is also around an evolved star and
has almost the same mass and radius within $1\sigma$ compared to \planetone{},
but receives five times less stellar irradiation.
Previous studies of giant planets have empirically derived a limit on stellar irradiation of
$ \langle S \rangle \approx \SI{2e8}{\erg\per\second\per\cm\squared} $,
below which planet inflation is not usually observed
\citep{2011ApJS..197...12D,2011ApJ...736L..29M}.
An uninflated Saturn orbiting a main-sequence host star
with irradiation of \circa \SI{6e7}{\erg\per\second\per\cm\squared},
\planettwo{}
fits this expectation.

However, that \planetone{} remains uninflated
around a moderately evolved star seems to contradict
proposed mechanisms that explain the inflated radii of gas giants in terms of the current value of the stellar irradiation 
\citep{2007ApJ...661..502B,2007ApJ...659.1661F}.
Based on data available at the time,
\cite{2016AJ....152..182H} found that the hot Jupiters with
radii exceeding \SI{1.5}{\radius\jupiter} tend to be found around
evolved stars rather than main-sequence stars,
possibly because the increased luminosity of evolved
stars causes the hot Jupiter to inflate.
At an age of \SI{6}{\giga\year},
which is more than $80\%$ of the way through its total life span,
\starone{} gives us a rare example of an evolved star with a dense hot Saturn.
One possible explanation for this apparent paradox is
that the reinflation mechanism is less effective for lower-mass giant 
planets, for which the core-to-envelope mass ratio is higher
\citep{2011ApJ...736L..29M,2012A&A...540A..99E}.
As uninflated hot Jupiters around evolved stars are precisely
the kind of planets that ground-based surveys would miss,
space-based exoplanet survey missions like \tess{} will be able to tell us
how common such planets are.

\subsection{Eccentricity and Tidal Circularization}
\label{sec:eccentricity}

Since \planetone{} and \planettwo{} differ in semimajor axes by a factor of 2
and are around stars at different life stages,
we would like to investigate whether their nearly circular orbits
are consistent with tidal dissipation.
The posterior distributions for the eccentricities of both planets
are consistent with that of a circular orbit
once we account for the bias toward higher eccentricities
described by \cite{1971AJ.....76..544L}.
Nevertheless, the posterior distributions suggest
that we cannot rule out a small but nonzero eccentricity at the $2\sigma$ level
for either of the two planets on a purely observational basis.

The timescale for eccentricity dampening is given by
\cite{1966Icar....5..375G} as
\begin{equation}
    \tauep = \frac{4}{63}
             \frac{\qmodplanet}{n}
             \frac{M_\mathrm{p}}{M_\star}
             \left( \frac{a}{R_\mathrm{p}} \right)^5 ,
\end{equation}
where $n = \sqrt{GM_\star / a^3}$ is the mean motion of the planet,
and \qmodplanet{} is the modified tidal quality factor.
Calculating or even estimating \qmodplanet{} is a notoriously hard problem.
Here we adopt a value of $\qmodplanet{} = 10^5$,
assuming that it is not too different from the present best estimate
for Saturn using Cassini observations of its moons
\citep{2017Icar..281..286L}.
Under this order-of-magnitude estimation,
we calculate that the \tauep{} for \planetone{} is \SI{0.04}{\giga\year}
and that for \planettwo{} is \SI{15}{\giga\year}. 
This is consistent with \planetone{} having circularized before reaching its current age, but the case for \planettwo{} warrants closer inspection.

There are two possible explanations that can reconcile the apparently circular orbit of \planettwo{}
with its eccentricity dampening timescale, which is on the order of the current age of the universe.
The first explanation is that the interior structure of \planettwo{} is dramatically different from Saturn's,
such that its \qmodplanet{} is an order of magnitude smaller,
which brings the \tauep{} in line with the estimated age of the star.
This is not implausible, given the huge uncertainties in estimating \qmodplanet{} from a theoretical basis \citep{1966Icar....5..375G}.
The other explanation is that some mechanism other than tidal dissipation is responsible for circularizing the orbit of \planettwo{}.
This scenario rules out high-eccentricity migration as the formation pathway for \planettwo{},
and instead points to disk migration or in situ formation as possible mechanisms.
Either way, only future investigations into the interior structures and formation pathways of short-period giant planets could provide us with a definitive answer.

\movetabledown 150pt
\begin{rotatetable*}
\begin{deluxetable*}{l|lllllllllll}
  \tablecaption{MCMC Samples of the Posterior Distribution of the \starone{} System Global Model \label{tab:mcmc_toi954}}
  \tabletypesize{\scriptsize}
  \tablehead{
    \colhead{$\ln(\mathrm{prob})$}
    & \colhead{$T_c - \num{2458000}$}
    & \colhead{$P$}
    & \colhead{$K$}
    & \colhead{$\sqrt{e} \cos \omega$}
    & \colhead{$\sqrt{e} \sin \omega$}
    & \colhead{$b$}
    & \colhead{$\si{\radius\planet} / \si{\radius\etoile}$}
    & \colhead{$q_{1,TESS}$}
    & \colhead{$q_{2,TESS}$}
    & \colhead{$q_{1,r}$}
    & \colhead{$q_{2,r}$} \\
    \colhead{}
    & \colhead{\bjdtdb}
    & \colhead{(day)}
    & \colhead{(\si{\m\per\s})}
  }
  \startdata
-14354.44023 &  1411.904941 &  3.684967688 &  17.93949680 & -0.1321959630 &  0.1994653707 &  0.4438686995 &  0.04485989839 &  0.5085029466 &  0.3006750291 &  0.5163011227 &  0.3154157370 \\
-14347.22724 &  1411.906643 &  3.684973192 &  20.75783768 & -0.1470678602 &  0.2588783622 &  0.05981187401 &  0.04338220271 &  0.5905204601 &  0.2547819277 &  0.2703767257 &  0.4323913481 \\
-14352.88142 &  1411.905847 &  3.684973180 &  19.21590553 &  0.1036840949 & -0.06051571668 &  0.06463619662 &  0.04338529032 &  0.7100074508 &  0.2643213422 &  0.3484777441 &  0.4038090189 \\
-14340.22022 &  1411.906707 &  3.684969887 &  22.60743336 & -0.03903125394 &  0.2787585995 &  0.1646960200 &  0.04378903058 &  0.4681846915 &  0.3345654345 &  0.2879253417 &  0.4313357366 \\
-14345.25183 &  1411.906510 &  3.684972419 &  18.54777627 &  0.08148884905 & -0.5760585941 &  0.5564955516 &  0.04769200456 &  0.2457396253 &  0.2319848422 &  0.4746678117 &  0.3392692825 \\
-14338.41898 &  1411.907719 &  3.684972127 &  21.84105097 &  0.1437721897 & -0.5550702656 &  0.5945629825 &  0.04803778206 &  0.09007431795 &  0.2880662791 &  0.4417166309 &  0.3586039049 \\
-14354.28326 &  1411.905692 &  3.684973177 &  22.61778834 &  0.03625643523 & -0.5737580300 &  0.5876941678 &  0.04774484122 &  0.05174179627 &  0.7234900406 &  0.3731319902 &  0.3931272665 \\
-14341.49854 &  1411.906795 &  3.684975454 &  21.31097357 &  0.08128176625 & -0.5406874196 &  0.5955651435 &  0.04835495284 &  0.1389292374 &  0.3041011897 &  0.5299276078 &  0.3853730486 \\
-14341.43762 &  1411.906658 &  3.684976182 &  19.82065313 & -0.06303504949 &  0.2370810200 &  0.2181803543 &  0.04375055243 &  0.5268741112 &  0.2835063251 &  0.4364446033 &  0.3662910117 \\
-14346.35703 &  1411.905627 &  3.684973458 &  17.23247567 &  0.03295028209 & -0.3071263492 &  0.5828023314 &  0.04638902008 &  0.1637512634 &  0.2059385337 &  0.3339720351 &  0.4543123464 \\
-14337.43893 &  1411.906118 &  3.684974089 &  18.81482023 &  0.06271621477 & -0.5058124843 &  0.5791370381 &  0.04733862278 &  0.2747636805 &  0.1011482824 &  0.4951416609 &  0.3499220283 \\
-14348.23500 &  1411.905989 &  3.684971157 &  25.59307727 & -0.1583322795 &  0.2436815449 &  0.1052972631 &  0.04302472352 &  0.7794911465 &  0.2336998563 &  0.3689845205 &  0.4013567264 \\
-14335.27738 &  1411.906735 &  3.684973915 &  20.22030064 &  0.1483815795 & -0.4049339034 &  0.5951231807 &  0.04730661203 &  0.1734650151 &  0.1827774447 &  0.4996120404 &  0.3184169612 \\
-14339.55455 &  1411.906185 &  3.684974464 &  19.67575117 & -0.07034729041 &  0.2257495232 &  0.4380603490 &  0.04488916414 &  0.6281338767 &  0.1316341167 &  0.3884916023 &  0.3840553854 \\
-14350.36375 &  1411.904869 &  3.684966998 &  20.68640930 & -0.003319478407 & -0.05019737051 &  0.3732592289 &  0.04393217416 &  0.6217080146 &  0.3005437417 &  0.4190346249 &  0.3799096822 \\
-14339.97917 &  1411.907245 &  3.684976999 &  22.27986690 &  0.005131115091 & -0.5147591338 &  0.5933582051 &  0.04774696096 &  0.3035273222 &  0.3267617896 &  0.5218413146 &  0.3723322275 \\
-14345.31232 &  1411.906206 &  3.684968273 &  20.99129342 &  0.02247086520 &  0.1007588562 &  0.4466256045 &  0.04410637996 &  0.5400836900 &  0.1511046601 &  0.3850686064 &  0.3878407228 \\
-14341.67816 &  1411.907309 &  3.684976490 &  17.78596649 &  0.06372446126 & -0.4557761007 &  0.5974025277 &  0.04719486840 &  0.1594767443 &  0.09728917632 &  0.3764136196 &  0.3683304861 \\
-14331.72344 &  1411.905958 &  3.684969672 &  21.76715677 &  0.09957060199 & -0.09439000641 &  0.4646776031 &  0.04480180311 &  0.5932157536 &  0.1384442778 &  0.3364832924 &  0.4089113585 \\
-14335.36406 &  1411.906224 &  3.684972072 &  21.55223063 &  0.1256627660 & -0.3802145670 &  0.5568186558 &  0.04600932199 &  0.2622893636 &  0.1541895825 &  0.5324306499 &  0.3748288960 \\
-14341.20333 &  1411.907222 &  3.684974876 &  21.79091895 &  0.1025398826 & -0.3367783723 &  0.5158244938 &  0.04584547138 &  0.2841291446 &  0.2727021669 &  0.3994056942 &  0.3689905959 \\
-14337.59830 &  1411.905951 &  3.684972413 &  21.06307807 &  0.0005353252144 &  0.08995562398 &  0.5070891553 &  0.04504412599 &  0.3177855513 &  0.2247290990 &  0.4047661945 &  0.3631466132 \\
-14338.03126 &  1411.906391 &  3.684970001 &  24.16027431 & -0.1617030780 &  0.2382183638 &  0.1964510917 &  0.04348256551 &  0.6857136671 &  0.1628751384 &  0.3704709731 &  0.3553858418 \\
-14336.68867 &  1411.906092 &  3.684970163 &  21.06430985 &  0.08456468853 &  0.08397766818 &  0.3751925416 &  0.04421586467 &  0.8280251055 &  0.1829022314 &  0.5206016071 &  0.3394633500 \\
-14347.09751 &  1411.906494 &  3.684968443 &  22.22161625 &  0.0009649445689 &  0.07164441523 &  0.2460708003 &  0.04336521144 &  0.5892065349 &  0.3694563621 &  0.5131371091 &  0.3506272133 \\
-14351.73395 &  1411.906676 &  3.684973375 &  17.96121667 & -0.04969993654 &  0.1982609904 &  0.1731649834 &  0.04329559104 &  0.8640311436 &  0.1509516960 &  0.5162851669 &  0.3735274039 \\
-14346.56720 &  1411.907471 &  3.684978017 &  23.44786465 &  0.1593834289 & -0.1610965562 &  0.4748392021 &  0.04503793264 &  0.4822695702 &  0.1665862166 &  0.3979304734 &  0.3918512060 \\
-14356.34227 &  1411.906892 &  3.684972014 &  20.61160384 &  0.1112040759 & -0.1466707139 &  0.3183631848 &  0.04389241941 &  0.6822461708 &  0.1684345403 &  0.3691392496 &  0.3574990317 \\
-14336.54598 &  1411.906537 &  3.684974359 &  20.57568508 &  0.09307898747 & -0.5356326155 &  0.5936841303 &  0.04766538989 &  0.1928450212 &  0.1362587927 &  0.3249889814 &  0.4345648770 \\
-14345.05049 &  1411.906287 &  3.684973235 &  18.62240006 & -0.08167320902 &  0.1328646244 &  0.1664352216 &  0.04370064023 &  0.3039776191 &  0.4557217266 &  0.4506124606 &  0.3527705190 \\
-14346.14623 &  1411.906705 &  3.684972632 &  24.33169243 & -0.1122754667 &  0.1183417164 &  0.3346694916 &  0.04426426509 &  0.5712488860 &  0.2989092867 &  0.4225259041 &  0.3730318492 \\
-14355.05078 &  1411.905681 &  3.684969354 &  20.59627822 &  0.1434136373 & -0.08272873859 &  0.2947494264 &  0.04342023442 &  0.7909987359 &  0.2561993288 &  0.3654090202 &  0.3612451668 \\
-14336.45289 &  1411.906863 &  3.684973913 &  19.52599955 & -0.07487505747 &  0.2503513018 &  0.2045794971 &  0.04373946603 &  0.6568896863 &  0.2424556110 &  0.3468749270 &  0.3919479925 \\
-14346.22757 &  1411.907521 &  3.684972801 &  20.00260768 &  0.1163266802 & -0.07066492419 &  0.4534207034 &  0.04475738492 &  0.7484458779 &  0.1642910050 &  0.3739162695 &  0.3775028523 \\
-14347.61110 &  1411.905376 &  3.684969990 &  20.70984310 & -0.05126948533 &  0.2697624584 &  0.2009483906 &  0.04296586054 &  0.8359839523 &  0.1255459343 &  0.4391808687 &  0.3412227685 \\
-14338.19926 &  1411.907811 &  3.684975892 &  20.47352672 &  0.07581075787 & -0.5869472772 &  0.6017090910 &  0.04810123185 &  0.01853912027 &  0.7815641718 &  0.3014104884 &  0.4442844274 \\
-14342.08726 &  1411.906187 &  3.684969084 &  20.02839433 & -0.004565593104 &  0.3097572041 &  0.09379264034 &  0.04274079017 &  0.5615150112 &  0.3132433019 &  0.4772986708 &  0.3020935814 \\
-14338.59795 &  1411.906291 &  3.684972221 &  21.80814275 & -0.2446106866 &  0.1427809932 &  0.3185367463 &  0.04391230083 &  0.4544389462 &  0.3327295658 &  0.4120598490 &  0.3396520775 \\
-14337.75724 &  1411.906274 &  3.684971486 &  22.90067581 &  0.1208747527 & -0.4932922494 &  0.5929823974 &  0.04767080487 &  0.1800976269 &  0.1285746232 &  0.3927563382 &  0.3990204496 \\
-14334.36458 &  1411.906477 &  3.684973082 &  18.90906159 &  0.04435296295 & -0.4752331848 &  0.5872245223 &  0.04725479068 &  0.1900903255 &  0.2033680790 &  0.4061012472 &  0.3735344208 \\
\enddata

  \tablecomments{%
    This table is available in its entirety in machine-readable form.
  }
\end{deluxetable*}
\end{rotatetable*}

\movetabledown 150pt
\begin{rotatetable*}
\begin{deluxetable*}{llllllllllllllllll}
  \savetablenum{10}
  \tablecaption{MCMC Samples of the Posterior Distribution of the \starone{} System Global Model, continued}
  \tabletypesize{\scriptsize}
  \tablehead{
     \colhead{$q_{1,\filterzs}$}
    & \colhead{$q_{2,\filterzs}$}
    & \colhead{$\si{\mass\etoile}_0$}
    & \colhead{Age}
    & \colhead{$\feh_0$}
    & \colhead{$D_{\mathrm{TESS}}$}
    & \colhead{$D_{\mathrm{HATS}}$}
    & \colhead{$C_{\mathrm{\lcogt}}$}
    & \colhead{$\gamma_{\mathrm{\chiron}}$}
    & \colhead{$\gamma_{\mathrm{\coralie}}$}
    & \colhead{$\gamma_{\mathrm{\harps}}$}
    & \colhead{$\gamma_{\mathrm{\pfs}}$} \\
    &
    & \colhead{(\si{\mass\sun})}
    & \colhead{(\si{\giga\year})}
    & \colhead{(dex)}
    &
    &
    &
    & \colhead{(\si{\m\per\s})}
    & \colhead{(\si{\m\per\s})}
    & \colhead{(\si{\m\per\s})}
    & \colhead{(\si{\m\per\s})}
  }
  \startdata
 0.3877730397 &  0.2758543226 &  1.223467441 &  5.683577481 &  0.2193189960 &  0.02528028062 &  1.040006626E-05 &  1.829901464 & -8841.035817 & -7347.298656 & -7320.362720 & -12.37423810 \\
 0.2644189443 &  0.3303547631 &  1.188821640 &  6.208531239 &  0.2275659284 &  0.03410290818 &  1.270651350E-05 &  1.828699023 & -8831.948588 & -7350.426027 & -7318.228388 & -12.07386531 \\
 0.2029125999 &  0.3375359681 &  1.285280887 &  4.492632011 &  0.2514601055 &  0.03379996276 &  3.863416147E-05 &  1.872203176 & -8797.505904 & -7345.867011 & -7322.612163 & -18.86731976 \\
 0.3433325600 &  0.3080209178 &  1.213206739 &  5.930744577 &  0.2484305461 &  0.04053131188 &  3.371767377E-06 &  1.789181241 & -8826.306308 & -7346.123978 & -7330.845584 & -14.28096754 \\
 0.3302481880 &  0.2879429666 &  1.328252919 &  4.179472397 &  0.3345612161 &  0.03761823635 &  4.572284952E-06 &  1.813761353 & -8830.647347 & -7348.127143 & -7318.355431 & -7.825585713 \\
 0.2734566166 &  0.2803173291 &  1.183986752 &  6.382777973 &  0.2152513601 &  0.03617816605 &  4.389649959E-05 &  1.855372886 & -8824.320814 & -7350.834176 & -7318.949805 & -15.88999885 \\
 0.2881986492 &  0.2394012679 &  1.188226123 &  6.431456155 &  0.2593972958 &  0.04613567764 &  8.275080300E-06 &  1.798427175 & -8826.195052 & -7350.928488 & -7331.530795 & -21.31681053 \\
 0.1490642148 &  0.3830242181 &  1.179978664 &  6.585374418 &  0.2189121719 &  0.04256104367 &  2.711080661E-05 &  1.859636055 & -8817.723714 & -7349.181549 & -7322.137215 & -16.24425184 \\
 0.4332264728 &  0.2473914107 &  1.177254900 &  6.383512003 &  0.2007912804 &  0.04264215163 &  9.301526159E-06 &  1.895952598 & -8825.037739 & -7348.757400 & -7323.186189 & -13.80774716 \\
 0.2867652405 &  0.2854867110 &  1.198882373 &  6.263346582 &  0.2337154064 &  0.03648821575 &  3.236805382E-06 &  1.795281062 & -8824.245563 & -7344.228504 & -7322.433812 & -13.60455006 \\
 0.2719484330 &  0.2868005955 &  1.183180165 &  6.186926196 &  0.1826721012 &  0.03168096077 &  8.270891914E-06 &  1.782988766 & -8823.999346 & -7350.223210 & -7322.506253 & -7.857324957 \\
 0.2437400710 &  0.2613229355 &  1.209218767 &  5.926610035 &  0.2426396049 &  0.04375856662 &  6.237687869E-07 &  1.778700832 & -8807.813278 & -7348.904125 & -7324.471983 & -19.78192044 \\
 0.2652828180 &  0.3145689859 &  1.180389894 &  6.428011124 &  0.2012333414 &  0.03134108589 &  2.379072381E-06 &  1.757917986 & -8819.080358 & -7343.840823 & -7323.352247 & -15.12826666 \\
 0.2493684719 &  0.3345163699 &  1.196126106 &  6.002800584 &  0.1757505290 &  0.02990726886 &  5.255591549E-06 &  1.833834096 & -8816.469214 & -7347.101321 & -7322.346843 & -9.411521681 \\
 0.4341037277 &  0.2431441726 &  1.201829662 &  6.133575993 &  0.2680511198 &  0.03820708618 &  8.347366308E-06 &  1.762129770 & -8846.467933 & -7348.388438 & -7321.988914 & -14.72824872 \\
 0.2788113579 &  0.2903011030 &  1.200524787 &  6.116670079 &  0.2185001677 &  0.03273905786 &  1.455566739E-05 &  1.859218979 & -8821.116022 & -7344.616593 & -7319.357724 & -19.09675290 \\
 0.4185901784 &  0.3070247347 &  1.209430860 &  6.241579701 &  0.2972515930 &  0.03547639672 &  1.075811673E-06 &  1.811841181 & -8828.553427 & -7350.494995 & -7319.691494 & -13.83627263 \\
 0.2289828990 &  0.2783163557 &  1.178754228 &  6.045644215 &  0.1272373430 &  0.03835144084 &  9.083398820E-06 &  1.818742114 & -8817.720341 & -7345.975817 & -7321.580034 & -16.63199429 \\
 0.2244471159 &  0.3578005554 &  1.176030063 &  6.552381370 &  0.2064540753 &  0.03136287168 &  1.072799654E-06 &  1.784404258 & -8825.235474 & -7345.619914 & -7320.992766 & -15.12524281 \\
 0.2799515568 &  0.3264811027 &  1.192363971 &  6.523716435 &  0.2769860323 &  0.04225660543 &  2.015248075E-05 &  1.802201924 & -8826.275478 & -7347.545539 & -7321.688829 & -14.45798964 \\
 0.3077526213 &  0.2949811906 &  1.190779839 &  6.393658256 &  0.2572537660 &  0.04430239788 &  2.542151003E-05 &  1.861104102 & -8820.378411 & -7345.021127 & -7323.060378 & -11.98473415 \\
 0.3019664942 &  0.3313821744 &  1.191697912 &  6.136562106 &  0.1785916202 &  0.02760536747 &  5.291938909E-06 &  1.889447201 & -8821.541270 & -7350.961392 & -7322.245760 & -15.80959800 \\
 0.3156493827 &  0.3209293954 &  1.183885632 &  6.263171080 &  0.1957587174 &  0.03883087319 &  8.342213507E-06 &  1.753360980 & -8830.090702 & -7344.771798 & -7320.059892 & -18.24218682 \\
 0.2899112629 &  0.2877722326 &  1.209869398 &  6.029909287 &  0.2679537143 &  0.03909113351 &  2.741967536E-06 &  1.790294172 & -8823.849201 & -7346.217782 & -7320.627159 & -16.42762405 \\
 0.3392360803 &  0.2915215742 &  1.194791344 &  6.172913430 &  0.2443477519 &  0.03072611789 &  1.215813918E-06 &  1.806075690 & -8814.378269 & -7351.560326 & -7322.870481 & -15.68620869 \\
 0.2105383765 &  0.3591274162 &  1.196521209 &  6.068105605 &  0.2338199196 &  0.03946363289 &  5.385968016E-06 &  1.763073114 & -8825.918967 & -7351.896770 & -7310.754319 & -11.80349166 \\
 0.3167096010 &  0.3013326068 &  1.182303528 &  6.403781488 &  0.2271149911 &  0.04109875804 &  7.678369020E-06 &  1.754320576 & -8831.952716 & -7351.754046 & -7331.379635 & -16.77918426 \\
 0.3821448520 &  0.2791165556 &  1.347317272 &  3.934392788 &  0.3014302672 &  0.02885023553 &  2.445509380E-06 &  1.737901248 & -8820.579677 & -7343.602254 & -7317.633215 & -7.867900055 \\
 0.2165943076 &  0.3635355571 &  1.181519840 &  6.393015042 &  0.2021487138 &  0.03002857506 &  1.705734758E-05 &  1.769764920 & -8824.453292 & -7345.506611 & -7324.273572 & -14.27407904 \\
 0.2427174123 &  0.3063163211 &  1.334882255 &  3.848882101 &  0.2379513865 &  0.03493853912 &  5.573697825E-06 &  1.828267379 & -8825.491700 & -7347.539412 & -7320.610149 & -15.17407663 \\
 0.3773304611 &  0.2495196712 &  1.157570412 &  6.736330536 &  0.1785851954 &  0.04507839940 &  1.965151408E-05 &  1.875677390 & -8814.345106 & -7341.668587 & -7321.174301 & -17.64596773 \\
 0.3677569042 &  0.3024836460 &  1.183369807 &  6.089580856 &  0.2151081942 &  0.03307871097 &  2.305554888E-05 &  1.761946770 & -8830.007230 & -7340.788464 & -7320.023374 & -12.58180985 \\
 0.2868644472 &  0.3075321202 &  1.199491167 &  6.322856512 &  0.2789884882 &  0.04124663405 &  8.062803614E-06 &  1.737093604 & -8830.999985 & -7347.908241 & -7320.778705 & -14.98488652 \\
 0.1518656025 &  0.3731067289 &  1.221202694 &  5.990927397 &  0.3239990863 &  0.03661275054 &  1.636545774E-05 &  1.846703598 & -8833.346249 & -7346.211631 & -7320.881671 & -11.94110430 \\
 0.3472821522 &  0.3278869853 &  1.209654020 &  6.017522498 &  0.2611074931 &  0.02951890790 &  2.524419403E-05 &  1.766513299 & -8829.835354 & -7345.097944 & -7313.115782 & -15.99873265 \\
 0.3302583453 &  0.2928952592 &  1.213187374 &  6.025413095 &  0.2822762448 &  0.03351253197 &  2.345895249E-05 &  1.794704359 & -8815.618509 & -7347.727419 & -7325.640264 & -11.54817883 \\
 0.3604860976 &  0.2372683889 &  1.161906207 &  6.602236404 &  0.1625857749 &  0.03611731799 &  6.051464290E-06 &  1.849068497 & -8828.534827 & -7348.772390 & -7320.397536 & -13.39325890 \\
 0.3055625411 &  0.2833887697 &  1.200616115 &  5.958183085 &  0.2081244246 &  0.03710032962 &  7.662052411E-06 &  1.889390757 & -8814.523964 & -7346.493705 & -7321.560691 & -16.39882944 \\
 0.2183407779 &  0.3334764758 &  1.167464281 &  6.595685530 &  0.1706446957 &  0.04465379713 &  1.188814320E-06 &  1.784200119 & -8829.638549 & -7345.749504 & -7331.837892 & -13.33772114 \\
 0.1984973341 &  0.3176988338 &  1.211401687 &  5.986831712 &  0.2574753237 &  0.04071822733 &  6.944352539E-06 &  1.738252774 & -8820.792068 & -7346.869812 & -7318.929360 & -18.62511727 \\
\enddata

  \tablecomments{%
    This table is available in its entirety in machine-readable form.
  }
\end{deluxetable*}
\end{rotatetable*}

\movetabledown 150pt
\begin{rotatetable*}
\begin{deluxetable*}{lllllll|lllll}
  \savetablenum{10}
  \tablecaption{MCMC Samples of the Posterior Distribution of the \starone{} System Global Model, continued}
  \tabletypesize{\scriptsize}
  \tablehead{
    \colhead{$\gamma_{\mathrm{Minerva}}$}
    & \colhead{$\sigma_{\mathrm{\chiron}}$}
    & \colhead{$\sigma_{\mathrm{\coralie}}$}
    & \colhead{$\sigma_{\mathrm{\harps}}$}
    & \colhead{$\sigma_{\mathrm{\pfs}}$}
    & \colhead{$\sigma_{\mathrm{Minerva}}$}
    & \colhead{$\sigma_{\mathrm{Gaia}}$}
    & \colhead{$\si{\mass\etoile}$}
    & \colhead{\tempeff}
    & \colhead{$\log g$}
    & \colhead{$\feh$}
    & \colhead{$A_V$} \\
    \colhead{(\si{\m\per\s})}
    & \colhead{(\si{\m\per\s})}
    & \colhead{(\si{\m\per\s})}
    & \colhead{(\si{\m\per\s})}
    & \colhead{(\si{\m\per\s})}
    & \colhead{(\si{\m\per\s})}
    & \colhead{(mag)}
    & \colhead{(\si{\mass\sun})}
    & \colhead{(K)}
    & \colhead{dex (cgs)}
    & \colhead{(dex)}
    & \colhead{(mag)}
  }
  \startdata
 4.913117106 &  11.80783509 &  2.931753686 &  1.878447865 &  4.094919393 &  10.29846034 &  0.04480028421 &  1.192716790 &  5731.284349 &  3.937569753 &  0.2045896312 &  0.1342397046 \\
 8.406623049 &  8.507686535 &  4.095164200 &  17.82453025 &  5.761413217 &  21.44129287 &  0.1387982466 &  1.312030934 &  5766.495460 &  3.996230807 &  0.2081991686 &  0.03136706608 \\
 14.08058982 &  35.00147701 &  6.976980739 &  0.006700955315 &  7.011269208 &  22.69133925 &  0.02813423692 &  1.355467134 &  5842.101206 &  4.072250717 &  0.2483260967 &  0.01103046308 \\
 14.07196053 &  6.516747088 &  3.641491102 &  6.648112996 &  3.600161745 &  20.46339681 &  0.02165150310 &  1.116850021 &  5738.920651 &  3.968491838 &  0.2352644563 &  0.09220223772 \\
 17.83442702 &  19.37917339 &  4.946833459 &  4.400141229 &  8.448334079 &  22.81689764 &  0.03678679989 &  1.409331713 &  5810.987252 &  4.052233097 &  0.3551318105 &  0.09655895941 \\
 0.02070557530 &  0.1838163759 &  0.01884944914 &  6.572272709 &  8.816697741 &  10.71240378 &  0.01771027079 &  1.197184830 &  5698.049438 &  3.966029128 &  0.1988657039 &  0.07204926153 \\
 19.33146403 &  2.119878200 &  7.496396568 &  7.244874987 &  13.82749980 &  31.20356312 &  0.03234060104 &  1.191986489 &  5701.755811 &  3.987839806 &  0.2470259390 &  0.008689247956 \\
 22.31817399 &  14.53796836 &  7.228652374 &  2.876562533 &  9.579588294 &  22.46036001 &  0.02344168207 &  1.121046806 &  5620.892916 &  3.944791607 &  0.2131999711 &  0.02219366333 \\
 3.563450731 &  0.6326510575 &  2.328799450 &  4.850152210 &  5.048973375 &  18.46258980 &  0.005206092754 &  1.148838044 &  5738.815236 &  3.981172492 &  0.1766869121 &  0.04003253252 \\
 5.211869777 &  22.68132680 &  0.3886761662 &  3.229266879 &  11.75815156 &  16.14897382 &  0.1318140004 &  1.207345586 &  5641.225872 &  3.940678100 &  0.2282473431 &  0.1142067055 \\
 5.867777159 &  6.027676764 &  3.871804703 &  2.440999499 &  10.79981040 &  25.20618901 &  0.03851062694 &  1.198638210 &  5757.133651 &  3.970316009 &  0.1563077667 &  0.08707005101 \\
 13.12721861 &  14.16861893 &  5.237943523 &  8.618002679 &  5.047610494 &  21.83843607 &  0.03807088465 &  1.225428141 &  5766.891370 &  3.980813485 &  0.2276395038 &  0.06939327379 \\
 3.553915186 &  6.697281174 &  3.790026877 &  1.436510543 &  4.988250649 &  11.21374102 &  0.001245845221 &  1.301749658 &  5677.895835 &  3.954064226 &  0.1872169392 &  0.04428581592 \\
 10.07683623 &  2.262685133 &  1.394595490 &  1.814924650 &  6.625348552 &  14.08536319 &  0.008265903646 &  1.236933927 &  5713.514355 &  3.935563801 &  0.1591434201 &  0.1572896005 \\
 15.38205475 &  148.9552268 &  4.947382397 &  1.646634530 &  5.750661050 &  19.84380454 &  0.04088647710 &  1.238186677 &  5740.948969 &  3.991439253 &  0.2577384914 &  0.05245038966 \\
 8.501750140 &  5.975969973 &  5.087348735 &  2.714587521 &  8.788953561 &  21.93928510 &  0.01907940284 &  1.152517280 &  5688.277418 &  3.946539235 &  0.2063048264 &  0.1028350231 \\
 7.620126703 &  6.389034882 &  3.048123872 &  5.409479019 &  6.240166004 &  13.83408160 &  0.06316473292 &  1.239610868 &  5622.325442 &  3.949449878 &  0.2989037720 &  0.08211061640 \\
 2.384338383 &  28.23580994 &  2.216949723 &  1.202927742 &  7.158482918 &  21.45862679 &  0.06710510119 &  1.204985283 &  5774.143575 &  3.947488034 &  0.09757203443 &  0.1519931079 \\
 9.847416448 &  2.954566110 &  4.078009569 &  0.7098452990 &  3.769144029 &  10.18785021 &  0.003184479935 &  1.170790712 &  5664.013015 &  3.956631710 &  0.1938552674 &  0.02689107850 \\
 7.858651224 &  4.358018399 &  4.332878809 &  0.2889593087 &  7.628840794 &  11.17817194 &  0.02357488530 &  1.127662373 &  5619.166822 &  3.958680453 &  0.2750989725 &  0.01441750085 \\
 12.79539951 &  3.656249098 &  11.00378023 &  4.810279063 &  8.262648896 &  15.63901817 &  0.01515699739 &  1.124039792 &  5696.251860 &  3.981667765 &  0.2449298045 &  0.01853188891 \\
 17.57572294 &  5.489734815 &  7.921830527 &  0.3917720695 &  6.117191837 &  19.87373193 &  0.01196837009 &  1.200820433 &  5674.034045 &  3.928531963 &  0.1674968515 &  0.1264494478 \\
 11.31171174 &  13.47250797 &  8.460501682 &  1.152617508 &  3.258410329 &  13.26054508 &  0.01082942966 &  1.113646202 &  5729.830713 &  3.966712649 &  0.1741093055 &  0.08288150331 \\
 17.63126710 &  1.324405317 &  6.424960035 &  0.7664411932 &  5.947849451 &  15.96548940 &  0.02915466899 &  1.233188304 &  5728.602003 &  3.976544583 &  0.2580902788 &  0.05346639719 \\
 10.29838012 &  20.59583343 &  14.72302473 &  2.699006146 &  9.260274359 &  15.05063889 &  0.02158589407 &  1.053021347 &  5758.553809 &  3.996251584 &  0.2293050828 &  0.05607924634 \\
 2.301710166 &  8.603827153 &  3.913863029 &  10.17692491 &  5.934916770 &  22.17351879 &  0.05496868073 &  1.125217368 &  5781.301533 &  3.997093024 &  0.2162152021 &  0.07765417492 \\
 5.734065828 &  10.66699673 &  0.6402156742 &  18.33114864 &  6.091916926 &  18.05614901 &  0.04624798310 &  1.303722370 &  5725.824839 &  3.986516216 &  0.2081678801 &  0.005525742263 \\
 7.495799525 &  2.298097149 &  13.48711243 &  4.226697186 &  24.88042920 &  14.60113047 &  0.06654212462 &  1.315684180 &  5835.910718 &  4.036930392 &  0.3250835609 &  0.1359233895 \\
 10.66596925 &  7.410046673 &  6.991311445 &  0.9115583912 &  10.15575511 &  14.61677579 &  0.03855745182 &  1.158683470 &  5689.631304 &  3.957702921 &  0.1864715297 &  0.05984142201 \\
 15.37069314 &  6.327783774 &  8.586967656 &  1.557868354 &  6.461052241 &  19.90169740 &  0.01386945092 &  1.323691854 &  5902.529793 &  4.047888287 &  0.2477031133 &  0.1296306728 \\
 11.61966651 &  6.274351121 &  4.661149511 &  2.270231225 &  6.276731826 &  12.95163133 &  0.02865177169 &  1.138278877 &  5705.807785 &  3.976050971 &  0.1563257835 &  0.01801473150 \\
 5.937221618 &  52.13329126 &  1.891143185 &  4.312582419 &  7.016582114 &  17.17910388 &  0.05383297394 &  1.195867742 &  5833.523333 &  4.026028345 &  0.1920930055 &  0.06029690521 \\
 3.630854451 &  4.300307167 &  7.368921843 &  1.501226267 &  5.893777414 &  15.99124839 &  0.02302487413 &  1.129131298 &  5665.420549 &  3.968470089 &  0.2730388398 &  0.004402878085 \\
 9.107783648 &  7.575949745 &  7.012400524 &  4.286923791 &  10.18420065 &  14.69097605 &  0.02751335604 &  1.249228026 &  5688.846060 &  3.970306306 &  0.3244272135 &  0.07695520214 \\
 11.57790494 &  1.488862253 &  8.140773542 &  5.585207704 &  5.353476122 &  20.37856312 &  0.05466481150 &  1.253790899 &  5732.584046 &  3.975589616 &  0.2500479839 &  0.1280071044 \\
 12.24469287 &  11.01476964 &  0.2445814819 &  5.203025939 &  6.937963082 &  13.21743100 &  0.01761201361 &  1.333395780 &  5711.395727 &  3.971654192 &  0.2751385976 &  0.08820161698 \\
 8.168354247 &  1.283724756 &  6.696077100 &  4.099055005 &  6.575974022 &  15.59417273 &  0.0002357175641 &  1.156060805 &  5696.956337 &  3.957756004 &  0.1429033286 &  0.03989873459 \\
 4.127732845 &  12.34611369 &  7.442641655 &  3.992511434 &  3.239300168 &  16.24363736 &  0.01395903093 &  1.165629618 &  5769.391641 &  3.970547838 &  0.1857060013 &  0.1105679583 \\
 2.955543186 &  0.2316528802 &  1.442505945 &  11.37524050 &  5.696059698 &  21.59925361 &  0.006289368820 &  1.030917382 &  5645.240341 &  3.938164470 &  0.1600860550 &  0.05562087269 \\
 7.877529543 &  8.557819461 &  4.429692884 &  4.696823526 &  9.174484819 &  13.49100335 &  0.01702212584 &  1.263610913 &  5731.465835 &  3.971634017 &  0.2458971457 &  0.09272161878 \\
\enddata

  \tablecomments{%
    This table is available in its entirety in machine-readable form.
  }
\end{deluxetable*}
\end{rotatetable*}

\movetabledown 200pt
\begin{rotatetable*}
\begin{deluxetable*}{l|lllllllllllll}
  \tablecaption{MCMC Chains from Global Model Fitting of \startwo{} \label{tab:mcmc_k2p}}
  \tabletypesize{\scriptsize}
  \tablehead{
      \colhead{$\ln(\mathrm{prob})$}
      & \colhead{$T_c - 2454833$}
      & \colhead{$P$}
      & \colhead{$K$}
      & \colhead{$\sqrt{e} \cos \omega$}
      & \colhead{$\sqrt{e} \sin \omega$}
      & \colhead{$b$}
      & \colhead{$\si{\radius\planet} / \si{\radius\etoile}$}
      & \colhead{$q_{1,Kp}$}
      & \colhead{$q_{2,Kp}$}
      & \colhead{$q_{1,\filterRc}$}
      & \colhead{$q_{2,\filterRc}$}
      & \colhead{$\si{\mass\etoile}_0$}
      & \colhead{Age} \\
    \colhead{}
    & \colhead{\bjdtdb}
    & \colhead{(day)}
    & \colhead{(\si{\m\per\s})}
    &
    &
    &
    &
    &
    &
    &
    &
    & \colhead{(\si{\mass\sun})}
    & \colhead{(\si{\giga\year})}
  }
  \startdata
 9030.186315 &  2940.157243 &  12.45512562 &  20.56601563 & -0.2720677634 & -0.05093490589 &  0.4086270170 &  0.09756262281 &  0.2793360027 &  0.4718267777 &  0.5584847432 &  0.3680928959 &  0.9164980254 &  0.2716261012 \\
 9028.646788 &  2940.157320 &  12.45512261 &  23.46413567 & -0.2637666133 &  0.1273939573 &  0.2424894930 &  0.09554129052 &  0.3482502978 &  0.4461379077 &  0.6860412212 &  0.3255358883 &  0.8886869605 &  1.966543666 \\
 9019.066395 &  2940.157442 &  12.45511398 &  25.82428284 & -0.2722661847 &  0.1566868578 &  0.3631822807 &  0.09685327914 &  0.2448005327 &  0.5840225232 &  0.4356197655 &  0.3986996156 &  0.8903151189 &  4.765447530 \\
 9030.701337 &  2940.157294 &  12.45512641 &  26.99085863 & -0.09567391021 & -0.05474587440 &  0.4265447408 &  0.09756948529 &  0.2762762805 &  0.5483463051 &  0.5310874441 &  0.3687939038 &  0.8975483490 &  2.599495029 \\
 9026.810889 &  2940.157308 &  12.45512165 &  20.90472432 &  0.1543200682 & -0.01250867347 &  0.3477047141 &  0.09659560606 &  0.3044600244 &  0.4661613346 &  0.6005465665 &  0.3612164923 &  0.9132745204 &  0.4123369852 \\
 9025.000904 &  2940.157275 &  12.45511796 &  23.88060574 &  0.01474868634 &  0.1946254453 &  0.3983025384 &  0.09741434271 &  0.2447636188 &  0.5497223171 &  0.5229856234 &  0.4178944872 &  0.8678692066 &  4.280864864 \\
 9028.810561 &  2940.157177 &  12.45513040 &  25.61321224 & -0.3055556357 &  0.1237041438 &  0.2263905754 &  0.09581476769 &  0.2714943517 &  0.5077245354 &  0.4916869276 &  0.3535972491 &  0.9061917477 &  0.2629036005 \\
 9033.741343 &  2940.157306 &  12.45512085 &  26.35508132 & -0.3309005399 &  0.03130792873 &  0.3727059201 &  0.09682899174 &  0.2808511621 &  0.5170698342 &  0.6271042593 &  0.3478391712 &  0.9043651351 &  1.435053620 \\
 9025.233071 &  2940.157211 &  12.45512511 &  22.28906522 & -0.1252359578 &  0.1998893268 &  0.2970575248 &  0.09611080562 &  0.2678139335 &  0.5215226029 &  0.5179156383 &  0.4034357411 &  0.8835300668 &  2.808092504 \\
 9034.396425 &  2940.157246 &  12.45512242 &  21.77867933 & -0.2506131206 &  0.1281167171 &  0.2826127292 &  0.09575630745 &  0.3357788869 &  0.4388244844 &  0.5498015043 &  0.3678371907 &  0.8775184508 &  1.574103279 \\
 9032.386749 &  2940.157316 &  12.45512263 &  26.42899528 & -0.2768530921 &  0.07079903188 &  0.3299538041 &  0.09622294241 &  0.3110059444 &  0.5017271267 &  0.6403464103 &  0.3103246988 &  0.8912041322 &  2.552825078 \\
 9034.456833 &  2940.157221 &  12.45512691 &  24.91658814 & -0.2638119273 &  0.07526987488 &  0.2891071232 &  0.09601246825 &  0.3100458964 &  0.4543586313 &  0.3901594156 &  0.4511326761 &  0.9249305225 &  0.3499671027 \\
 9031.027030 &  2940.157314 &  12.45512234 &  22.44398240 & -0.2488323130 &  0.07222002323 &  0.2695096248 &  0.09538851156 &  0.3762204668 &  0.4150962327 &  0.5607598861 &  0.3426925057 &  0.8818576064 &  1.465937607 \\
 9031.855324 &  2940.157209 &  12.45512545 &  25.00653167 & -0.2254374465 &  0.03336957080 &  0.3541544370 &  0.09703573586 &  0.2558602704 &  0.5446425393 &  0.4915776937 &  0.3872660785 &  0.8915138663 &  0.8675418747 \\
 9032.505365 &  2940.157257 &  12.45512606 &  26.36736386 & -0.2088444834 &  0.1577836761 &  0.3038029833 &  0.09642644372 &  0.2500208689 &  0.5620857896 &  0.5431548657 &  0.3634151987 &  0.8945128174 &  2.121103643 \\
 9030.162700 &  2940.157200 &  12.45512239 &  22.29810913 & -0.3104654293 &  0.1372487790 &  0.4560738799 &  0.09758243208 &  0.2911223756 &  0.4772347213 &  0.4734006756 &  0.4020026277 &  0.8608677332 &  8.055674129 \\
 9032.424064 &  2940.157296 &  12.45511854 &  25.12159573 &  0.09363066716 & -0.04489802256 &  0.3570176985 &  0.09680647551 &  0.2739056856 &  0.5424325476 &  0.5833737838 &  0.3831716148 &  0.8962886433 &  2.122750419 \\
 9033.194525 &  2940.157271 &  12.45512145 &  24.68954975 & -0.2603135493 &  0.1801376044 &  0.3717805490 &  0.09679105956 &  0.2638948301 &  0.5619077814 &  0.5266685973 &  0.3870735356 &  0.9027255962 &  3.514625915 \\
 9032.799839 &  2940.157137 &  12.45512147 &  23.76513865 & -0.2576906509 &  0.2176437180 &  0.3797844534 &  0.09674006111 &  0.2879732501 &  0.5020657098 &  0.5644662528 &  0.3560344196 &  0.8817978513 &  6.537262268 \\
 9033.814696 &  2940.157292 &  12.45512364 &  24.63792087 & -0.2867964316 & -0.05471525640 &  0.4146227391 &  0.09748247703 &  0.2765910557 &  0.5207609601 &  0.6195410945 &  0.3332648100 &  0.8967332385 &  0.9867299870 \\
 9015.384090 &  2940.157109 &  12.45512644 &  24.62322400 &  0.04149317211 &  0.06146205384 &  0.4072335019 &  0.09727450932 &  0.2760995736 &  0.5806179483 &  0.7447304519 &  0.3536503512 &  0.9273634518 &  0.4748768228 \\
 9028.917069 &  2940.157259 &  12.45512262 &  22.53611809 &  0.09255373477 &  0.06760447856 &  0.4232350960 &  0.09749448057 &  0.2897817478 &  0.4778429471 &  0.4267162038 &  0.4051381754 &  0.9308320111 &  0.9381137145 \\
 9030.007819 &  2940.157223 &  12.45512491 &  26.30801239 & -0.1558783232 &  0.02716873102 &  0.3370871441 &  0.09678340167 &  0.2788298063 &  0.4785606472 &  0.4809590921 &  0.4119755656 &  0.8902685780 &  0.6500916401 \\
 9030.012865 &  2940.157277 &  12.45511946 &  22.78156296 & -0.3563693872 &  0.07563848018 &  0.3016102345 &  0.09602873156 &  0.3200209880 &  0.4363747572 &  0.5999623716 &  0.3243059883 &  0.8819976807 &  2.131245224 \\
 9032.482524 &  2940.157240 &  12.45511867 &  24.63500373 & -0.2175260190 &  0.05978725035 &  0.3794037471 &  0.09689903311 &  0.2992957087 &  0.4833220392 &  0.4521017948 &  0.3934466932 &  0.9176458733 &  1.139045188 \\
 9030.723132 &  2940.157379 &  12.45511849 &  25.41361476 & -0.2519305301 &  0.008347532640 &  0.3741333634 &  0.09683966210 &  0.3060060131 &  0.4663451566 &  0.5640508466 &  0.3547036568 &  0.9015023516 &  0.7895645913 \\
 9029.086963 &  2940.157300 &  12.45512090 &  26.38714860 & -0.2493640922 &  0.2110184519 &  0.3642955543 &  0.09684462518 &  0.2713298432 &  0.5013325377 &  0.5807193296 &  0.3601693407 &  0.9085089798 &  6.009454530 \\
 9033.202281 &  2940.157362 &  12.45512161 &  26.67510025 & -0.2875575853 &  0.03945424767 &  0.4333934181 &  0.09772790083 &  0.2607997907 &  0.5401638276 &  0.5613693380 &  0.3797207846 &  0.8903806385 &  3.274889093 \\
 9033.926934 &  2940.157129 &  12.45512598 &  23.69906171 & -0.2759662093 &  0.02425994092 &  0.3721233293 &  0.09708015676 &  0.2776312146 &  0.5068067620 &  0.6121307921 &  0.3544870153 &  0.8957547579 &  1.504118921 \\
 9028.090482 &  2940.157223 &  12.45512023 &  20.98101253 & -0.1437736863 & -0.04874003464 &  0.4332163478 &  0.09807795540 &  0.2356842113 &  0.5874802170 &  0.7578131104 &  0.3145929614 &  0.8984002184 &  1.851117382 \\
 9025.367350 &  2940.157321 &  12.45512297 &  18.82915680 & -0.2003746794 &  0.1449210640 &  0.3515362579 &  0.09644904672 &  0.2945613680 &  0.5025204833 &  0.5126778124 &  0.4207223701 &  0.9189836025 &  2.142203765 \\
 9031.228110 &  2940.157251 &  12.45511789 &  23.19467994 & -0.3208909375 &  0.009818677124 &  0.3757268704 &  0.09698040918 &  0.2798154577 &  0.4722182089 &  0.5185740787 &  0.3707205450 &  0.9158902728 &  0.3880841795 \\
 9024.205210 &  2940.157454 &  12.45511933 &  23.67576431 & -0.1972182063 & -0.04899478901 &  0.3575259452 &  0.09678886140 &  0.2804870844 &  0.5216895123 &  0.4481496301 &  0.4147153130 &  0.8658790848 &  1.157401863 \\
 9030.452297 &  2940.157310 &  12.45511720 &  22.54384054 & -0.1876029699 &  0.04047476942 &  0.3714148586 &  0.09720247667 &  0.2402272524 &  0.5610116016 &  0.4818513913 &  0.3914228497 &  0.8864300686 &  1.889068301 \\
 9030.669104 &  2940.157200 &  12.45512683 &  21.67559349 & -0.2604466501 &  0.1442098216 &  0.3794101331 &  0.09654716779 &  0.3443920236 &  0.4261694710 &  0.5402810197 &  0.3588858644 &  0.8904461498 &  4.063394757 \\
 9033.161702 &  2940.157321 &  12.45511900 &  24.89990735 & -0.1610580905 &  0.1969612781 &  0.3847155735 &  0.09713458515 &  0.2563181872 &  0.5395767857 &  0.5001970516 &  0.4123744672 &  0.8561404028 &  6.827279998 \\
 9025.936565 &  2940.157391 &  12.45512021 &  26.45843996 & -0.2858815902 &  0.1481158535 &  0.3985160567 &  0.09689635131 &  0.3048857082 &  0.4792874894 &  0.7173478060 &  0.3387693557 &  0.8963835788 &  5.697772098 \\
 9027.303352 &  2940.157353 &  12.45512282 &  25.53851303 & -0.1395790216 &  0.07863888710 &  0.3518060376 &  0.09669658522 &  0.2686282596 &  0.5302908409 &  0.6752654643 &  0.3624906978 &  0.8889396480 &  1.521362850 \\
 9033.458382 &  2940.157294 &  12.45512183 &  23.08605913 & -0.08917373111 &  0.2422073072 &  0.2145733987 &  0.09558820407 &  0.2891356013 &  0.4986695935 &  0.5103258634 &  0.3734914973 &  0.8774202935 &  2.167557525 \\
 9032.954743 &  2940.157302 &  12.45512007 &  25.83384412 & -0.3116018376 &  0.08838955294 &  0.3769955032 &  0.09691491219 &  0.2948222948 &  0.4811202733 &  0.5906567344 &  0.3332350332 &  0.8976326928 &  2.919940091 \\
\enddata

  \tablecomments{%
    This table is available in its entirety in machine-readable form.
  }
\end{deluxetable*}
\end{rotatetable*}

\movetabledown 205pt
\begin{rotatetable*}
\begin{deluxetable*}{lllllllll|lllll}
  \savetablenum{11}
  \tablecaption{MCMC Chains from Global Model Fitting of \startwo{}, continued}
  \tabletypesize{\scriptsize}
  \tablehead{
    \colhead{$\feh_0$}
    & \colhead{$\sigma_{\mathrm{\pest}}$}
    & \colhead{$\gamma_{\mathrm{\feros}}$}
    & \colhead{$\gamma_{\mathrm{\harps}}$}
    & \colhead{$\gamma_{\mathrm{\pfs}}$}
    & \colhead{$\sigma_{\mathrm{\feros}}$}
    & \colhead{$\sigma_{\mathrm{\harps}}$}
    & \colhead{$\sigma_{\mathrm{\pfs}}$}
    & \colhead{$\sigma_{\mathrm{Gaia}}$}
    & \colhead{$\si{\mass\etoile}$}
    & \colhead{\tempeff}
    & \colhead{$\log g$}
    & \colhead{$\feh$}
    & \colhead{$A_V$} \\
    \colhead{(dex)}
    &
    & \colhead{(\si{\m\per\s})}
    & \colhead{(\si{\m\per\s})}
    & \colhead{(\si{\m\per\s})}
    & \colhead{(\si{\m\per\s})}
    & \colhead{(\si{\m\per\s})}
    & \colhead{(\si{\m\per\s})}
    & \colhead{(mag)}
    & \colhead{(\si{\mass\sun})}
    & \colhead{(K)}
    & \colhead{dex (cgs)}
    & \colhead{(dex)}
    & \colhead{(mag)}
  }
  \startdata
 0.1067472846 &  0.002295298344 & -17000.20888 & -16980.01836 & -0.1349523945 &  17.54211334 &  4.347131013 &  4.902039183 &  0.1030793367 &  0.9468298273 &  5262.157394 &  4.579652770 &  0.1544892802 &  0.05603475074 \\
 0.04006723270 &  0.002012772325 & -16993.48997 & -16984.57233 &  0.2317459247 &  18.83289560 &  1.534166414 &  3.690704396 &  0.06754909507 &  0.8430637766 &  5279.781709 &  4.567779742 &  0.06761418847 &  0.09774834280 \\
 0.1772763964 &  0.003735519623 & -16997.31034 & -16983.48886 &  1.919175728 &  11.13976590 &  3.455383736 &  1.504493500 &  0.02382188014 &  0.8798104862 &  5167.821368 &  4.540016004 &  0.1940271931 &  0.09933061924 \\
 0.08056193366 &  0.002129071992 & -16998.62380 & -16984.20833 &  1.391843624 &  19.52987952 &  3.652537582 &  3.597829329 &  0.05602356562 &  0.8856640205 &  5281.637757 &  4.557093263 &  0.1055326539 &  0.1317277318 \\
 0.08868968073 &  0.002299946889 & -17001.29128 & -16985.61597 &  0.7763066214 &  23.30827709 &  1.426333615 &  6.095073278 &  0.08453888953 &  0.8477593447 &  5283.435685 &  4.578515825 &  0.1338508866 &  0.1021427142 \\
-0.004141030805 &  0.002346784897 & -16995.40113 & -16979.05105 &  1.311991174 &  17.90302183 &  11.36634868 &  3.860725887 &  0.1141486634 &  0.9179740089 &  5298.076979 &  4.547418398 & -0.0005308913197 &  0.1019990887 \\
 0.008289072968 &  0.002565125175 & -17003.80185 & -16985.30257 &  2.519088845 &  17.74207034 &  0.5588223166 &  5.271083855 &  0.05966098790 &  0.9053959538 &  5357.729187 &  4.581942062 &  0.04918065132 &  0.1475731608 \\
 0.05546687583 &  0.002539916708 & -17002.01870 & -16980.39011 &  1.450221779 &  20.00197191 &  11.17886818 &  2.131267261 &  0.07009740759 &  0.9183952833 &  5321.167628 &  4.567819302 &  0.08870814425 &  0.1153523402 \\
 0.04581985619 &  0.002758082449 & -16992.66663 & -16985.88286 &  2.887513566 &  17.32002788 &  4.034844223 &  5.146304685 &  0.05090912150 &  0.8895325817 &  5267.780473 &  4.560220625 &  0.06646148225 &  0.1024913294 \\
-0.01509570930 &  0.001890767687 & -16999.79665 & -16981.83529 &  2.429900422 &  16.83394168 &  6.351907733 &  4.270526647 &  0.01167836270 &  0.9303034786 &  5296.391365 &  4.576645815 &  0.01232667236 &  0.07443200349 \\
 0.09612472436 &  0.002508905460 & -16998.08015 & -16981.18945 &  0.6168003521 &  17.85561584 &  1.597753950 &  2.554074201 &  0.005745753486 &  0.8437302395 &  5227.477134 &  4.561728062 &  0.1232046745 &  0.06809470113 \\
 0.1304501914 &  0.002771594868 & -16992.73436 & -16983.40014 &  1.173137588 &  15.66371296 &  3.385892671 &  3.429236081 &  0.003691080904 &  0.8474482544 &  5271.017599 &  4.575242366 &  0.1791012990 &  0.1322967636 \\
-0.05074210170 &  0.002383689254 & -17001.02069 & -16983.06698 &  0.6986417891 &  16.71875041 &  2.235542694 &  3.015000769 &  0.1252313983 &  0.8439107707 &  5368.577630 &  4.576038209 & -0.02427191132 &  0.1579018341 \\
-0.005627884190 &  0.001943777833 & -16996.70622 & -16985.08332 & -1.987181959 &  17.05251966 &  3.722273604 &  4.288478453 &  0.04547944312 &  0.9095405827 &  5331.277816 &  4.579139593 &  0.02862337326 &  0.1032990534 \\
 0.09435851674 &  0.001928710168 & -16996.66106 & -16980.36597 &  0.8101121825 &  18.40927171 &  0.3236596374 &  3.767146628 &  0.02661271315 &  0.8783141709 &  5235.424990 &  4.565227352 &  0.1249373512 &  0.07320728311 \\
 0.07799380927 &  0.002181404360 & -16999.68548 & -16983.22231 & -0.7188172432 &  14.79691617 &  4.092690362 &  5.891297018 &  0.03960571032 &  0.8979726180 &  5239.997996 &  4.507772032 &  0.05999886079 &  0.1216115401 \\
 0.07171633357 &  0.002504800558 & -16994.46714 & -16981.64704 &  1.069107252 &  15.79932201 &  8.181928232 &  4.356120293 &  0.01562893266 &  0.7795995005 &  5276.001312 &  4.563426352 &  0.1002557634 &  0.1283475843 \\
 0.08729300400 &  0.002064813951 & -16998.36053 & -16982.06165 &  0.9177453928 &  16.58169916 &  2.365137525 &  5.090594402 &  0.1807509487 &  0.9526603293 &  5319.081738 &  4.542107982 &  0.1043709740 &  0.1231837109 \\
 0.1384384107 &  0.002269691633 & -17009.45953 & -16983.48104 &  0.1365339380 &  18.10370755 &  5.990631893 &  2.927023662 &  0.001127107908 &  0.9066321354 &  5223.868344 &  4.519030996 &  0.1375899594 &  0.1516727496 \\
 0.02934686316 &  0.002557659470 & -17004.79339 & -16979.99727 &  0.7387377089 &  20.85353474 &  8.241606335 &  2.180475664 &  0.04059190063 &  0.9459432814 &  5309.261728 &  4.576166787 &  0.06488567854 &  0.1033271373 \\
 0.1586177077 &  0.001433918498 & -16990.85191 & -16981.77419 &  3.306513082 &  14.22088662 &  6.376785289 &  6.838085311 &  0.01957658071 &  1.000827330 &  5246.104031 &  4.572659421 &  0.2081309095 &  0.1261102351 \\
 0.1793458391 &  0.002842203553 & -16996.82011 & -16986.54076 &  0.9860413207 &  16.71288883 &  11.60123858 &  5.719895773 &  0.01464752931 &  0.9900017622 &  5246.006749 &  4.565257187 &  0.2262153818 &  0.1202009892 \\
 0.04522864857 &  0.001869724455 & -16998.49912 & -16978.98948 &  0.1754043530 &  20.11707252 &  9.481499089 &  2.282640809 &  0.03966895999 &  0.8779575214 &  5247.867426 &  4.583904527 &  0.08524151381 &  0.08033685379 \\
-0.01340667690 &  0.002526608780 & -17000.84149 & -16983.56724 & -1.373368934 &  16.60880492 &  6.768414365 &  6.117204170 &  0.04591400546 &  0.8816788224 &  5330.430457 &  4.567411498 &  0.008646459610 &  0.1203286173 \\
 0.09443408168 &  0.002122534113 & -17001.51209 & -16987.91182 & -0.6894916436 &  16.95140513 &  3.605853050 &  4.244751372 &  0.09244849896 &  0.9178887677 &  5315.676454 &  4.566458138 &  0.1331700344 &  0.09951815448 \\
 0.05572101837 &  0.002393842976 & -17002.03930 & -16985.22820 &  1.253240848 &  17.43990175 &  18.27769974 &  4.291414687 &  0.03535438633 &  0.9235192589 &  5288.707122 &  4.577468452 &  0.09505569778 &  0.1016837287 \\
 0.2312026258 &  0.002888195009 & -16998.41017 & -16981.86424 &  0.7488585413 &  14.77685458 &  2.202477121 &  3.957219429 &  0.1479616017 &  0.8202541275 &  5208.453418 &  4.515072170 &  0.2426581929 &  0.1054168545 \\
 0.1029434406 &  0.001845108734 & -16995.74005 & -16983.79678 &  0.3327538097 &  15.90309904 &  4.518923864 &  4.470092914 &  0.04991446888 &  0.9610246800 &  5232.269991 &  4.553605626 &  0.1244391522 &  0.08252748404 \\
 0.05309628686 &  0.002159469178 & -16996.09624 & -16986.47596 &  0.3984961007 &  17.57705383 &  10.57424995 &  3.082352703 &  0.03979904231 &  0.8962388817 &  5283.287930 &  4.570628536 &  0.08572520448 &  0.08586189814 \\
 0.07292565430 &  0.002425734914 & -16997.64063 & -16984.34796 & -0.3262286635 &  19.63841522 &  3.463988103 &  4.111668042 &  0.03006232855 &  0.9396074699 &  5277.562405 &  4.565863053 &  0.1039768636 &  0.1115405137 \\
 0.1234817428 &  0.001498037503 & -17004.89146 & -16983.50921 &  1.109900642 &  14.18506845 &  3.174678990 &  5.599390764 &  0.1179672361 &  0.8846147004 &  5303.529953 &  4.553176384 &  0.1553047431 &  0.1375370376 \\
 0.1130282428 &  0.002515319646 & -16998.50921 & -16985.69089 & -1.171428016 &  11.84676700 &  7.984478888 &  4.968205019 &  0.02187878308 &  0.9435140594 &  5256.696000 &  4.578012726 &  0.1601410838 &  0.09396631619 \\
-0.03919969376 &  0.001820250084 & -16996.71556 & -16985.78730 &  2.630264883 &  18.00543276 &  4.981436803 &  5.401721010 &  0.02940716843 &  0.8784696565 &  5264.150793 &  4.587142375 & -0.008743736223 &  0.06005346434 \\
-0.01190615515 &  0.002003258966 & -16997.94580 & -16982.21389 &  1.195357517 &  18.68726760 &  4.570444939 &  1.974482539 &  0.09421797659 &  0.8644565404 &  5343.346538 &  4.568256617 &  0.01237824158 &  0.1339318471 \\
 0.09453776765 &  0.003014052825 & -17001.00203 & -16981.52754 &  2.919802198 &  16.32589820 &  4.483059850 &  3.620512861 &  0.09197313342 &  0.9338102348 &  5264.162009 &  4.542873638 &  0.1084143501 &  0.1165613845 \\
 0.05540738354 &  0.002250593708 & -16996.92826 & -16980.95885 &  2.365003412 &  17.60498443 &  8.321640063 &  2.967705295 &  0.03719175450 &  0.8316403786 &  5216.250324 &  4.526815783 &  0.04492130087 &  0.1057559596 \\
 0.1358063835 &  0.001750942534 & -17001.28867 & -16983.61164 &  4.518031896 &  11.16718412 &  4.205925033 &  3.362333781 &  0.1147203230 &  0.8400701722 &  5276.013631 &  4.519275371 &  0.1399245331 &  0.1338361434 \\
 0.03844058226 &  0.002556962704 & -17000.71313 & -16978.74640 & -1.936519929 &  20.27875801 &  1.042637648 &  2.304653351 &  0.01622794540 &  0.9002545818 &  5272.430456 &  4.573132162 &  0.06991915110 &  0.1095326859 \\
-0.01200913928 &  0.002133379963 & -16997.46321 & -16979.46720 & -0.5809969008 &  21.43096704 &  4.256258402 &  4.444883139 &  0.002071142173 &  0.9123416778 &  5306.098662 &  4.569250316 &  0.009976464461 &  0.1331505863 \\
 0.1289033098 &  0.002546789921 & -17002.01502 & -16986.41576 &  0.8333170442 &  13.37606126 &  0.2282957654 &  3.766475566 &  0.04115436671 &  0.9294784578 &  5221.552232 &  4.555389704 &  0.1554884223 &  0.06978906375 \\
\enddata

  \tablecomments{%
    This table is available in its entirety in machine-readable form.
  }
\end{deluxetable*}
\end{rotatetable*}

\vspace{3em}
\section*{Acknowledgments}

This paper includes data collected by the \tess{} mission, which are publicly available from the Mikulski Archive for Space Telescopes (MAST).
Funding for the \tess{} mission is provided by NASA's Science Mission directorate.
Resources supporting this work were provided by the NASA High-End Computing (HEC) Program through the NASA Advanced Supercomputing (NAS) Division at Ames Research Center for the production of the SPOC data products.

This paper includes data collected by the \ktwo{} mission.
Funding for the \ktwo{} mission is provided by the NASA Science Mission directorate.

This research has made use of the Exoplanet Follow-up Observation Program website, which is operated by the California Institute of Technology, under contract with the National Aeronautics and Space Administration under the Exoplanet Exploration Program.

This work makes use of observations from the LCOGT network.

The HATSouth network is operated by a collaboration consisting of Princeton University (PU), the Max-Planck-Institut für Astronomie (MPIA), the Australian National University (ANU), and the Pontificia Universidad Católica de Chile (PUC).  The station at Las Campanas Observatory (LCO) of the Carnegie Institute is operated by PU in conjunction with PUC, the station at the High Energy Spectroscopic Survey (H.E.S.S.) site is operated in conjunction with MPIA, and the station at Siding Spring Observatory (SSO) is operated jointly with ANU. Development of the HATSouth project was funded by NSF MRI grant NSF/AST-0723074, and operations have been supported by NASA grants NNX09AB29G, NNX12AH91H, and NNX17AB61G.

This paper includes observations made with the Nordic Optical Telescope (Program ID: 55-019), operated by the Nordic Optical Telescope Scientific Association at the Observatorio del Roque de los Muchachos, La Palma, Spain, of the Instituto de Astrofísica de Canarias.

This paper includes data gathered with the 6.5\thinspace m Magellan Telescopes located at Las Campanas Observatory, Chile.

This paper includes observations made with {\textsc{Minerva}}-Australis. {\textsc{Minerva}}-Australis is supported by Australian Research Council LIEF Grant LE160100001, Discovery Grant DP180100972, Mount Cuba Astronomical Foundation, and institutional partners University of Southern Queensland, UNSW Australia, MIT, Nanjing University, George Mason University, University of Louisville, University of California Riverside, University of Florida, and The University of Texas at Austin.
We respectfully acknowledge the traditional custodians of all lands throughout Australia and recognize their continued cultural and spiritual connection to the land, waterways, cosmos, and community. We pay our deepest respects to all Elders, ancestors, and descendants of the Giabal, Jarowair, and Kambuwal nations, upon whose lands the {\textsc{Minerva}}-Australis facility at Mt.\ Kent is situated.

This research is based on observations collected with the CORALIE echelle spectrograph mounted on the 1.2\thinspace m Swiss telescope and with the HARPS spectrograph on the ESO 3.6\thinspace m telescope at La Silla Observatory of the European Organisation for Astronomical Research in the Southern Hemisphere under ESO program 0103.C-0874(A).

NESSI was funded by the NASA Exoplanet Exploration Program and the NASA Ames Research Center. NESSI was built at the Ames Research Center by Steve B. Howell, Nic Scott, Elliott P. Horch, and Emmett Quigley. The authors are honored to be permitted to conduct observations on Iolkam Du'ag (Kitt Peak), a mountain within the Tohono O'odham Nation with particular significance to the Tohono O'odham people.

We  thank  the  Swiss  National  Science  Foundation  (SNSF) and the Geneva University for their continuous support of our planet search programs. This work has been carried out under the framework of the National Centre for Competence in Research ``PlanetS'' supported by SNSF. This publication makes use of the Data \& Analysis Center for Exoplanets (DACE), which is a facility based at the University of Geneva dedicated to extrasolar planet data visualization, exchange, and analysis. DACE is a platform of the Swiss National Centre of Competence in Research (NCCR) PlanetS, federating the Swiss expertise in Exoplanet research. The DACE platform is available at \url{https://dace.unige.ch}.

This work is done under the framework of the KESPRINT collaboration (\url{http://kesprint.science}). KESPRINT is an international consortium devoted to the characterization and research of exoplanets discovered with space-based missions.

A.J. and R.B. acknowledge support from FONDECYT Postdoctoral Fellowship Projects 1171208 and 3180246, respectively, and ANID---Millennium Science Initiative---ICN12\_009.

D.D. and J.K.T. acknowledge support provided by NASA through Hubble Fellowship grants HST-HF2-51372.001-A and HST-HF2-51399.001, respectively, awarded by the Space Telescope Science Institute, which is operated by the Association of Universities for Research in Astronomy, Inc., for NASA, under contract NAS5-26555.

I.J.M.C. acknowledges support from the NSF through grant AST-1824644.

J.H. is supported by SNSF through Ambizione grant No.\ PZ00P2\_180098.

K.P. acknowledges support from NASA grant 80NSSC18K1009.

M.R.D. acknowledges the support of CONICYT/\-PFCHA-Doctorado Nacional-2014 21140646 Chile and Proyecto Basal AFB-170002.

M.T. is supported by MEXT/JSPS KAKENHI grant Nos.\ 18H05442, 15H02063, and 22000005.

This material is based upon work supported by the National Science Foundation Graduate Research Fellowship Program under grant No.\ DGE-1746045. Any opinions, findings, and conclusions or recommendations expressed in this material are those of the authors and do not necessarily reflect the views of the National Science Foundation. 

Part of this research was carried out at the Jet Propulsion Laboratory, California Institute of Technology, under a contract with the National Aeronautics and Space Administration (NASA).

This work is partly supported by JSPS KAKENHI grant No.\ JP18H01265 and JP18H05439, and JST PRESTO grant No.\ JPMJPR1775.

L.S. would like to thank William S. Moses and MIT CSAIL for providing computing resources.

\facilities{%
    ATT (echelle spectrograph),
    CTIO:1.5\thinspace m (\chiron{}),
    ESO:3.6\thinspace m (\harps{}),
    Euler:1.2\thinspace m (\coralie{}),
    FLWO:1.5\thinspace m (\tres),
    \gaia{},
    HATSouth,
    \kepler{} (\ktwo{}),
    LCOGT (SAAO),
    Magellan:Clay (Planet Finder Spectrograph),
    Max Planck:2.2\thinspace m (\feros{}),
    \minervaaus{},
    MKO CDK700,
    NOT:2.56\thinspace m (\fies{}),
    Perth Exoplanet Survey Telescope (\pest{}),
    Peter van der Kamp:0.6\thinspace m,
    SOAR (HRCam),
    \tess{},
    VLT:Antu (NaCo),
    WIYN (NESSI).
}

\software{%
    AstroImageJ \citep{Collins:2017},
    Astropy \citep{astr13,astr18},
    Batman \citep{batman_paper},
    Emcee \citep{emcee_paper,emcee_zenodo},
    EXOFASTv2 \citep{Eastman:2013,Eastman:2019},
    H5py,
    Isochrones \citep{isochrones_python},
    Matplotlib \citep{matplotlibpaper},
    Numpy \citep{harris2020array},
    MIT Quick Look Pipeline \citep{qlp_rnaas_1,qlp_rnaas_2},
    \tess{} SPOC Pipeline \citep{Jenkins:2016,li18,twic18},
    Pandas \citep{pandas_zenodo},
    Radvel \citep{radvel_paper,radvel_zenodo},
    Scikit-learn \citep{scikit-learn},
    Scipy \citep{2020SciPy-NMeth},
    Tapir \citep{Jensen_2013},
    TRES SPC \citep{2012Natur.486..375B},
    Vartools \citep{hartman_vartools_2016}. 
}

\bibliographystyle{aasjournal}
\bibliography{toi954_k2329}

\begin{thebibliography}{}
\expandafter\ifx\csname natexlab\endcsname\relax\def\natexlab#1{#1}\fi
\providecommand{\url}[1]{\href{#1}{#1}}
\providecommand{\dodoi}[1]{doi:~\href{http://doi.org/#1}{\nolinkurl{#1}}}
\providecommand{\doeprint}[1]{\href{http://ascl.net/#1}{\nolinkurl{http://ascl.net/#1}}}
\providecommand{\doarXiv}[1]{\href{https://arxiv.org/abs/#1}{\nolinkurl{https://arxiv.org/abs/#1}}}

\bibitem[{{Addison} {et~al.}(2019){Addison}, {Wright}, {Wittenmyer}, {Horner},
  {Mengel}, {Johns}, {Marti}, {Nicholson}, {Soutter}, {Bowler}, {Crossfield},
  {Kane}, {Kielkopf}, {Plavchan}, {Tinney}, {Zhang}, {Clark}, {Clerte},
  {Eastman}, {Swift}, {Bottom}, {Muirhead}, {McCrady}, {Herzig}, {Hogstrom},
  {Wilson}, {Sliski}, {Johnson}, {Wright}, {Johnson}, {Blake}, {Riddle}, {Lin},
  {Cornachione}, {Bedding}, {Stello}, {Huber}, {Marsden}, \&
  {Carter}}]{2019PASP..131k5003A}
{Addison}, B., {Wright}, D.~J., {Wittenmyer}, R.~A., {et~al.} 2019, \pasp, 131,
  115003, \dodoi{10.1088/1538-3873/ab03aa}

\bibitem[{{Addison} {et~al.}(2020){Addison}, {Wright}, {Nicholson}, {Cale},
  {Mocnik}, {Huber}, {Plavchan}, {Wittenmyer}, {Vanderburg}, {Chaplin},
  {Chontos}, {Clark}, {Eastman}, {Ziegler}, {Brahm}, {Carter}, {Clerte},
  {Espinoza}, {Horner}, {Bentley}, {Kane}, {Kielkopf}, {Laychock}, {Mengel},
  {Okumura}, {Stassun}, {Bedding}, {Bowler}, {Burnelis}, {Collins},
  {Crossfield}, {Davis}, {Evensberget}, {Heitzmann}, {Howell}, {Law}, {Mann},
  {Marsden}, {O'Connor}, {Shporer}, {Stevens}, {Tinney}, {Tylor}, {Wang},
  {Zhang}, {Henning}, {Kossakowski}, {Ricker}, {Sarkis}, {Vanderspek},
  {Latham}, {Seager}, {Winn}, {Jenkins}, {Mireles}, {Rowden}, {Pepper},
  {Daylan}, {Schlieder}, {Collins}, {Collins}, {Tan}, {Ball}, {Basu}, {Buzasi},
  {Campante}, {Corsaro}, {Gonz{\'a}lez-Cuesta}, {Davies}, {A. Garc{\'\i} a},
  {Guo}, {Handberg}, {Hekker}, {Hey}, {Kallinger}, {Kawaler}, {Kayhan},
  {Kuszlewicz}, {Lund}, {Lyttle}, {Mathur}, {Miglio}, {Mosser}, {Nielsen},
  {Serenelli}, {Silva Aguirre}, \& {Themessl}}]{2020arXiv200107345A}
{Addison}, B.~C., {Wright}, D.~J., {Nicholson}, B.~A., {et~al.} 2020, arXiv
  e-prints, arXiv:2001.07345.
\newblock \doarXiv{2001.07345}

\bibitem[{{Akeson} {et~al.}(2013){Akeson}, {Chen}, {Ciardi}, {Crane}, {Good},
  {Harbut}, {Jackson}, {Kane}, {Laity}, {Leifer}, {Lynn}, {McElroy}, {Papin},
  {Plavchan}, {Ram{\'\i}rez}, {Rey}, {von Braun}, {Wittman}, {Abajian}, {Ali},
  {Beichman}, {Beekley}, {Berriman}, {Berukoff}, {Bryden}, {Chan}, {Groom},
  {Lau}, {Payne}, {Regelson}, {Saucedo}, {Schmitz}, {Stauffer}, {Wyatt}, \&
  {Zhang}}]{akeson2013}
{Akeson}, R.~L., {Chen}, X., {Ciardi}, D., {et~al.} 2013, \pasp, 125, 989,
  \dodoi{10.1086/672273}

\bibitem[{{Astropy Collaboration} {et~al.}(2013){Astropy Collaboration},
  {Robitaille}, {Tollerud}, {Greenfield}, {Droettboom}, {Bray}, {Aldcroft},
  {Davis}, {Ginsburg}, {Price-Whelan}, {Kerzendorf}, {Conley}, {Crighton},
  {Barbary}, {Muna}, {Ferguson}, {Grollier}, {Parikh}, {Nair}, {Unther},
  {Deil}, {Woillez}, {Conseil}, {Kramer}, {Turner}, {Singer}, {Fox}, {Weaver},
  {Zabalza}, {Edwards}, {Azalee Bostroem}, {Burke}, {Casey}, {Crawford},
  {Dencheva}, {Ely}, {Jenness}, {Labrie}, {Lim}, {Pierfederici}, {Pontzen},
  {Ptak}, {Refsdal}, {Servillat}, \& {Streicher}}]{astr13}
{Astropy Collaboration}, {Robitaille}, T.~P., {Tollerud}, E.~J., {et~al.} 2013,
  \aap, 558, A33, \dodoi{10.1051/0004-6361/201322068}

\bibitem[{{Astropy Collaboration} {et~al.}(2018){Astropy Collaboration},
  {Price-Whelan}, {Sip{\H o}cz}, {G{\"u}nther}, {Lim}, {Crawford}, {Conseil},
  {Shupe}, {Craig}, {Dencheva}, {Ginsburg}, {VanderPlas}, {Bradley},
  {P{\'e}rez-Su{\'a}rez}, {de Val-Borro}, {Aldcroft}, {Cruz}, {Robitaille},
  {Tollerud}, {Ardelean}, {Babej}, {Bach}, {Bachetti}, {Bakanov}, {Bamford},
  {Barentsen}, {Barmby}, {Baumbach}, {Berry}, {Biscani}, {Boquien}, {Bostroem},
  {Bouma}, {Brammer}, {Bray}, {Breytenbach}, {Buddelmeijer}, {Burke},
  {Calderone}, {Cano Rodr{\'{\i}}guez}, {Cara}, {Cardoso}, {Cheedella},
  {Copin}, {Corrales}, {Crichton}, {D'Avella}, {Deil}, {Depagne}, {Dietrich},
  {Donath}, {Droettboom}, {Earl}, {Erben}, {Fabbro}, {Ferreira}, {Finethy},
  {Fox}, {Garrison}, {Gibbons}, {Goldstein}, {Gommers}, {Greco}, {Greenfield},
  {Groener}, {Grollier}, {Hagen}, {Hirst}, {Homeier}, {Horton}, {Hosseinzadeh},
  {Hu}, {Hunkeler}, {Ivezi{\'c}}, {Jain}, {Jenness}, {Kanarek}, {Kendrew},
  {Kern}, {Kerzendorf}, {Khvalko}, {King}, {Kirkby}, {Kulkarni}, {Kumar},
  {Lee}, {Lenz}, {Littlefair}, {Ma}, {Macleod}, {Mastropietro}, {McCully},
  {Montagnac}, {Morris}, {Mueller}, {Mumford}, {Muna}, {Murphy}, {Nelson},
  {Nguyen}, {Ninan}, {N{\"o}the}, {Ogaz}, {Oh}, {Parejko}, {Parley}, {Pascual},
  {Patil}, {Patil}, {Plunkett}, {Prochaska}, {Rastogi}, {Reddy Janga},
  {Sabater}, {Sakurikar}, {Seifert}, {Sherbert}, {Sherwood-Taylor}, {Shih},
  {Sick}, {Silbiger}, {Singanamalla}, {Singer}, {Sladen}, {Sooley},
  {Sornarajah}, {Streicher}, {Teuben}, {Thomas}, {Tremblay}, {Turner},
  {Terr{\'o}n}, {van Kerkwijk}, {de la Vega}, {Watkins}, {Weaver}, {Whitmore},
  {Woillez}, {Zabalza}, \& {Astropy Contributors}}]{astr18}
{Astropy Collaboration}, {Price-Whelan}, A.~M., {Sip{\H o}cz}, B.~M., {et~al.}
  2018, \aj, 156, 123, \dodoi{10.3847/1538-3881/aabc4f}

\bibitem[{{Bakos} {et~al.}(2010){Bakos}, {Torres}, {P{\'a}l}, {Hartman},
  {Kov{\'a}cs}, {Noyes}, {Latham}, {Sasselov}, {Sip{\H{o}}cz}, {Esquerdo},
  {Fischer}, {Johnson}, {Marcy}, {Butler}, {Isaacson}, {Howard}, {Vogt},
  {Kov{\'a}cs}, {Fernandez}, {Mo{\'o}r}, {Stefanik}, {L{\'a}z{\'a}r}, {Papp},
  \& {S{\'a}ri}}]{2010ApJ...710.1724B}
{Bakos}, G.~{\'A}., {Torres}, G., {P{\'a}l}, A., {et~al.} 2010, \apj, 710,
  1724, \dodoi{10.1088/0004-637X/710/2/1724}

\bibitem[{{Bakos} {et~al.}(2013){Bakos}, {Csubry}, {Penev}, {Bayliss},
  {Jord{\'a}n}, {Afonso}, {Hartman}, {Henning}, {Kov{\'a}cs}, {Noyes},
  {B{\'e}ky}, {Suc}, {Cs{\'a}k}, {Rabus}, {L{\'a}z{\'a}r}, {Papp}, {S{\'a}ri},
  {Conroy}, {Zhou}, {Sackett}, {Schmidt}, {Mancini}, {Sasselov}, \&
  {Ueltzhoeffer}}]{Bakos:2013}
{Bakos}, G.~{\'A}., {Csubry}, Z., {Penev}, K., {et~al.} 2013, \pasp, 125, 154,
  \dodoi{10.1086/669529}

\bibitem[{{Barnes} {et~al.}(2012){Barnes}, {Gibson}, {Nield}, \&
  {Cochrane}}]{2012SPIE.8446E..88B}
{Barnes}, S.~I., {Gibson}, S., {Nield}, K., \& {Cochrane}, D. 2012, in
  \procspie, Vol. 8446, 844688, \dodoi{10.1117/12.926527}

\bibitem[{{Bonomo} {et~al.}(2017){Bonomo}, {Desidera}, {Benatti}, {Borsa},
  {Crespi}, {Damasso}, {Lanza}, {Sozzetti}, {Lodato}, {Marzari}, {Boccato},
  {Claudi}, {Cosentino}, {Covino}, {Gratton}, {Maggio}, {Micela}, {Molinari},
  {Pagano}, {Piotto}, {Poretti}, {Smareglia}, {Affer}, {Biazzo}, {Bignamini},
  {Esposito}, {Giacobbe}, {H{\'e}brard}, {Malavolta}, {Maldonado}, {Mancini},
  {Martinez Fiorenzano}, {Masiero}, {Nascimbeni}, {Pedani}, {Rainer}, \& {Scand
  ariato}}]{2017A&A...602A.107B}
{Bonomo}, A.~S., {Desidera}, S., {Benatti}, S., {et~al.} 2017, \aap, 602, A107,
  \dodoi{10.1051/0004-6361/201629882}

\bibitem[{{Borucki} {et~al.}(2010){Borucki}, {Koch}, {Basri}, {Batalha},
  {Brown}, {Caldwell}, {Caldwell}, {Christensen-Dalsgaard}, {Cochran},
  {DeVore}, {Dunham}, {Dupree}, {Gautier}, {Geary}, {Gilliland}, {Gould},
  {Howell}, {Jenkins}, {Kondo}, {Latham}, {Marcy}, {Meibom}, {Kjeldsen},
  {Lissauer}, {Monet}, {Morrison}, {Sasselov}, {Tarter}, {Boss}, {Brownlee},
  {Owen}, {Buzasi}, {Charbonneau}, {Doyle}, {Fortney}, {Ford}, {Holman},
  {Seager}, {Steffen}, {Welsh}, {Rowe}, {Anderson}, {Buchhave}, {Ciardi},
  {Walkowicz}, {Sherry}, {Horch}, {Isaacson}, {Everett}, {Fischer}, {Torres},
  {Johnson}, {Endl}, {MacQueen}, {Bryson}, {Dotson}, {Haas}, {Kolodziejczak},
  {Van Cleve}, {Chandrasekaran}, {Twicken}, {Quintana}, {Clarke}, {Allen},
  {Li}, {Wu}, {Tenenbaum}, {Verner}, {Bruhweiler}, {Barnes}, \&
  {Prsa}}]{2010Sci...327..977B}
{Borucki}, W.~J., {Koch}, D., {Basri}, G., {et~al.} 2010, Science, 327, 977,
  \dodoi{10.1126/science.1185402}

\bibitem[{{Brahm} {et~al.}(2017){Brahm}, {Jord{\'a}n}, \&
  {Espinoza}}]{2017PASP..129c4002B}
{Brahm}, R., {Jord{\'a}n}, A., \& {Espinoza}, N. 2017, \pasp, 129, 034002,
  \dodoi{10.1088/1538-3873/aa5455}

\bibitem[{{Brahm} {et~al.}(2018){Brahm}, {Hartman}, {Jord{\'a}n}, {Bakos},
  {Espinoza}, {Rabus}, {Bhatti}, {Penev}, {Sarkis}, {Suc}, {Csubry}, {Bayliss},
  {Bento}, {Zhou}, {Mancini}, {Henning}, {Ciceri}, {de Val-Borro}, {Shectman},
  {Crane}, {Arriagada}, {Butler}, {Teske}, {Thompson}, {Osip}, {D{\'\i}az},
  {Schmidt}, {L{\'a}z{\'a}r}, {Papp}, \& {S{\'a}ri}}]{2018AJ....155..112B}
{Brahm}, R., {Hartman}, J.~D., {Jord{\'a}n}, A., {et~al.} 2018, \aj, 155, 112,
  \dodoi{10.3847/1538-3881/aaa898}

\bibitem[{{Brasseur} {et~al.}(2019){Brasseur}, {Phillip}, {Fleming},
  {Mullally}, \& {White}}]{tesscut}
{Brasseur}, C.~E., {Phillip}, C., {Fleming}, S.~W., {Mullally}, S.~E., \&
  {White}, R.~L. 2019, {Astrocut: Tools for creating cutouts of TESS images}.
\newblock \doeprint{1905.007}

\bibitem[{{Buchhave} {et~al.}(2012){Buchhave}, {Latham}, {Johansen},
  {Bizzarro}, {Torres}, {Rowe}, {Batalha}, {Borucki}, {Brugamyer}, {Caldwell},
  {Bryson}, {Ciardi}, {Cochran}, {Endl}, {Esquerdo}, {Ford}, {Geary},
  {Gilliland}, {Hansen}, {Isaacson}, {Laird}, {Lucas}, {Marcy}, {Morse},
  {Robertson}, {Shporer}, {Stefanik}, {Still}, \&
  {Quinn}}]{2012Natur.486..375B}
{Buchhave}, L.~A., {Latham}, D.~W., {Johansen}, A., {et~al.} 2012, \nat, 486,
  375, \dodoi{10.1038/nature11121}

\bibitem[{{Burrows} {et~al.}(2007){Burrows}, {Hubeny}, {Budaj}, \&
  {Hubbard}}]{2007ApJ...661..502B}
{Burrows}, A., {Hubeny}, I., {Budaj}, J., \& {Hubbard}, W.~B. 2007, \apj, 661,
  502, \dodoi{10.1086/514326}

\bibitem[{{Butler} {et~al.}(1996){Butler}, {Marcy}, {Williams}, {McCarthy},
  {Dosanjh}, \& {Vogt}}]{But96}
{Butler}, R.~P., {Marcy}, G.~W., {Williams}, E., {et~al.} 1996, \pasp, 108,
  500, \dodoi{10.1086/133755}

\bibitem[{{Castelli} \& {Kurucz}(2004)}]{Castelli:2004}
{Castelli}, F., \& {Kurucz}, R.~L. 2004, ArXiv Astrophysics e-prints.
\newblock \doarXiv{astro-ph/0405087}

\bibitem[{{Chen} \& {Kipping}(2017)}]{2017ApJ...834...17C}
{Chen}, J., \& {Kipping}, D. 2017, \apj, 834, 17,
  \dodoi{10.3847/1538-4357/834/1/17}

\bibitem[{{Choi} {et~al.}(2016){Choi}, {Dotter}, {Conroy}, {Cantiello},
  {Paxton}, \& {Johnson}}]{2016ApJ...823..102C}
{Choi}, J., {Dotter}, A., {Conroy}, C., {et~al.} 2016, \apj, 823, 102,
  \dodoi{10.3847/0004-637X/823/2/102}

\bibitem[{{Claret}(2017)}]{2017AnA...600A..30C}
{Claret}, A. 2017, \aap, 600, A30, \dodoi{10.1051/0004-6361/201629705}

\bibitem[{{Claret} {et~al.}(2013){Claret}, {Hauschildt}, \&
  {Witte}}]{2013AnA...552A..16C}
{Claret}, A., {Hauschildt}, P.~H., \& {Witte}, S. 2013, \aap, 552, A16,
  \dodoi{10.1051/0004-6361/201220942}

\bibitem[{{Collins} {et~al.}(2017){Collins}, {Kielkopf}, {Stassun}, \&
  {Hessman}}]{Collins:2017}
{Collins}, K.~A., {Kielkopf}, J.~F., {Stassun}, K.~G., \& {Hessman}, F.~V.
  2017, \aj, 153, 77, \dodoi{10.3847/1538-3881/153/2/77}

\bibitem[{{Collins} {et~al.}(2014){Collins}, {Eastman}, {Beatty}, {Siverd},
  {Gaudi}, {Pepper}, {Kielkopf}, {Johnson}, {Howard}, {Fischer}, {Manner},
  {Bieryla}, {Latham}, {Fulton}, {Gregorio}, {Buchhave}, {Jensen}, {Stassun},
  {Penev}, {Crepp}, {Hinkley}, {Street}, {Cargile}, {Mack}, {Oberst}, {Avril},
  {Mellon}, {McLeod}, {Penny}, {Stefanik}, {Berlind}, {Calkins}, {Mao},
  {Richert}, {DePoy}, {Esquerdo}, {Gould}, {Marshall}, {Oelkers}, {Pogge},
  {Trueblood}, \& {Trueblood}}]{2014AJ....147...39C}
{Collins}, K.~A., {Eastman}, J.~D., {Beatty}, T.~G., {et~al.} 2014, \aj, 147,
  39, \dodoi{10.1088/0004-6256/147/2/39}

\bibitem[{{Crane} {et~al.}(2006){Crane}, {Shectman}, \& {Butler}}]{Cra06}
{Crane}, J.~D., {Shectman}, S.~A., \& {Butler}, R.~P. 2006, in Proceedings of
  the SPIE, Vol. 6269, \dodoi{10.1117/12.672339}

\bibitem[{{Crane} {et~al.}(2010){Crane}, {Shectman}, {Butler}, {Thompson},
  {Birk}, {Jones}, \& {Burley}}]{Cra10}
{Crane}, J.~D., {Shectman}, S.~A., {Butler}, R.~P., {et~al.} 2010, in
  Proceedings of the SPIE, Vol. 7735, \dodoi{10.1117/12.857792}

\bibitem[{{Crane} {et~al.}(2008){Crane}, {Shectman}, {Butler}, {Thompson}, \&
  {Burley}}]{Cra08}
{Crane}, J.~D., {Shectman}, S.~A., {Butler}, R.~P., {Thompson}, I.~B., \&
  {Burley}, G.~S. 2008, in Proceedings of the SPIE, Vol. 7014,
  \dodoi{10.1117/12.789637}

\bibitem[{{Demory} \& {Seager}(2011)}]{2011ApJS..197...12D}
{Demory}, B.-O., \& {Seager}, S. 2011, \apjs, 197, 12,
  \dodoi{10.1088/0067-0049/197/1/12}

\bibitem[{{Dobbs-Dixon} {et~al.}(2004){Dobbs-Dixon}, {Lin}, \&
  {Mardling}}]{dobbsdixon_spin_2004}
{Dobbs-Dixon}, I., {Lin}, D.~N.~C., \& {Mardling}, R.~A. 2004, \apj, 610, 464,
  \dodoi{10.1086/421510}

\bibitem[{{Dong} {et~al.}(2018){Dong}, {Xie}, {Zhou}, {Zheng}, \&
  {Luo}}]{2018PNAS..115..266D}
{Dong}, S., {Xie}, J.-W., {Zhou}, J.-L., {Zheng}, Z., \& {Luo}, A. 2018,
  Proceedings of the National Academy of Science, 115, 266,
  \dodoi{10.1073/pnas.1711406115}

\bibitem[{{Dotter}(2016)}]{2016ApJS..222....8D}
{Dotter}, A. 2016, \apjs, 222, 8, \dodoi{10.3847/0067-0049/222/1/8}

\bibitem[{{Eastman} {et~al.}(2013){Eastman}, {Gaudi}, \& {Agol}}]{Eastman:2013}
{Eastman}, J., {Gaudi}, B.~S., \& {Agol}, E. 2013, \pasp, 125, 83,
  \dodoi{10.1086/669497}

\bibitem[{{Eastman} {et~al.}(2010){Eastman}, {Siverd}, \&
  {Gaudi}}]{2010PASP..122..935E}
{Eastman}, J., {Siverd}, R., \& {Gaudi}, B.~S. 2010, \pasp, 122, 935,
  \dodoi{10.1086/655938}

\bibitem[{{Eastman} {et~al.}(2019){Eastman}, {Rodriguez}, {Agol}, {Stassun},
  {Beatty}, {Vanderburg}, {Gaudi}, {Collins}, \& {Luger}}]{Eastman:2019}
{Eastman}, J.~D., {Rodriguez}, J.~E., {Agol}, E., {et~al.} 2019, arXiv
  e-prints.
\newblock \doarXiv{1907.09480}

\bibitem[{{Enoch} {et~al.}(2012){Enoch}, {Collier Cameron}, \&
  {Horne}}]{2012A&A...540A..99E}
{Enoch}, B., {Collier Cameron}, A., \& {Horne}, K. 2012, \aap, 540, A99,
  \dodoi{10.1051/0004-6361/201117317}

\bibitem[{Fausnaugh {et~al.}(2019{\natexlab{a}})Fausnaugh, Burke, Caldwell,
  Jenkins, Smith, Twicken, Vanderspek, Doty, Li, Ting, \&
  Villasenor}]{tess_dr05_s04}
Fausnaugh, M.~M., Burke, C.~J., Caldwell, D.~A., {et~al.} 2019{\natexlab{a}}.
\newblock
  \url{https://archive.stsci.edu/missions/tess/doc/tess_drn/tess_sector_04_drn05_v04.pdf}

\bibitem[{Fausnaugh {et~al.}(2019{\natexlab{b}})Fausnaugh, Burke, Caldwell,
  Jenkins, Smith, Twicken, Vanderspek, Doty, Ting, \&
  Villasenor}]{tess_dr07_s05}
---. 2019{\natexlab{b}}.
\newblock
  \url{https://archive.stsci.edu/missions/tess/doc/tess_drn/tess_sector_05_drn07_v02.pdf}

\bibitem[{{Foreman-Mackey} {et~al.}(2013){Foreman-Mackey}, {Hogg}, {Lang}, \&
  {Goodman}}]{emcee_paper}
{Foreman-Mackey}, D., {Hogg}, D.~W., {Lang}, D., \& {Goodman}, J. 2013, \pasp,
  125, 306, \dodoi{10.1086/670067}

\bibitem[{Foreman-Mackey {et~al.}(2019)Foreman-Mackey, Farr, Sinha, Archibald,
  Hogg, Sanders, Zuntz, Williams, Nelson, de~Val-Borro, Erhardt, Pashchenko, \&
  Pla}]{emcee_zenodo}
Foreman-Mackey, D., Farr, W.~M., Sinha, M., {et~al.} 2019, {emcee v3: A Python
  ensemble sampling toolkit for affine-invariant MCMC}, v3.0.2,  Zenodo,
  \dodoi{10.5281/zenodo.3543502}

\bibitem[{{Fortney} {et~al.}(2007){Fortney}, {Marley}, \&
  {Barnes}}]{2007ApJ...659.1661F}
{Fortney}, J.~J., {Marley}, M.~S., \& {Barnes}, J.~W. 2007, \apj, 659, 1661,
  \dodoi{10.1086/512120}

\bibitem[{{Frandsen} \& {Lindberg}(1999)}]{Frandsen1999}
{Frandsen}, S., \& {Lindberg}, B. 1999, in Astrophysics with the NOT, ed.
  H.~{Karttunen} \& V.~{Piirola}, 71

\bibitem[{Fulton {et~al.}(2019)Fulton, Blunt, Hurt, Mills, cpsdoppler,
  sinukoff, Hadden, Bouma, zhexingli, iancrossfield, dos Santos, Benneke,
  Weiss, Yee, Rosenthal, Petigura, \& Foreman-Mackey}]{radvel_zenodo}
Fulton, B., Blunt, S., Hurt, S., {et~al.} 2019,
  California-Planet-Search/radvel, v1.3.2,  Zenodo,
  \dodoi{10.5281/zenodo.3586003}

\bibitem[{{Fulton} {et~al.}(2018){Fulton}, {Petigura}, {Blunt}, \&
  {Sinukoff}}]{radvel_paper}
{Fulton}, B.~J., {Petigura}, E.~A., {Blunt}, S., \& {Sinukoff}, E. 2018, \pasp,
  130, 044504, \dodoi{10.1088/1538-3873/aaaaa8}

\bibitem[{{Gaia Collaboration} {et~al.}(2018){Gaia Collaboration}, {Brown},
  {Vallenari}, {Prusti}, {de Bruijne}, {Babusiaux}, {Bailer-Jones}, {Biermann},
  {Evans}, {Eyer}, {Jansen}, {Jordi}, {Klioner}, {Lammers}, {Lindegren},
  {Luri}, {Mignard}, {Panem}, {Pourbaix}, {Randich}, {Sartoretti}, {Siddiqui},
  {Soubiran}, {van Leeuwen}, {Walton}, {Arenou}, {Bastian}, {Cropper},
  {Drimmel}, {Katz}, {Lattanzi}, {Bakker}, {Cacciari}, {Casta{\~n}eda},
  {Chaoul}, {Cheek}, {De Angeli}, {Fabricius}, {Guerra}, {Holl}, {Masana},
  {Messineo}, {Mowlavi}, {Nienartowicz}, {Panuzzo}, {Portell}, {Riello},
  {Seabroke}, {Tanga}, {Th{\'e}venin}, {Gracia-Abril}, {Comoretto},
  {Garcia-Reinaldos}, {Teyssier}, {Altmann}, {Andrae}, {Audard},
  {Bellas-Velidis}, {Benson}, {Berthier}, {Blomme}, {Burgess}, {Busso},
  {Carry}, {Cellino}, {Clementini}, {Clotet}, {Creevey}, {Davidson}, {De
  Ridder}, {Delchambre}, {Dell'Oro}, {Ducourant},
  {Fern{\'a}ndez-Hern{\'a}ndez}, {Fouesneau}, {Fr{\'e}mat}, {Galluccio},
  {Garc{\'\i}a-Torres}, {Gonz{\'a}lez-N{\'u}{\~n}ez}, {Gonz{\'a}lez-Vidal},
  {Gosset}, {Guy}, {Halbwachs}, {Hambly}, {Harrison}, {Hern{\'a}ndez},
  {Hestroffer}, {Hodgkin}, {Hutton}, {Jasniewicz}, {Jean-Antoine-Piccolo},
  {Jordan}, {Korn}, {Krone-Martins}, {Lanzafame}, {Lebzelter}, {L{\"o}ffler},
  {Manteiga}, {Marrese}, {Mart{\'\i}n-Fleitas}, {Moitinho}, {Mora}, {Muinonen},
  {Osinde}, {Pancino}, {Pauwels}, {Petit}, {Recio-Blanco}, {Richards},
  {Rimoldini}, {Robin}, {Sarro}, {Siopis}, {Smith}, {Sozzetti}, {S{\"u}veges},
  {Torra}, {van Reeven}, {Abbas}, {Abreu Aramburu}, {Accart}, {Aerts},
  {Altavilla}, {{\'A}lvarez}, {Alvarez}, {Alves}, {Anderson}, {Andrei},
  {Anglada Varela}, {Antiche}, {Antoja}, {Arcay}, {Astraatmadja}, {Bach},
  {Baker}, {Balaguer-N{\'u}{\~n}ez}, {Balm}, {Barache}, {Barata}, {Barbato},
  {Barblan}, {Barklem}, {Barrado}, {Barros}, {Barstow}, {Bartholom{\'e}
  Mu{\~n}oz}, {Bassilana}, {Becciani}, {Bellazzini}, {Berihuete}, {Bertone},
  {Bianchi}, {Bienaym{\'e}}, {Blanco-Cuaresma}, {Boch}, {Boeche}, {Bombrun},
  {Borrachero}, {Bossini}, {Bouquillon}, {Bourda}, {Bragaglia}, {Bramante},
  {Breddels}, {Bressan}, {Brouillet}, {Br{\"u}semeister}, {Brugaletta},
  {Bucciarelli}, {Burlacu}, {Busonero}, {Butkevich}, {Buzzi}, {Caffau},
  {Cancelliere}, {Cannizzaro}, {Cantat-Gaudin}, {Carballo}, {Carlucci},
  {Carrasco}, {Casamiquela}, {Castellani}, {Castro-Ginard}, {Charlot},
  {Chemin}, {Chiavassa}, {Cocozza}, {Costigan}, {Cowell}, {Crifo}, {Crosta},
  {Crowley}, {Cuypers}, {Dafonte}, {Damerdji}, {Dapergolas}, {David}, {David},
  {de Laverny}, {De Luise}, {De March}, {de Martino}, {de Souza}, {de Torres},
  {Debosscher}, {del Pozo}, {Delbo}, {Delgado}, {Delgado}, {Di Matteo},
  {Diakite}, {Diener}, {Distefano}, {Dolding}, {Drazinos}, {Dur{\'a}n},
  {Edvardsson}, {Enke}, {Eriksson}, {Esquej}, {Eynard Bontemps}, {Fabre},
  {Fabrizio}, {Faigler}, {Falc{\~a}o}, {Farr{\`a}s Casas}, {Federici},
  {Fedorets}, {Fernique}, {Figueras}, {Filippi}, {Findeisen}, {Fonti},
  {Fraile}, {Fraser}, {Fr{\'e}zouls}, {Gai}, {Galleti}, {Garabato},
  {Garc{\'\i}a-Sedano}, {Garofalo}, {Garralda}, {Gavel}, {Gavras}, {Gerssen},
  {Geyer}, {Giacobbe}, {Gilmore}, {Girona}, {Giuffrida}, {Glass}, {Gomes},
  {Granvik}, {Gueguen}, {Guerrier}, {Guiraud}, {Guti{\'e}rrez-S{\'a}nchez},
  {Haigron}, {Hatzidimitriou}, {Hauser}, {Haywood}, {Heiter}, {Helmi}, {Heu},
  {Hilger}, {Hobbs}, {Hofmann}, {Holland}, {Huckle}, {Hypki}, {Icardi},
  {Jan{\ss}en}, {Jevardat de Fombelle}, {Jonker}, {Juh{\'a}sz}, {Julbe},
  {Karampelas}, {Kewley}, {Klar}, {Kochoska}, {Kohley}, {Kolenberg},
  {Kontizas}, {Kontizas}, {Koposov}, {Kordopatis}, {Kostrzewa-Rutkowska},
  {Koubsky}, {Lambert}, {Lanza}, {Lasne}, {Lavigne}, {Le Fustec}, {Le
  Poncin-Lafitte}, {Lebreton}, {Leccia}, {Leclerc}, {Lecoeur-Taibi},
  {Lenhardt}, {Leroux}, {Liao}, {Licata}, {Lindstr{\o}m}, {Lister}, {Livanou},
  {Lobel}, {L{\'o}pez}, {Managau}, {Mann}, {Mantelet}, {Marchal}, {Marchant},
  {Marconi}, {Marinoni}, {Marschalk{\'o}}, {Marshall}, {Martino}, {Marton},
  {Mary}, {Massari}, {Matijevi{\v{c}}}, {Mazeh}, {McMillan}, {Messina},
  {Michalik}, {Millar}, {Molina}, {Molinaro}, {Moln{\'a}r}, {Montegriffo},
  {Mor}, {Morbidelli}, {Morel}, {Morris}, {Mulone}, {Muraveva}, {Musella},
  {Nelemans}, {Nicastro}, {Noval}, {O'Mullane}, {Ord{\'e}novic},
  {Ord{\'o}{\~n}ez-Blanco}, {Osborne}, {Pagani}, {Pagano}, {Pailler},
  {Palacin}, {Palaversa}, {Panahi}, {Pawlak}, {Piersimoni}, {Pineau}, {Plachy},
  {Plum}, {Poggio}, {Poujoulet}, {Pr{\v{s}}a}, {Pulone}, {Racero}, {Ragaini},
  {Rambaux}, {Ramos-Lerate}, {Regibo}, {Reyl{\'e}}, {Riclet}, {Ripepi}, {Riva},
  {Rivard}, {Rixon}, {Roegiers}, {Roelens}, {Romero-G{\'o}mez}, {Rowell},
  {Royer}, {Ruiz-Dern}, {Sadowski}, {Sagrist{\`a} Sell{\'e}s}, {Sahlmann},
  {Salgado}, {Salguero}, {Sanna}, {Santana-Ros}, {Sarasso}, {Savietto},
  {Schultheis}, {Sciacca}, {Segol}, {Segovia}, {S{\'e}gransan}, {Shih},
  {Siltala}, {Silva}, {Smart}, {Smith}, {Solano}, {Solitro}, {Sordo}, {Soria
  Nieto}, {Souchay}, {Spagna}, {Spoto}, {Stampa}, {Steele},
  {Steidelm{\"u}ller}, {Stephenson}, {Stoev}, {Suess}, {Surdej}, {Szabados},
  {Szegedi-Elek}, {Tapiador}, {Taris}, {Tauran}, {Taylor}, {Teixeira},
  {Terrett}, {Teyssand ier}, {Thuillot}, {Titarenko}, {Torra Clotet}, {Turon},
  {Ulla}, {Utrilla}, {Uzzi}, {Vaillant}, {Valentini}, {Valette}, {van Elteren},
  {Van Hemelryck}, {van Leeuwen}, {Vaschetto}, {Vecchiato}, {Veljanoski},
  {Viala}, {Vicente}, {Vogt}, {von Essen}, {Voss}, {Votruba}, {Voutsinas},
  {Walmsley}, {Weiler}, {Wertz}, {Wevers}, {Wyrzykowski}, {Yoldas},
  {{\v{Z}}erjal}, {Ziaeepour}, {Zorec}, {Zschocke}, {Zucker}, {Zurbach}, \&
  {Zwitter}}]{gaia_dr2}
{Gaia Collaboration}, {Brown}, A.~G.~A., {Vallenari}, A., {et~al.} 2018, \aap,
  616, A1, \dodoi{10.1051/0004-6361/201833051}

\bibitem[{{Gandolfi} {et~al.}(2017){Gandolfi}, {Barrag{\'a}n}, {Hatzes},
  {Fridlund}, {Fossati}, {Donati}, {Johnson}, {Nowak}, {Prieto-Arranz},
  {Albrecht}, {Dai}, {Deeg}, {Endl}, {Grziwa}, {Hjorth}, {Korth}, {Nespral},
  {Saario}, {Smith}, {Antoniciello}, {Alarcon}, {Bedell}, {Blay}, {Brems},
  {Cabrera}, {Csizmadia}, {Cusano}, {Cochran}, {Eigm{\"u}ller}, {Erikson},
  {Gonz{\'a}lez Hern{\'a}ndez}, {Guenther}, {Hirano}, {Su{\'a}rez
  Mascare{\~n}o}, {Narita}, {Palle}, {Parviainen}, {P{\"a}tzold}, {Persson},
  {Rauer}, {Saviane}, {Schmidtobreick}, {Van Eylen}, {Winn}, \&
  {Zakhozhay}}]{Gandolfi2017}
{Gandolfi}, D., {Barrag{\'a}n}, O., {Hatzes}, A.~P., {et~al.} 2017, \aj, 154,
  123, \dodoi{10.3847/1538-3881/aa832a}

\bibitem[{{Goldreich} \& {Soter}(1966)}]{1966Icar....5..375G}
{Goldreich}, P., \& {Soter}, S. 1966, \icarus, 5, 375,
  \dodoi{10.1016/0019-1035(66)90051-0}

\bibitem[{Goodman \& Weare(2010)}]{goodman2010}
Goodman, J., \& Weare, J. 2010, Communications in Applied Mathematics and
  Computational Science, 5, 65, \dodoi{10.2140/camcos.2010.5.65}

\bibitem[{Guerrero {et~al.}(2020)Guerrero, Seager, Huang, Vanderburg, Soto, \&
  Mireles}]{tess_toi_paper}
Guerrero, N., Seager, S., Huang, C.~X., {et~al.} 2020, The {TESS} Objects of
  Interest Catalog from the {TESS} Prime Mission

\bibitem[{Harris {et~al.}(2020)Harris, Millman, van~der Walt, Gommers,
  Virtanen, Cournapeau, Wieser, Taylor, Berg, Smith, Kern, Picus, Hoyer, van
  Kerkwijk, Brett, Haldane, del R{'{\i}}o, Wiebe, Peterson,
  G{'{e}}rard-Marchant, Sheppard, Reddy, Weckesser, Abbasi, Gohlke, \&
  Oliphant}]{harris2020array}
Harris, C.~R., Millman, K.~J., van~der Walt, S.~J., {et~al.} 2020, Nature, 585,
  357, \dodoi{10.1038/s41586-020-2649-2}

\bibitem[{Hartman \& Bakos(2016)}]{hartman_vartools_2016}
Hartman, J.~D., \& Bakos, G.~A. 2016, A\&C, 17, 1,
  \dodoi{10.1016/j.ascom.2016.05.006}

\bibitem[{{Hartman} {et~al.}(2016){Hartman}, {Bakos}, {Bhatti}, {Penev},
  {Bieryla}, {Latham}, {Kov{\'a}cs}, {Torres}, {Csubry}, {de Val-Borro},
  {Buchhave}, {Kov{\'a}cs}, {Quinn}, {Howard}, {Isaacson}, {Fulton}, {Everett},
  {Esquerdo}, {B{\'e}ky}, {Szklenar}, {Falco}, {Santerne}, {Boisse},
  {H{\'e}brard}, {Burrows}, {L{\'a}z{\'a}r}, {Papp}, \&
  {S{\'a}ri}}]{2016AJ....152..182H}
{Hartman}, J.~D., {Bakos}, G.~{\'A}., {Bhatti}, W., {et~al.} 2016, \aj, 152,
  182, \dodoi{10.3847/0004-6256/152/6/182}

\bibitem[{{Hatzes} \& {Rauer}(2015)}]{2015ApJ...810L..25H}
{Hatzes}, A.~P., \& {Rauer}, H. 2015, \apjl, 810, L25,
  \dodoi{10.1088/2041-8205/810/2/L25}

\bibitem[{{Huang} {et~al.}(2020{\natexlab{a}}){Huang}, {Vanderburg}, {P{\'a}l},
  {Sha}, {Yu}, {Fong}, {Fausnaugh}, {Shporer}, {Guerrero}, {Vanderspek}, \&
  {Ricker}}]{qlp_rnaas_1}
{Huang}, C.~X., {Vanderburg}, A., {P{\'a}l}, A., {et~al.} 2020{\natexlab{a}},
  RNAAS, 4, 204, \dodoi{10.3847/2515-5172/abca2e}

\bibitem[{{Huang} {et~al.}(2020{\natexlab{b}}){Huang}, {Vanderburg}, {P{\'a}l},
  {Sha}, {Yu}, {Fong}, {Fausnaugh}, {Shporer}, {Guerrero}, {Vanderspek}, \&
  {Ricker}}]{qlp_rnaas_2}
---. 2020{\natexlab{b}}, RNAAS, 4, 206, \dodoi{10.3847/2515-5172/abca2d}

\bibitem[{{Huber} {et~al.}(2016){Huber}, {Bryson}, {Haas}, {Barclay},
  {Barentsen}, {Howell}, {Sharma}, {Stello}, \& {Thompson}}]{epic}
{Huber}, D., {Bryson}, S.~T., {Haas}, M.~R., {et~al.} 2016, \apjs, 224, 2,
  \dodoi{10.3847/0067-0049/224/1/2}

\bibitem[{{Huber} {et~al.}(2019){Huber}, {Chaplin}, {Chontos}, {Kjeldsen},
  {Christensen-Dalsgaard}, {Bedding}, {Ball}, {Brahm}, {Espinoza}, {Henning},
  {Jord{\'a}n}, {Sarkis}, {Knudstrup}, {Albrecht}, {Grundahl}, {Fredslund
  Andersen}, {Pall{\'e}}, {Crossfield}, {Fulton}, {Howard}, {Isaacson},
  {Weiss}, {Handberg}, {Lund}, {Serenelli}, {R{\o}rsted Mosumgaard},
  {Stokholm}, {Bieryla}, {Buchhave}, {Latham}, {Quinn}, {Gaidos}, {Hirano},
  {Ricker}, {Vanderspek}, {Seager}, {Jenkins}, {Winn}, {Antia}, {Appourchaux},
  {Basu}, {Bell}, {Benomar}, {Bonanno}, {Buzasi}, {Campante}, {{\c{C}}elik
  Orhan}, {Corsaro}, {Cunha}, {Davies}, {Deheuvels}, {Grunblatt}, {Hasanzadeh},
  {Di Mauro}, {Garc{\'\i}a}, {Gaulme}, {Girardi}, {Guzik}, {Hon}, {Jiang},
  {Kallinger}, {Kawaler}, {Kuszlewicz}, {Lebreton}, {Li}, {Lucas}, {Lundkvist},
  {Mann}, {Mathis}, {Mathur}, {Mazumdar}, {Metcalfe}, {Miglio}, {Monteiro},
  {Mosser}, {Noll}, {Nsamba}, {Ong}, {{\"O}rtel}, {Pereira}, {Ranadive},
  {R{\'e}gulo}, {Rodrigues}, {Roxburgh}, {Silva Aguirre}, {Smalley},
  {Schofield}, {Sousa}, {Stassun}, {Stello}, {Tayar}, {White}, {Verma},
  {Vrard}, {Y{\i}ld{\i}z}, {Baker}, {Bazot}, {Beichmann}, {Bergmann}, {Bugnet},
  {Cale}, {Carlino}, {Cartwright}, {Christiansen}, {Ciardi}, {Creevey},
  {Dittmann}, {Do Nascimento}, {Van Eylen}, {F{\"u}r{\'e}sz}, {Gagn{\'e}},
  {Gao}, {Gazeas}, {Giddens}, {Hall}, {Hekker}, {Ireland }, {Latouf}, {LeBrun},
  {Levine}, {Matzko}, {Natinsky}, {Page}, {Plavchan}, {Mansouri-Samani},
  {McCauliff}, {Mullally}, {Orenstein}, {Garcia Soto}, {Paegert}, {van Saders},
  {Schnaible}, {Soderblom}, {Szab{\'o}}, {Tanner}, {Tinney}, {Teske}, {Thomas},
  {Trampedach}, {Wright}, {Yuan}, \& {Zohrabi}}]{2019AJ....157..245H}
{Huber}, D., {Chaplin}, W.~J., {Chontos}, A., {et~al.} 2019, \aj, 157, 245,
  \dodoi{10.3847/1538-3881/ab1488}

\bibitem[{{Hunter}(2007)}]{matplotlibpaper}
{Hunter}, J.~D. 2007, Computing in Science Engineering, 9, 90,
  \dodoi{10.1109/MCSE.2007.55}

\bibitem[{{Jenkins} {et~al.}(2016){Jenkins}, {Twicken}, {McCauliff},
  {Campbell}, {Sanderfer}, {Lung}, {Mansouri-Samani}, {Girouard}, {Tenenbaum},
  {Klaus}, {Smith}, {Caldwell}, {Chacon}, {Henze}, {Heiges}, {Latham},
  {Morgan}, {Swade}, {Rinehart}, \& {Vanderspek}}]{Jenkins:2016}
{Jenkins}, J.~M., {Twicken}, J.~D., {McCauliff}, S., {et~al.} 2016, in
  \procspie, Vol. 9913, Software and Cyberinfrastructure for Astronomy IV,
  99133E, \dodoi{10.1117/12.2233418}

\bibitem[{{Jensen}(2013)}]{Jensen_2013}
{Jensen}, E. 2013, {Tapir: A web interface for transit/eclipse observability}.
\newblock \doeprint{1306.007}

\bibitem[{{K2 Science Office}(2020)}]{k2_dr_c19}
{K2 Science Office}. 2020.
\newblock
  \url{https://keplerscience.arc.nasa.gov/k2-data-release-notes.html#k2-campaign-19}

\bibitem[{{Kaufer} {et~al.}(1999){Kaufer}, {Stahl}, {Tubbesing},
  {N{\o}rregaard}, {Avila}, {Francois}, {Pasquini}, \&
  {Pizzella}}]{1999Msngr..95....8K}
{Kaufer}, A., {Stahl}, O., {Tubbesing}, S., {et~al.} 1999, The Messenger, 95,
  8.
\newblock \url{http://articles.adsabs.harvard.edu/full/1999Msngr..95....8K}

\bibitem[{{Kipping}(2013)}]{2013MNRAS.435.2152K}
{Kipping}, D.~M. 2013, \mnras, 435, 2152, \dodoi{10.1093/mnras/stt1435}

\bibitem[{{Kov{\'a}cs} {et~al.}(2005){Kov{\'a}cs}, {Bakos}, \&
  {Noyes}}]{Kovacs:2005}
{Kov{\'a}cs}, G., {Bakos}, G., \& {Noyes}, R.~W. 2005, \mnras, 356, 557,
  \dodoi{10.1111/j.1365-2966.2004.08479.x}

\bibitem[{{Kov{\'a}cs} {et~al.}(2002){Kov{\'a}cs}, {Zucker}, \&
  {Mazeh}}]{kovacs_bls_2002}
{Kov{\'a}cs}, G., {Zucker}, S., \& {Mazeh}, T. 2002, \aap, 391, 369,
  \dodoi{10.1051/0004-6361:20020802}

\bibitem[{{Kreidberg}(2015)}]{batman_paper}
{Kreidberg}, L. 2015, \pasp, 127, 1161, \dodoi{10.1086/683602}

\bibitem[{{Lainey} {et~al.}(2017){Lainey}, {Jacobson}, {Tajeddine}, {Cooper},
  {Murray}, {Robert}, {Tobie}, {Guillot}, {Mathis}, {Remus}, {Desmars},
  {Arlot}, {De Cuyper}, {Dehant}, {Pascu}, {Thuillot}, {Le Poncin-Lafitte}, \&
  {Zahn}}]{2017Icar..281..286L}
{Lainey}, V., {Jacobson}, R.~A., {Tajeddine}, R., {et~al.} 2017, \icarus, 281,
  286, \dodoi{10.1016/j.icarus.2016.07.014}

\bibitem[{Lenzen {et~al.}(2003)Lenzen, Hartung, Brandner, Finger, Hubin,
  Lacombe, Lagrange, Lehnert, Moorwood, \& Mouillet}]{lenzen_naos-conica_2003}
Lenzen, R., Hartung, M., Brandner, W., {et~al.} 2003, in Proceedings of the
  {SPIE}, Vol. 4841 (SPIE), 944--952, \dodoi{10.1117/12.460044}

\bibitem[{{Li} {et~al.}(2018){Li}, {Caldwell}, {Jenkins}, {Twicken}, {Wohler},
  {Chen}, {Rose}, \& {TESS Science Processing Operations Center}}]{li18}
{Li}, J., {Caldwell}, D.~A., {Jenkins}, J.~M., {et~al.} 2018, in American
  Astronomical Society Meeting Abstracts, Vol. 232, 120.03

\bibitem[{{Lucy} \& {Sweeney}(1971)}]{1971AJ.....76..544L}
{Lucy}, L.~B., \& {Sweeney}, M.~A. 1971, \aj, 76, 544, \dodoi{10.1086/111159}

\bibitem[{{Mayor} {et~al.}(2003){Mayor}, {Pepe}, {Queloz}, {Bouchy},
  {Rupprecht}, {Lo Curto}, {Avila}, {Benz}, {Bertaux}, {Bonfils}, {Dall},
  {Dekker}, {Delabre}, {Eckert}, {Fleury}, {Gilliotte}, {Gojak}, {Guzman},
  {Kohler}, {Lizon}, {Longinotti}, {Lovis}, {Megevand}, {Pasquini}, {Reyes},
  {Sivan}, {Sosnowska}, {Soto}, {Udry}, {van Kesteren}, {Weber}, \&
  {Weilenmann}}]{2003Msngr.114...20M}
{Mayor}, M., {Pepe}, F., {Queloz}, D., {et~al.} 2003, The Messenger, 114, 20.
\newblock
  \url{http://www.eso.org/sci/publications/messenger/archive/no.114-dec03/messenger-no114-20-24.pdf}

\bibitem[{{Miller} \& {Fortney}(2011)}]{2011ApJ...736L..29M}
{Miller}, N., \& {Fortney}, J.~J. 2011, \apjl, 736, L29,
  \dodoi{10.1088/2041-8205/736/2/L29}

\bibitem[{{Morton}(2015)}]{isochrones_python}
{Morton}, T.~D. 2015, {isochrones: Stellar model grid package}, v2.1.
\newblock \doeprint{1503.010}

\bibitem[{{Mulders} {et~al.}(2016){Mulders}, {Pascucci}, {Apai}, {Frasca}, \&
  {Molenda-{\.Z}akowicz}}]{2016AJ....152..187M}
{Mulders}, G.~D., {Pascucci}, I., {Apai}, D., {Frasca}, A., \&
  {Molenda-{\.Z}akowicz}, J. 2016, \aj, 152, 187,
  \dodoi{10.3847/0004-6256/152/6/187}

\bibitem[{{Pandas Development Team}(2020)}]{pandas_zenodo}
{Pandas Development Team}. 2020, pandas-dev/pandas: Pandas 1.0.3, v1.0.3,
  Zenodo, \dodoi{10.5281/zenodo.3715232}

\bibitem[{Pedregosa {et~al.}(2011)Pedregosa, Varoquaux, Gramfort, Michel,
  Thirion, Grisel, Blondel, Prettenhofer, Weiss, Dubourg, Vanderplas, Passos,
  Cournapeau, Brucher, Perrot, \& Duchesnay}]{scikit-learn}
Pedregosa, F., Varoquaux, G., Gramfort, A., {et~al.} 2011, Journal of Machine
  Learning Research, 12, 2825.
\newblock \url{https://dl.acm.org/doi/10.5555/1953048.2078195}

\bibitem[{{Pepe} {et~al.}(2002){Pepe}, {Mayor}, {Rupprecht}, {Avila},
  {Ballester}, {Beckers}, {Benz}, {Bertaux}, {Bouchy}, {Buzzoni}, {Cavadore},
  {Deiries}, {Dekker}, {Delabre}, {D'Odorico}, {Eckert}, {Fischer}, {Fleury},
  {George}, {Gilliotte}, {Gojak}, {Guzman}, {Koch}, {Kohler}, {Kotzlowski},
  {Lacroix}, {Le Merrer}, {Lizon}, {Lo Curto}, {Longinotti}, {Megevand},
  {Pasquini}, {Petitpas}, {Pichard}, {Queloz}, {Reyes}, {Richaud}, {Sivan},
  {Sosnowska}, {Soto}, {Udry}, {Ureta}, {van Kesteren}, {Weber}, {Weilenmann},
  {Wicenec}, {Wieland}, {Christensen-Dalsgaard}, {Dravins}, {Hatzes},
  {K{\"u}rster}, {Paresce}, \& {Penny}}]{Pepe2002}
{Pepe}, F., {Mayor}, M., {Rupprecht}, G., {et~al.} 2002, The Messenger, 110, 9.
\newblock \url{http://articles.adsabs.harvard.edu/full/2002Msngr.110....9P}

\bibitem[{{Petigura} {et~al.}(2017){Petigura}, {Sinukoff}, {Lopez},
  {Crossfield}, {Howard}, {Brewer}, {Fulton}, {Isaacson}, {Ciardi}, {Howell},
  {Everett}, {Horch}, {Hirsch}, {Weiss}, \& {Schlieder}}]{2017AJ....153..142P}
{Petigura}, E.~A., {Sinukoff}, E., {Lopez}, E.~D., {et~al.} 2017, \aj, 153,
  142, \dodoi{10.3847/1538-3881/aa5ea5}

\bibitem[{{Petigura} {et~al.}(2018){Petigura}, {Marcy}, {Winn}, {Weiss},
  {Fulton}, {Howard}, {Sinukoff}, {Isaacson}, {Morton}, \&
  {Johnson}}]{petigura2018}
{Petigura}, E.~A., {Marcy}, G.~W., {Winn}, J.~N., {et~al.} 2018, \aj, 155, 89,
  \dodoi{10.3847/1538-3881/aaa54c}

\bibitem[{{Pr{\v{s}}a} {et~al.}(2016){Pr{\v{s}}a}, {Harmanec}, {Torres},
  {Mamajek}, {Asplund}, {Capitaine}, {Christensen-Dalsgaard}, {Depagne},
  {Haberreiter}, {Hekker}, {Hilton}, {Kopp}, {Kostov}, {Kurtz}, {Laskar},
  {Mason}, {Milone}, {Montgomery}, {Richards}, {Schmutz}, {Schou}, \&
  {Stewart}}]{iau2015b3}
{Pr{\v{s}}a}, A., {Harmanec}, P., {Torres}, G., {et~al.} 2016, \aj, 152, 41,
  \dodoi{10.3847/0004-6256/152/2/41}

\bibitem[{{Queloz} {et~al.}(2001){Queloz}, {Mayor}, {Udry}, {Burnet},
  {Carrier}, {Eggenberger}, {Naef}, {Santos}, {Pepe}, {Rupprecht}, {Avila},
  {Baeza}, {Benz}, {Bertaux}, {Bouchy}, {Cavadore}, {Delabre}, {Eckert},
  {Fischer}, {Fleury}, {Gilliotte}, {Goyak}, {Guzman}, {Kohler}, {Lacroix},
  {Lizon}, {Megevand}, {Sivan}, {Sosnowska}, \&
  {Weilenmann}}]{2001Msngr.105....1Q}
{Queloz}, D., {Mayor}, M., {Udry}, S., {et~al.} 2001, The Messenger, 105, 1.
\newblock \url{http://articles.adsabs.harvard.edu/full/2001Msngr.105....1Q}

\bibitem[{{Ricker} {et~al.}(2015){Ricker}, {Winn}, {Vanderspek}, {Latham},
  {Bakos}, {Bean}, {Berta-Thompson}, {Brown}, {Buchhave}, {Butler}, {Butler},
  {Chaplin}, {Charbonneau}, {Christensen-Dalsgaard}, {Clampin}, {Deming},
  {Doty}, {De Lee}, {Dressing}, {Dunham}, {Endl}, {Fressin}, {Ge}, {Henning},
  {Holman}, {Howard}, {Ida}, {Jenkins}, {Jernigan}, {Johnson}, {Kaltenegger},
  {Kawai}, {Kjeldsen}, {Laughlin}, {Levine}, {Lin}, {Lissauer}, {MacQueen},
  {Marcy}, {McCullough}, {Morton}, {Narita}, {Paegert}, {Palle}, {Pepe},
  {Pepper}, {Quirrenbach}, {Rinehart}, {Sasselov}, {Sato}, {Seager},
  {Sozzetti}, {Stassun}, {Sullivan}, {Szentgyorgyi}, {Torres}, {Udry}, \&
  {Villasenor}}]{tess_mission_paper}
{Ricker}, G.~R., {Winn}, J.~N., {Vanderspek}, R., {et~al.} 2015, Journal of
  Astronomical Telescopes, Instruments, and Systems, 1, 014003,
  \dodoi{10.1117/1.JATIS.1.1.014003}

\bibitem[{Rousset {et~al.}(2003)Rousset, Lacombe, Puget, Hubin, Gendron, Fusco,
  Arsenault, Charton, Feautrier, Gigan, Kern, Lagrange, Madec, Mouillet,
  Rabaud, Rabou, Stadler, \& Zins}]{rousset_naos_2003}
Rousset, G., Lacombe, F., Puget, P., {et~al.} 2003, in \procspie, Vol. 4839,
  140--149, \dodoi{10.1117/12.459332}

\bibitem[{{Schlegel} {et~al.}(1998){Schlegel}, {Finkbeiner}, \&
  {Davis}}]{Schlegel:1998}
{Schlegel}, D.~J., {Finkbeiner}, D.~P., \& {Davis}, M. 1998, \apj, 500, 525,
  \dodoi{10.1086/305772}

\bibitem[{{Scott} {et~al.}(2018){Scott}, {Howell}, {Horch}, \&
  {Everett}}]{2018PASP..130e4502S}
{Scott}, N.~J., {Howell}, S.~B., {Horch}, E.~P., \& {Everett}, M.~E. 2018,
  \pasp, 130, 054502, \dodoi{10.1088/1538-3873/aab484}

\bibitem[{{Stassun} {et~al.}(2017){Stassun}, {Collins}, \&
  {Gaudi}}]{Stassun:2017}
{Stassun}, K.~G., {Collins}, K.~A., \& {Gaudi}, B.~S. 2017, \aj, 153, 136,
  \dodoi{10.3847/1538-3881/aa5df3}

\bibitem[{{Stassun} {et~al.}(2018){Stassun}, {Corsaro}, {Pepper}, \&
  {Gaudi}}]{Stassun:2018}
{Stassun}, K.~G., {Corsaro}, E., {Pepper}, J.~A., \& {Gaudi}, B.~S. 2018, \aj,
  155, 22, \dodoi{10.3847/1538-3881/aa998a}

\bibitem[{{Stassun} \& {Torres}(2016)}]{Stassun:2016}
{Stassun}, K.~G., \& {Torres}, G. 2016, \aj, 152, 180,
  \dodoi{10.3847/0004-6256/152/6/180}

\bibitem[{{Stassun} \& {Torres}(2018)}]{StassunTorres:2018}
---. 2018, \apj, 862, 61, \dodoi{10.3847/1538-4357/aacafc}

\bibitem[{{Stassun} {et~al.}(2019){Stassun}, {Oelkers}, {Paegert}, {Torres},
  {Pepper}, {De Lee}, {Collins}, {Latham}, {Muirhead}, {Chittidi},
  {Rojas-Ayala}, {Fleming}, {Rose}, {Tenenbaum}, {Ting}, {Kane}, {Barclay},
  {Bean}, {Brassuer}, {Charbonneau}, {Ge}, {Lissauer}, {Mann}, {McLean},
  {Mullally}, {Narita}, {Plavchan}, {Ricker}, {Sasselov}, {Seager}, {Sharma},
  {Shiao}, {Sozzetti}, {Stello}, {Vanderspek}, {Wallace}, \& {Winn}}]{tic8}
{Stassun}, K.~G., {Oelkers}, R.~J., {Paegert}, M., {et~al.} 2019, \aj, 158,
  138, \dodoi{10.3847/1538-3881/ab3467}

\bibitem[{Szentgyorgyi(2004)}]{tres_design}
Szentgyorgyi, A. 2004.
\newblock \url{http://www.sao.arizona.edu/FLWO/60/TRES/TRESdesign.pdf}

\bibitem[{Tan(n.d.)}]{pest_pipeline}
Tan, T.-G. n.d.
\newblock \url{http://pestobservatory.com/the-pest-pipeline/}

\bibitem[{{Telting} {et~al.}(2014){Telting}, {Avila}, {Buchhave}, {Frandsen},
  {Gandolfi}, {Lindberg}, {Stempels}, {Prins}, \& {NOT staff}}]{Telting2014}
{Telting}, J.~H., {Avila}, G., {Buchhave}, L., {et~al.} 2014, Astronomische
  Nachrichten, 335, 41, \dodoi{10.1002/asna.201312007}

\bibitem[{{Tokovinin}(2018)}]{2018PASP..130c5002T}
{Tokovinin}, A. 2018, \pasp, 130, 035002, \dodoi{10.1088/1538-3873/aaa7d9}

\bibitem[{{Tokovinin} {et~al.}(2013){Tokovinin}, {Fischer}, {Bonati},
  {Giguere}, {Moore}, {Schwab}, {Spronck}, \&
  {Szymkowiak}}]{2013PASP..125.1336T}
{Tokovinin}, A., {Fischer}, D.~A., {Bonati}, M., {et~al.} 2013, \pasp, 125,
  1336, \dodoi{10.1086/674012}

\bibitem[{{Twicken} {et~al.}(2018){Twicken}, {Catanzarite}, {Clarke},
  {Girouard}, {Jenkins}, {Klaus}, {Li}, {McCauliff}, {Seader}, {Tenenbaum},
  {Wohler}, {Bryson}, {Burke}, {Caldwell}, {Haas}, {Henze}, \&
  {Sanderfer}}]{twic18}
{Twicken}, J.~D., {Catanzarite}, J.~H., {Clarke}, B.~D., {et~al.} 2018, \pasp,
  130, 064502, \dodoi{10.1088/1538-3873/aab694}

\bibitem[{Vanderburg \& Johnson(2014)}]{vanderburg_technique_2014}
Vanderburg, A., \& Johnson, J.~A. 2014, \pasp, 126, 948, \dodoi{10.1086/678764}

\bibitem[{Vanderburg {et~al.}(2016)Vanderburg, Latham, Buchhave, Bieryla,
  Berlind, Calkins, Esquerdo, Welsh, \& Johnson}]{vanderburg_planetary_2016}
Vanderburg, A., Latham, D.~W., Buchhave, L.~A., {et~al.} 2016, \apjs, 222, 14,
  \dodoi{10.3847/0067-0049/222/1/14}

\bibitem[{Virtanen {et~al.}(2020)Virtanen, Gommers, Oliphant, Haberland, Reddy,
  Cournapeau, Burovski, Peterson, Weckesser, Bright, {van der Walt}, Brett,
  Wilson, Millman, Mayorov, Nelson, Jones, Kern, Larson, Carey, Polat, Feng,
  Moore, {VanderPlas}, Laxalde, Perktold, Cimrman, Henriksen, Quintero, Harris,
  Archibald, Ribeiro, Pedregosa, {van Mulbregt}, \& {SciPy 1.0
  Contributors}}]{2020SciPy-NMeth}
Virtanen, P., Gommers, R., Oliphant, T.~E., {et~al.} 2020, Nature Methods, 17,
  261, \dodoi{10.1038/s41592-019-0686-2}

\bibitem[{{Wakeford} {et~al.}(2018){Wakeford}, {Sing}, {Deming}, {Lewis},
  {Goyal}, {Wilson}, {Barstow}, {Kataria}, {Drummond}, {Evans}, {Carter},
  {Nikolov}, {Knutson}, {Ballester}, \& {Mand ell}}]{wakeford2018}
{Wakeford}, H.~R., {Sing}, D.~K., {Deming}, D., {et~al.} 2018, \aj, 155, 29,
  \dodoi{10.3847/1538-3881/aa9e4e}

\bibitem[{{Winn} {et~al.}(2008){Winn}, {Holman}, {Torres}, {McCullough},
  {Johns-Krull}, {Latham}, {Shporer}, {Mazeh}, {Garcia-Melendo}, {Foote},
  {Esquerdo}, \& {Everett}}]{2008ApJ...683.1076W}
{Winn}, J.~N., {Holman}, M.~J., {Torres}, G., {et~al.} 2008, \apj, 683, 1076,
  \dodoi{10.1086/589737}

\bibitem[{{Ziegler} {et~al.}(2020){Ziegler}, {Tokovinin}, {Brice{\~n}o},
  {Mang}, {Law}, \& {Mann}}]{2020AJ....159...19Z}
{Ziegler}, C., {Tokovinin}, A., {Brice{\~n}o}, C., {et~al.} 2020, \aj, 159, 19,
  \dodoi{10.3847/1538-3881/ab55e9}

\end{thebibliography}

\end{document}